\documentclass[11pt]{article}

\usepackage{fullpage}
\usepackage{comment}
\usepackage{amsthm}
\usepackage{tikz}
\usetikzlibrary{positioning,calc,arrows.meta}
\usepackage{pgfplots}
\pgfplotsset{compat=1.18}
\usetikzlibrary{patterns}
\usepgfplotslibrary{fillbetween}
\usetikzlibrary{intersections}

\usepackage{csquotes}

\usepackage[dvipsnames]{xcolor}

\usepackage{bigints}
\usepackage{ifthen}
\usepackage{amsmath,amssymb}
\usepackage{tcolorbox}

\usepackage{thm-restate}

\usepackage[backend=biber, bibstyle=acmnumeric, citestyle=alphabetic,
  labelalpha=true, sorting=anyt, sortcites=true, natbib=true, isbn=false,
  datamodel=acmdatamodel]{biblatex}
\DeclareFieldFormat{titlecase}{#1}
\AtEveryBibitem{\clearfield{month}}
\AtBeginDocument{\urlstyle{same}}
\renewbibmacro*{organization+location+date}{%
  \printlist{location}%
  \iflistundef{organization}
    {\setunit*{\addcomma\space}}
    {\setunit*{\addcolon\space}}%
  \printlist{organization}}
\DeclareFieldFormat{eprint:arxiv}{%
  arXiv\addcolon
  \href{https://arxiv.org/abs/#1}{\nolinkurl{#1}%
    \iffieldundef{eprintclass}{}{\addspace\mkbibbrackets{\thefield{eprintclass}}}}}
\DeclareNolabel{\nolabel{\regexp{[\p{P}\p{S}\p{C}\s]+}}}
\DeclareFieldFormat{labelalphawidth}{\mkbibbrackets{#1}}
\defbibenvironment{bibliography}
  {\list
     {\printtext[labelalphawidth]{%
        \printfield{labelprefix}%
        \printfield{labelalpha}%
        \printfield{extraalpha}}}
     {\setlength{\labelwidth}{\labelalphawidth}%
      \setlength{\leftmargin}{\labelwidth}%
      \setlength{\labelsep}{\biblabelsep}%
      \addtolength{\leftmargin}{\labelsep}%
      \setlength{\itemsep}{\bibitemsep}%
      \setlength{\parsep}{\bibparsep}}%
      }
  {\endlist}
  {\item}

\usepackage{hyperref}       %
\definecolor{linkblue}{cmyk}{1,0.58,0,0.21}
\hypersetup{colorlinks=true,allcolors=linkblue}
\hypersetup{pdfview=FitH}

\newcommand{\xhdr}[1]{\vspace{2mm}\noindent{\bf {#1}\ }}
\renewcommand{\paragraph}{\xhdr}

\usepackage{url}            %
\usepackage{booktabs}       %
\usepackage{tabularx}
\usepackage{threeparttable}
\usepackage{pdflscape}
\usepackage{amsfonts}       %
\usepackage{nicefrac}       %
\usepackage{microtype}      %
\usepackage{dsfont}

\usepackage{mathtools, bm}

\usepackage{algorithm}

\usepackage[noEnd=true,indLines=true]{algpseudocodex}
\tikzset{algpxIndentLine/.style={draw=black!30,very thin}}
\makeatletter
\renewcommand{\algpx@commentFormat}[1]{{\footnotesize#1}}
\makeatother

\usepackage{subcaption}
\usepackage{multirow}
\usepackage{float}
\usepackage{xspace}

\usepackage{enumitem}

\usepackage{xfrac}

\usepackage{cleveref}
\crefname{enumi}{part}{parts}
\newlist{lemmaitems}{enumerate}{1}
\setlist[lemmaitems]{label=\textup{(\roman*)},ref=\thecorrelatedlemma\,\textup{(\roman*)},leftmargin=*}
\crefname{lemmaitemsi}{Lemma}{Lemmas}
\Crefname{lemmaitemsi}{Lemma}{Lemmas}
\crefname{equation}{eq.}{eqs.}
\Crefname{equation}{Eq.}{Eqs.}

\theoremstyle{plain}

\newtheorem{adversarialtheorem}{Theorem}
\newtheorem{adversariallemma}{Lemma}
\crefname{adversarialtheorem}{theorem}{theorems}
\Crefname{adversarialtheorem}{Theorem}{Theorems}
\crefname{adversariallemma}{lemma}{lemmas}
\Crefname{adversariallemma}{Lemma}{Lemmas}

\newtheorem{correlatedtheorem}[adversarialtheorem]{Theorem}
\newtheorem{correlatedlemma}[adversariallemma]{Lemma}
\crefname{correlatedtheorem}{theorem}{theorems}
\Crefname{correlatedtheorem}{Theorem}{Theorems}
\crefname{correlatedlemma}{lemma}{lemmas}
\Crefname{correlatedlemma}{Lemma}{Lemmas}
\newtheorem{correlatedobservation}{Observation}
\crefname{correlatedobservation}{observation}{observations}
\Crefname{correlatedobservation}{Observation}{Observations}
\newtheorem{correlatedcorollary}{Corollary}

\newtheoremstyle{introexample}%
  {.5\baselineskip plus .2\baselineskip minus .2\baselineskip}%
  {.5\baselineskip plus .2\baselineskip minus .2\baselineskip}%
  {\itshape}{\parindent}{\bfseries}{.}{.5em}{}
\theoremstyle{introexample}
\newtheorem*{example*}{Example}
\newtheorem{example}{Example}
\theoremstyle{plain}
\crefname{correlatedcorollary}{corollary}{corollaries}
\Crefname{correlatedcorollary}{Corollary}{Corollaries}

\allowdisplaybreaks

\usepackage{etoolbox}
\preto\part{\setcounter{section}{0}}

\newcommand{\mr}[1]{\mbox{\footnotesize $\triangleright\;$#1}\quad}
\newcommand{\stepmark}[2]{\overset{\textnormal{\scriptsize(#1)}}{#2}}

\newcommand{\tiH}{T}
\newcommand{\ti}{t}
\newcommand{\intrate}{r_{\tiH}}
\newcommand{\rateexponent}{\theta}

\newcommand{\sval}[1]{s_{#1}}
\newcommand{\bval}[1]{b_{#1}}
\newcommand{\sprice}[1]{p_{#1}}
\newcommand{\bprice}[1]{q_{#1}}
\newcommand{\tradeindicator}[1]{Z_{#1}}
\newcommand{\account}[1]{\Pi_{#1}}              %
\newcommand{\GFT}{\mathsf{GFT}}
\newcommand{\Profit}{\mathsf{PRO}}
\newcommand{\reg}{\mathsf{REG}}
\newcommand{\fixedpricebenchmark}{\GFT^{\star}}      %
\newcommand{\distributionbenchmark}{\GFT^{\star}_{\mathrm{dist}}}    %

\newcommand{\blockgft}[1]{\GFT^{\star}_{#1}}         %
\newcommand{\valuelaw}{\mu}
\newcommand{\history}[1]{\mathcal H_{#1}}
\newcommand{\softO}[1]{\widetilde{\mathcal{O}}\parent{#1}}
\newcommand{\softTheta}[1]{\widetilde{\Theta}\bigl(#1\bigr)}
\newcommand{\Unif}{\operatorname{Unif}}
\newcommand{\Bern}{\operatorname{Bern}}
\newcommand{\Dirac}{\operatorname{Dirac}}

\newcommand{\seedend}{\tau_{1}}
\newcommand{\bridgeend}{\tau_{2}}
\newcommand{\gftmax}{\textnormal{\textsc{GFT-Max}}}
\newcommand{\subsidizedgft}{\textnormal{\textsc{Subsidized-GFT}}}

\newcommand{\advregret}[1]{\reg^{\mathrm{adv}}(#1)}
\newcommand{\advminimaxregret}{\advregret{\intrate}}

\newcommand{\learner}{\mathcal M}

\newcommand{\chenfamilysize}{K}
\newcommand{\chenperturb}{\delta}
\newcommand{\uppersupport}{\mathcal V_{\mathrm{UL}}}
\newcommand{\lowersupport}{\mathcal V_{\mathrm{LR}}}
\newcommand{\cornersupport}{\mathcal V_{\mathrm{cor}}}
\newcommand{\majorityatom}{v_{\mathrm{maj}}}

\newcommand{\actiongrid}{\mathcal A}
\newcommand{\goodset}{\mathcal G}
\newcommand{\badset}{\mathcal B}
\newcommand{\goodrow}[1]{\mathcal G^{#1}}
\newcommand{\infoset}[1]{\mathcal I^{#1}}
\newcommand{\refprob}{\mathrm{Pr}^{\star}}           %

\newcommand{\corrregret}[1]{\reg^{\mathrm{cor}}(#1)}
\newcommand{\corrminimaxregret}{\corrregret{\intrate}}

\newcommand{\revsupport}{\mathcal V_{\mathrm{rev}}}
\newcommand{\revmass}{m}
\newcommand{\revspanindex}{J}

\newcommand{\firstbest}{\mathsf{FB}}
\newcommand{\scoremark}{\mathrm{UB}}               %
\newcommand{\accscale}{\varepsilon}
\newcommand{\hprob}[1]{\widehat{\mathrm{Pr}}[#1]}
\newcommand{\auxprice}{Y}                          %
\newcommand{\gftest}{\textnormal{\textsc{GFT-Est}}}
\newcommand{\gftplusest}{\textnormal{\textsc{GFT}$^{\scoremark}$\textsc{-Est}}}
\newcommand{\safepost}{\textnormal{\textsc{Safe-Post}}}
\newcommand{\Tscore}{\mathcal T_{\mathrm{score}}}
\newcommand{\Tpair}{\mathcal T_{\mathrm{pair}}}
\newcommand{\Texplore}{\mathcal T_{\mathrm{explore}}}
\newcommand{\Texploit}{\mathcal T_{\mathrm{exploit}}}
\newcommand{\Tsave}{\mathcal T_{\mathrm{save}}}
\newcommand{\fbest}{\textnormal{\textsc{FB-Est}}}

\newtheorem{alphatheorem}[adversarialtheorem]{Theorem}
\newtheorem{alphalemma}[adversariallemma]{Lemma}
\crefname{alphatheorem}{theorem}{theorems}
\Crefname{alphatheorem}{Theorem}{Theorems}
\crefname{alphalemma}{lemma}{lemmas}
\Crefname{alphalemma}{Lemma}{Lemmas}
\makeatletter
\renewcommand*{\theHALG@line}{\thealgorithm.\arabic{ALG@line}}
\makeatother
\newcommand{\regretfactor}{\alpha}
\newcommand{\criticalfactor}{\regretfactor_{\star}}
\newcommand{\critfactor}[1]{\bar{\regretfactor}\parent{#1}}
\newcommand{\comparatordist}{\gamma}
\newcommand{\alpharegret}[2]{{#1}\text{-}\reg(#2)}
\newcommand{\needlewidth}{\varepsilon}
\newcommand{\learnrate}{\eta}
\newcommand{\exploreprob}{\varepsilon}

\newcommand{\suffixrounds}{\mathcal T}         %

\definecolor{chenred}{rgb}{1,0,0}
\definecolor{chengreen}{rgb}{0,1,0}
\definecolor{chenbrickred}{cmyk}{0,0.89,0.94,0.28}
\definecolor{chenskyblue}{cmyk}{0.62,0,0.12,0}
\definecolor{chenteal}{rgb}{0,0.5,0.5}
\definecolor{chendarkgray}{rgb}{0.25,0.25,0.25}
\definecolor{chenorange}{rgb}{1,0.5,0}
\definecolor{chenblue}{rgb}{0,0,1}

\definecolor{oursblue}{HTML}{0072B2}
\definecolor{neutralgold}{HTML}{E69F00}

\tikzset{
    alb picture/.style={line cap=round,line join=round},
    chen axis/.style={draw=black,line width=.75pt},
    chen grid/.style={draw=black,dotted,draw opacity=.5,line width=.75pt},
    chen diagonal/.style={draw=black,line width=.75pt},
    chen bad region/.style={fill=chenred,fill opacity=.2,draw=none},
    chen good region/.style={fill=chengreen,fill opacity=.2,draw=none},
    chen upper large/.style={circle,draw=black,fill=black,line width=.75pt,minimum size=4pt,inner sep=0},
    chen lower large/.style={circle,draw=black,fill=white,line width=.75pt,minimum size=4pt,inner sep=0},
    chen corner large/.style={circle,draw=black,fill=chenbrickred,line width=.75pt,minimum size=4pt,inner sep=0},
    chen majority large/.style={circle,draw=black,fill=chenskyblue,line width=.75pt,minimum size=4pt,inner sep=0},
    chen upper small/.style={circle,draw=black,fill=black,line width=.75pt,minimum size=2.4pt,inner sep=0},
    chen lower small/.style={circle,draw=black,fill=white,line width=.75pt,minimum size=2.4pt,inner sep=0},
    chen corner small/.style={circle,draw=black,fill=chenbrickred,line width=.75pt,minimum size=2.4pt,inner sep=0},
    chen majority small/.style={circle,draw=black,fill=chenskyblue,line width=.75pt,minimum size=2.4pt,inner sep=0},
    chen informative/.style={draw=chendarkgray,line width=5pt,draw opacity=.5,line cap=butt},
    chen good line/.style={draw=chenteal,line width=5pt,draw opacity=.75,line cap=butt},
    chen added solid/.style={circle,draw=black,fill=chenblue,line width=.75pt,minimum size=4pt,inner sep=0},
    chen added open/.style={circle,draw=chenblue,fill=white,line width=1pt,minimum size=4pt,inner sep=0},
    chen removed solid/.style={circle,draw=black,fill=chenorange,line width=.75pt,minimum size=4pt,inner sep=0},
    chen removed open/.style={circle,draw=chenorange,fill=white,line width=1pt,minimum size=4pt,inner sep=0},
    chen transfer arrow/.style={->,draw=chendarkgray,line width=1pt,shorten <=3pt,shorten >=3pt},
    chen legend/.style={font=\footnotesize,inner sep=1pt},
    weighted box/.style={draw=oursblue,line width=1pt,rounded corners=2pt,fill=white,inner sep=5pt}
}

\definecolor{corred}{rgb}{1,0,0}
\definecolor{corgreen}{rgb}{0,1,0}
\definecolor{corbrickred}{cmyk}{0,0.89,0.94,0.28}
\definecolor{corskyblue}{cmyk}{0.62,0,0.12,0}
\definecolor{corteal}{rgb}{0,0.5,0.5}
\definecolor{cordarkgray}{rgb}{0.25,0.25,0.25}
\definecolor{corblue}{HTML}{0072B2}
\definecolor{corviolet}{HTML}{7B3294}
\definecolor{corange}{HTML}{E69F00}

\tikzset{
    cor picture/.style={line cap=round,line join=round},
    cor axis/.style={draw=black,line width=.75pt},
    cor grid/.style={draw=black,dotted,draw opacity=.5,line width=.75pt},
    cor diagonal/.style={draw=black,line width=.75pt},
    cor bad region/.style={fill=corred,fill opacity=.2,draw=none},
    cor good region/.style={fill=corgreen,fill opacity=.2,draw=none},
    cor upper large/.style={circle,draw=black,fill=black,line width=.75pt,minimum size=4pt,inner sep=0},
    cor lower large/.style={circle,draw=black,fill=white,line width=.75pt,minimum size=4pt,inner sep=0},
    cor corner large/.style={circle,draw=black,fill=corbrickred,line width=.75pt,minimum size=4pt,inner sep=0},
    cor majority large/.style={circle,draw=black,fill=corskyblue,line width=.75pt,minimum size=4pt,inner sep=0},
    cor upper small/.style={circle,draw=black,fill=black,line width=.75pt,minimum size=2.4pt,inner sep=0},
    cor lower small/.style={circle,draw=black,fill=white,line width=.75pt,minimum size=2.4pt,inner sep=0},
    cor corner small/.style={circle,draw=black,fill=corbrickred,line width=.75pt,minimum size=2.4pt,inner sep=0},
    cor majority small/.style={circle,draw=black,fill=corskyblue,line width=.75pt,minimum size=2.4pt,inner sep=0},
    cor reversal large/.style={rectangle,draw=black,fill=corviolet,line width=.75pt,minimum size=3.6pt,inner sep=0},
    cor reversal small/.style={rectangle,draw=black,fill=corviolet,line width=.6pt,minimum size=2.6pt,inner sep=0},
    cor reversal line/.style={draw=corviolet,line width=.7pt,draw opacity=.75,dash pattern=on 2.4pt off 1.6pt},
    cor span arrow/.style={<->,draw=corviolet,line width=.8pt},
    cor informative/.style={draw=cordarkgray,line width=5pt,draw opacity=.5,line cap=butt},
    cor good line/.style={draw=corteal,line width=5pt,draw opacity=.75,line cap=butt},
    cor plus/.style={circle,draw=black,fill=corblue,line width=.75pt,minimum size=4.4pt,inner sep=0},
    cor minus/.style={circle,draw=black,fill=corange,line width=.75pt,minimum size=4.4pt,inner sep=0},
    cor legend/.style={font=\footnotesize,inner sep=1pt},
    cor box/.style={draw=corblue,line width=1pt,rounded corners=2pt,fill=white,inner sep=5pt}
}

\definecolor{alrred}{rgb}{1,0,0}
\definecolor{alrgreen}{rgb}{0,1,0}
\definecolor{alrblue}{HTML}{0072B2}
\definecolor{alrbrickred}{cmyk}{0,0.89,0.94,0.28}
\definecolor{alrskyblue}{cmyk}{0.62,0,0.12,0}
\definecolor{alrviolet}{HTML}{7B3294}
\definecolor{alrdarkgray}{rgb}{0.25,0.25,0.25}

\tikzset{
    alr picture/.style={line cap=round,line join=round},
    alr axis/.style={draw=black,line width=.75pt},
    alr diagonal/.style={draw=black,line width=.75pt},
    alr bad region/.style={fill=alrred,fill opacity=.2,draw=none},
    alr good region/.style={fill=alrgreen,fill opacity=.2,draw=none},
    alr good region j/.style={fill=alrblue,fill opacity=.22,draw=none},
    alr earning/.style={circle,draw=black,fill=black,line width=.75pt,minimum size=4pt,inner sep=0},
    alr earning j/.style={circle,draw=black,fill=white,line width=.75pt,minimum size=4pt,inner sep=0},
    alr central/.style={circle,draw=black,fill=alrskyblue,line width=.75pt,minimum size=4pt,inner sep=0},
    alr trap/.style={rectangle,draw=black,fill=alrviolet,line width=.75pt,minimum size=3.6pt,inner sep=0},
    alr trap j/.style={rectangle,draw=alrviolet,fill=white,line width=.9pt,minimum size=3.6pt,inner sep=0},
    alr residual/.style={circle,draw=black,fill=alrbrickred,line width=.75pt,minimum size=4pt,inner sep=0},
    alr guide/.style={draw=black!60,line width=.6pt,dash pattern=on 1.8pt off 1.6pt},
    alr dim/.style={|<->|,draw=alrdarkgray,line width=.7pt},
    alr legend/.style={font=\footnotesize,inner sep=1pt}
}

\newcommand{\set}[1]{\left\{#1\right\}}
\newcommand{\parent}[1]{\left(#1\right)}
\newcommand{\card}[1]{\left|#1\right|}
\newcommand{\deq}{\triangleq}

\newcommand{\loitv}[1]{(#1]}
\newcommand{\roitv}[1]{[#1)}
\newcommand{\tinterval}[1]{\left[#1\right]}

\DeclareMathOperator*{\argmax}{arg\,max}
\DeclareMathOperator*{\argmin}{arg\,min}

\newcommand{\condition}{\,\mid\,}

\DeclareMathOperator*{\prb}{\mathrm{Pr}}
\newcommand{\prob}[2][]{\prb\ifthenelse{\not\equal{}{#1}}{\nolimits_{#1}}{}\!\left[{\def\givenn{\middle|}#2}\right]}
\newcommand{\expect}[2][]{\mathbb{E}\ifthenelse{\not\equal{}{#1}}{_{#1}}{}\!\left[{\def\givenn{\middle|}#2}\right]}

\newcommand{\dd}{{\mathrm d}}
\newcommand{\ee}{{\mathrm e}}
\newcommand{\bigO}[2]{\mathcal{O}_{#1}\parent{#2}}
\newcommand{\smallO}[2]{o_{#1}\parent{#2}}
\newcommand{\bO}[1]{\mathcal{O}\parent{#1}}
\newcommand{\btheta}[1]{\Theta\parent{#1}}
\newcommand{\bomega}[1]{\Omega\parent{#1}}

\newcommand{\ceil}[1]{\lceil #1 \rceil}
\newcommand{\floor}[1]{\lfloor #1 \rfloor}
\newcommand{\bceil}[1]{\left\lceil #1 \right\rceil}

\newcommand{\indicator}[1]{{\mathds{1}\left\{ #1 \right\}}}

\newcommand{\plus}[1]{{\left( #1 \right)^+}}

\title{Optimal Online Learning in Bilateral Trade Through
  Interest\thanks{Authors are listed in alphabetical order.}}

\author{%
  Mengfan Ma\thanks{Central China Normal University, China.
    Email: \texttt{mengfanma1@gmail.com}}
  \and
  Haoyu Song\thanks{Purdue University, USA.
    Email: \texttt{song522@purdue.edu}}
  \and
  Yiran Tong\thanks{Central China Normal University, China.
    Email: \texttt{yirantong1@gmail.com}}%
}
\date{}

\begin{document}

\maketitle

\begin{abstract}

We consider online learning in bilateral trade on platforms such as Uber, Airbnb, and eBay.
In each round, a platform posts a payment to a seller and a price to a buyer, both with private valuations, and observes only whether trade occurs.
Its objective is to maximize cumulative gains-from-trade, the sum of the differences between buyers' and sellers' valuations across completed trades.
We measure the platform's performance by its regret against the best fixed common price.
Financial self-sufficiency motivates \emph{global budget balance}~\citep{bernasconi2024}, which allows profits from some trades to subsidize others, provided that the final balance is nonnegative.
Such financial planning should also account for the \emph{time value of money}: profits earned earlier can accumulate interest, providing additional funds for future subsidies.
Motivated by this observation, we introduce \emph{global budget balance with interest}, under which a platform starting with no funds can use both trading profits and accrued interest to finance subsidies.

For a horizon of $\tiH$ rounds and a fixed exponent $\rateexponent\ge0$, we consider the per-round interest rate $\intrate=\tiH^{-\rateexponent}$.
We study two classic valuation models, \emph{adversarial} and \emph{correlated i.i.d.}, and obtain tight bounds on regret against the best fixed common price in both.
As the interest rate grows, optimal regret in each model drops from $\softTheta{\tiH^{3/4}}$ to $\softTheta{\tiH^{5/7}}$, where $\widetilde{\Theta}$ suppresses polylogarithmic factors.
The intermediate rates differ: $\softTheta{\tiH^{3\rateexponent/4}}$ for adversarial valuations when $\tfrac{20}{21}<\rateexponent<1$, and $\softTheta{\tiH^{(2\rateexponent+3)/7}}$ for correlated i.i.d.\ valuations when $1<\rateexponent<\tfrac98$.
Finally, as an extension, we study $\regretfactor$-regret, with approximation factor $\regretfactor\ge1$, for adversarial valuations against the best feasible distribution over price pairs, and show that the optimal competitive ratio rises from $2$ to $3$ as interest strengthens.

\end{abstract}

\newpage
{\hypersetup{linktoc=page}\tableofcontents}

\newpage

\section{Introduction}
\label{sec:intro}
\emph{Bilateral trade} is a classical and fundamental economic model of exchange between a seller who owns an item and a buyer who may wish to acquire it~\citep{myerson1983,chatterjee1983,hagerty1987}.
Each trader has a private valuation: the seller's valuation is the least payment the seller is willing to accept, and the buyer's valuation is the most the buyer is willing to pay.
An intermediary offers a payment to the seller and charges a price to the buyer, and trade occurs when both accept.
The intermediary aims to facilitate transactions by posting prices and payments that enable trade whenever the buyer's valuation exceeds the seller's.

With the growth of the digital economy, bilateral trade takes place repeatedly on platforms like Uber, Airbnb, and eBay.
These platforms act as intermediaries and can collect user data to inform pricing decisions and facilitate future trade, motivating the study of bilateral trade as an \emph{online learning} problem~\citep{cesabianchi2021extended,azar2022,cesabianchi2023colt,bernasconi2024,chen2025,jin2026}.
In online bilateral trade, a new seller--buyer pair with private valuations arrives in each of $\tiH$ rounds.
On each arrival, the platform posts a payment to the seller and a price to the buyer based on feedback from past rounds, and then observes only whether trade occurs.

Financial sustainability is essential for online trading platforms to operate in the long run. To ensure such stability, most earlier work imposes \emph{budget balance}, requiring the payment collected from the buyer to cover the payment made to the seller in \textit{every} round~\citep{cesabianchi2021extended,azar2022,cesabianchi2023colt}.
By completely ruling out non-profitable transactions, this stringent constraint makes sublinear regret impossible under adversarial valuations.
\citet{bernasconi2024} relaxes this requirement to \emph{global budget balance}, which requires only that total revenue from buyers cover total payments to sellers over the \textit{entire} horizon.
Crucially, the platform can then use profits earned in earlier rounds to finance subsidies in later rounds, enabling $\widetilde{O}(\tiH^{3/4})$ regret.
Going one step further, \citet{lunghi2025} allow a controlled \emph{budget violation}, which can be interpreted as an external subsidy or investment to the platform.
They show that a subsidy of order $\tiH^{6/7}$ suffices to further improve the regret to $\widetilde{O}(\tiH^{5/7})$.

Although external subsidies or investments can improve learning performance, online trading platforms may struggle to secure sufficient funding. Even when such funding is available, accepting it may be undesirable due to potentially stringent conditions that could restrict the platform's operational flexibility. It is therefore natural to seek alternative sources of financing. Indeed, the \emph{time value of money} offers such financial flexibility: profits earned from earlier trades can be saved in interest-bearing accounts or invested in interest-bearing assets, allowing interest to accumulate and provide additional funds for future transactions.
Such forward-looking decisions are central to the work recognized by the 2026 Nobel Prize in Economics~\citep{nobel2026}.
The following examples show that interest income can be substantial for online platforms.

\begin{example}
Airbnb collects payments from guests at booking but generally pays hosts only after check-in, earning interest on funds held in the meantime as well as on its own cash and investments. Its interest income increased from $186$ million in 2022 to $721$ million in 2023~\citep{airbnb2023form10k}.
\end{example}

\begin{example}
Shopify enables businesses to run online stores and manage sales and payments. It invests cash reserves in interest-bearing instruments, including short-term securities. In 2025, it reported $331$ million in interest income, compared with $308$ million in 2024~\citep{shopify2025form10k}.
\end{example}

While interest income can be substantial, it remains unclear how to use it effectively to improve the platform's learning performance. The platform must decide not only how much to earn and spend, but also when to do so. Using accumulated profits to finance loss-making trades may immediately increase gains-from-trade, whereas saving or investing them lets interest accumulate and funds more trades in the future. This trade-off is further complicated by limited feedback, as the platform must learn effective prices without observing the agents' valuations or the resulting gains-from-trade. Designing an optimal algorithm therefore requires jointly balancing learning, trading performance, and the timing of financial decisions. With those difficulties in mind, we ask
\begin{quote}
\itshape
How much can interest on the platform's own earnings improve its learning performance without external funding?
How does the best achievable performance depend on the interest rate?
\end{quote}

\subsection{Our Contribution and Techniques}
We first formalize online learning in bilateral trade with interest over a horizon of $\tiH$ rounds.
The platform starts with no funds and receives no external funding.
Following the treatment of interest as accruing between discrete decision dates in \citet[p.~65]{woodford2003interest}, we let the platform's account earn interest between trading rounds at a fixed per-round rate $\intrate$.
The platform may subsidize trades provided that its final account balance is nonnegative with probability one.
For a fixed exponent $\rateexponent\ge0$, we set the per-round interest rate $\intrate\deq\tiH^{-\rateexponent}$ to study how small the rate can be, relative to the horizon $\tiH$, while still improving learning.

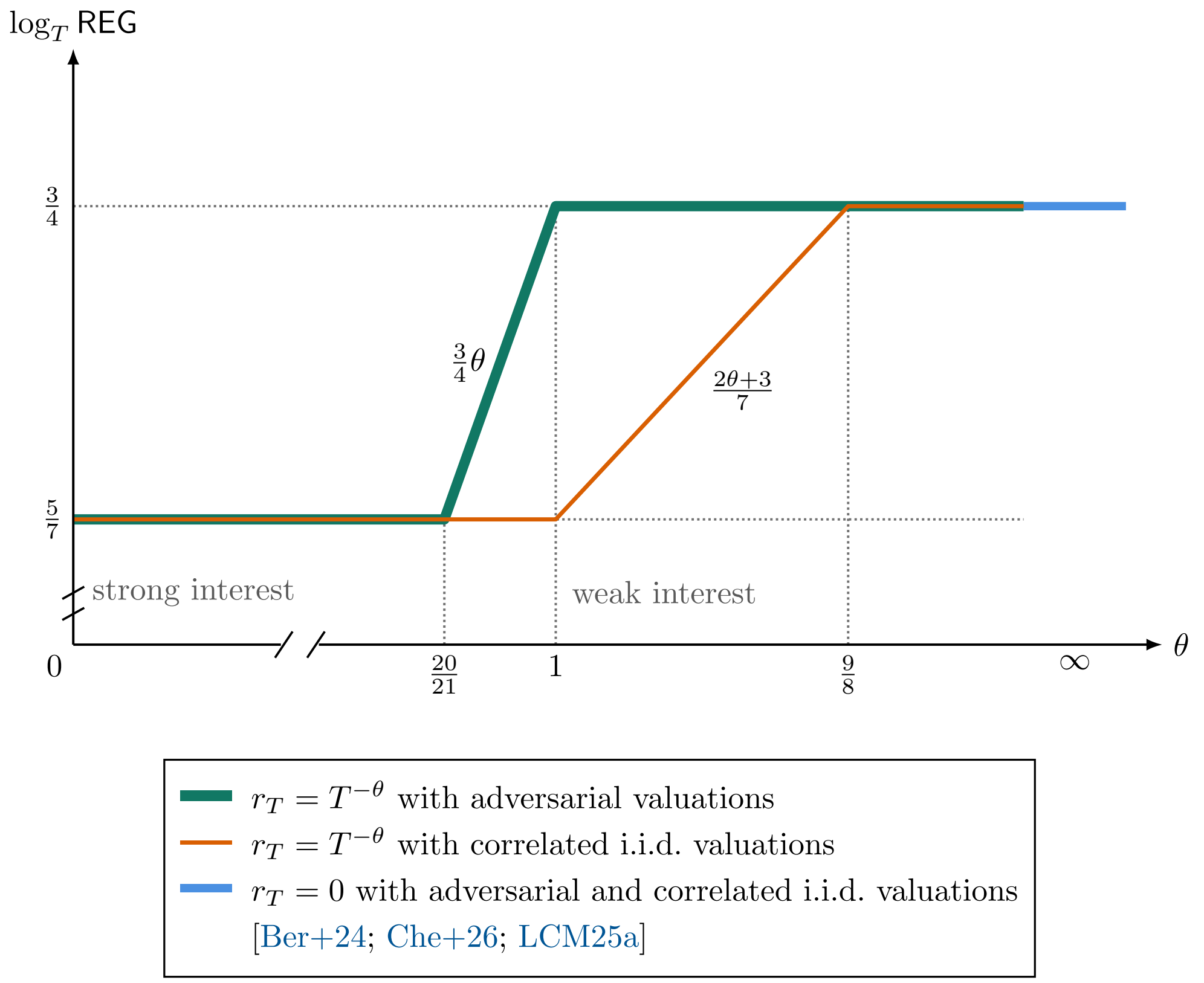
\begin{figure}[t]
\centering
\definecolor{gapAdv}{RGB}{17,120,100}
\definecolor{gapCor}{RGB}{217,95,2}
\definecolor{gapGbb}{RGB}{74,144,226}
\begin{tikzpicture}[line width=0.75pt,font=\small]
\pgfmathsetmacro{\ylow}{0.700}
\pgfmathsetmacro{\yfive}{5/7-\ylow}
\pgfmathsetmacro{\yfour}{3/4-\ylow}
\pgfmathsetmacro{\ytop}{0.7625-\ylow}
\pgfmathsetmacro{\xL}{1.4}
\pgfmathsetmacro{\xR}{1.7}
\pgfmathsetmacro{\sR}{16}
\newcommand{\XR}[1]{\xR+((#1)-0.9)*\sR}
\pgfmathsetmacro{\xtwentyone}{\XR{20/21}}
\pgfmathsetmacro{\xone}{\XR{1}}
\pgfmathsetmacro{\xnineeight}{\XR{9/8}}
\pgfmathsetmacro{\xend}{\XR{1.2}}
\pgfmathsetmacro{\xsa}{\xend}
\pgfmathsetmacro{\xsb}{\xsa+0.7}

\begin{scope}[x=1.6cm,y=96cm]
    \draw[gapGbb,line width=2.6pt] (\xsa,\yfour) -- (\xsb,\yfour);

    \draw[densely dotted,black!55]
        (0,\yfive) -- (\xend,\yfive)
        (0,\yfour) -- (\xend,\yfour);
    \draw[densely dotted,black!55]
        (\xtwentyone,0) -- (\xtwentyone,\yfive)
        (\xone,0) -- (\xone,\yfour)
        (\xnineeight,0) -- (\xnineeight,\yfour);

    \draw[gapAdv,line width=3.2pt,line join=round]
        (0,\yfive) -- (\xtwentyone,\yfive) -- (\xone,\yfour) -- (\xend,\yfour);
    \draw[gapCor,line width=1.3pt,line join=round]
        (0,\yfive) -- (\xone,\yfive) -- (\xnineeight,\yfour) -- (\xend,\yfour);

    \path (\xtwentyone,\yfive) -- (\xone,\yfour)
        node[midway,left,inner sep=4pt]
        {$\tfrac34\rateexponent$};
    \path (\xone,\yfive) -- (\xnineeight,\yfour)
        node[midway,below right,inner sep=2pt]
        {$\tfrac{2\rateexponent+3}{7}$};

    \draw[-{Latex[length=5pt]}] (0,0) -- (0,\ytop+0.0055);
    \draw[-{Latex[length=5pt]}] (0,0) -- (\xsb+0.25,0);
    \draw[line width=0.7pt]
        (-0.075,0.0028) -- (0.075,0.0042)
        (-0.075,0.0052) -- (0.075,0.0066);
    \draw[white,line width=3pt] (\xL+0.02,0) -- (\xR-0.02,0);
    \draw[line width=0.7pt]
        (\xL-0.02,-0.0015) -- (\xL+0.10,0.0015)
        (\xR-0.10,-0.0015) -- (\xR+0.02,0.0015);

    \node[anchor=east,inner sep=3pt] at (0,\yfive) {$\tfrac57$};
    \node[anchor=east,inner sep=3pt] at (0,\yfour) {$\tfrac34$};
    \node[anchor=north east,inner sep=3pt] at (0,0) {$0$};
    \node[anchor=north,inner sep=3pt] at (\xtwentyone,0) {$\tfrac{20}{21}$};
    \node[anchor=north,inner sep=3pt] at (\xone,0) {$1$};
    \node[anchor=north,inner sep=3pt] at (\xnineeight,0) {$\tfrac98$};
    \node[anchor=north,inner sep=3pt] at (0.5*\xsa+0.5*\xsb,0) {$\infty$};

    \node[anchor=south,inner sep=2pt] at (0,\ytop+0.0055)
        {$\log_{\tiH}\reg$};
    \node[anchor=west,inner sep=3pt] at (\xsb+0.25,0)
        {$\rateexponent$};

    \node[anchor=west,inner sep=1pt,text=black!65]
        at (0.10,0.0060) {strong interest};
    \node[anchor=west,inner sep=1pt,text=black!65]
        at (\xone+0.08,0.0060) {weak interest};
\end{scope}

\node[draw,line width=0.6pt,anchor=north,inner sep=5pt] at ({0.8*\xsb},-1.25)
{\renewcommand{\arraystretch}{1.2}%
\begin{tabular}{@{}l@{\hspace{.6em}}l@{}}
    \textcolor{gapAdv}{\rule[.22em]{1.6em}{3.2pt}}
        & $\intrate=\tiH^{-\rateexponent}$ with adversarial valuations\\
    \textcolor{gapCor}{\rule[.3em]{1.6em}{1.3pt}}
        & $\intrate=\tiH^{-\rateexponent}$ with correlated i.i.d.\ valuations\\
    \textcolor{gapGbb}{\rule[.25em]{1.6em}{2.6pt}}
        & $\intrate=0$ with adversarial and correlated i.i.d.\ valuations\\
        & \citep{bernasconi2024,chen2025,lunghi2025}
\end{tabular}};
\end{tikzpicture}
\caption{Exponent of the optimal regret, up to polylogarithmic factors, at
the interest rate $\intrate=\tiH^{-\rateexponent}$
(\Cref{thm:adv-upper,thm:adv-lower} for adversarial and
\Cref{thm:corr-upper,thm:corr-lower} for correlated i.i.d.\ valuations); the
segment at $\infty$ is the setting without interest.}
\label{fig:regret-gap}
\end{figure}

We then answer both questions in the two classic settings, adversarial and correlated i.i.d.\ valuations: for every such interest rate, we determine the optimal regret up to polylogarithmic factors with matching upper and lower bounds (\Cref{thm:adv-upper,thm:adv-lower} for adversarial and \Cref{thm:corr-upper,thm:corr-lower} for correlated i.i.d.\ valuations), and \Cref{fig:regret-gap} shows its exponent.
Interestingly, when the per-round interest rate satisfies $\intrate>\tiH^{-20/21}$, our bounds match the optimal regret $\softO{\tiH^{5/7}}$ attained by \citet{lunghi2025} with external subsidies of $\bO{\tiH^{6/7}}$.
This shows that interest on the platform's own earnings can be as effective as external funding in improving learning.

As an extension, we study the $\regretfactor$-regret for adversarial valuations against the stronger benchmark of \citet{bernasconi2024}, the best feasible distribution over price pairs.
For every interest rate, \Cref{thm:alpha-main} determines the optimal competitive ratio, and in particular closes the gap between $\tfrac{36}{35}$ and $2$ that \citet{bernasconi2024} left open without interest.

For the regret against the best fixed price, our results reveal a gap between the two settings for $\frac{20}{21}<\rateexponent<\frac98$, although they share the optimal regret without interest~\citep{bernasconi2024,chen2025}.
Our techniques rest on one structural observation about the relaxation of budget balance, and we explain the gap by the \emph{stationarity} of correlated i.i.d.\ valuations, which lets the intermediary take fuller advantage of compound interest.

\paragraph{Competing without locating the best price.}
Consider the map that sends each price pair to its gains-from-trade.
Under strong or weak budget balance, every feasible pair is dominated by each common price between its two prices, so no convex combination of their gains-from-trade exceeds those of the best fixed price.
The only way to compete with this benchmark is thus to locate a nearly optimal fixed price, which is essentially impossible without a density assumption~\citep{cesabianchi2021extended}.
Global budget balance lifts this per-round constraint, so the best fixed price no longer dominates every feasible pair: a subsidy pair below the diagonal can gain more than it.
Whenever it does, the gains-from-trade of the best fixed price lie strictly inside the convex hull of the gains-from-trade of feasible pairs, so combining different price pairs instead of locating the best fixed price opens a new route to competing with the benchmark, one that attains sublinear regret even for adversarial valuations~\citep{bernasconi2024}.

\paragraph{Why interest separates the two settings.}
Since every round draws from the same distribution under correlated i.i.d.\ valuations, the map from price pairs to their expected gains-from-trade and profit is \emph{stationary}, staying the same over the whole horizon, but this property is lost under adversarial valuations.
Without interest, this difference is masked by the budget, which weighs a profit equally whenever it is earned: exploiting explored pairs does not accumulate enough funds to lower the regret, so the two settings share the same optimal regret under the global budget balance of \citet{bernasconi2024}.
Sufficient interest changes the picture.
The stochastic intermediary can now keep exploiting the pairs it has explored and let their profit compound to fund subsidized trades with high gains-from-trade in many later rounds.
Lacking this stationarity, the adversarial intermediary cannot rely on the pairs it has explored.
Both settings benefit from the growth of the balance, but only the stochastic one from exploiting explored pairs stably, which is what makes even weak interest useful.

\paragraph{Learning algorithm for adversarial valuations.}
As a warm-up, the adversarial upper bound (\Cref{sec:adv-upper}) shows the power of interest most directly.
Weak interest ($\rateexponent\ge1$) brings almost no gain, and the algorithm $\gftmax$ of \citet{bernasconi2024} is already optimal.
Under strong interest, the intermediary only needs to earn a seed of $\tfrac1\tiH$ at a regret of $\bO{1}$, wait for interest to compound it into a reserve of order $\tiH$, and then spend this reserve on the subsidized algorithm of \citet{lunghi2025}.
The wait can be long, so we run $\gftmax$ during it to keep the regret of these rounds under control.
Running $\gftmax$ throughout would only keep its guarantee of $\softO{\tiH^{3/4}}$, because it ignores the interest that its profit earns.

\paragraph{Learning algorithm for correlated i.i.d.\ valuations.}
For correlated i.i.d.\ valuations (\Cref{sec:corr-upper}), the algorithm exploits stationarity.
Since the regret is measured in expectation, it suffices to combine two pairs over time, posting an earning pair in the first $\lambda\tiH$ rounds and a subsidy pair in the remaining $(1-\lambda)\tiH$ rounds.
\citet{bernasconi2024} find such pairs in two steps, first maximizing the profit and then maximizing the gains-from-trade with the profit collected, which attains regret $\softO{\tiH^{3/4}}$.
Interest lets us do better, because the earning pair serves only to collect money, whereas the subsidy pair carries the high gains-from-trade: the profit of the earning rounds grows before it is spent, so fewer earning rounds fund the same subsidy and more rounds can post the subsidy pair.
To exploit this, our algorithm chooses the two pairs and $\lambda$ jointly in a single optimization problem under global budget balance with interest.
Every candidate pair in this problem needs estimates that cost exploration rounds, yet the feasible set must contain enough candidates to include a good combination, so it has to be built economically.
We achieve this by adaptively growing a forest of binary trees over price intervals along the diagonal, guided by a new quantity \hyperref[eq:corr-gft-ub]{$\GFT^{\scoremark}$} of each interval $\tinterval{p,q}$, and then pairing the bisected intervals into candidates.
This new quantity does double duty (\Cref{lem:corr-ub}), and both duties are essential to building the feasible set economically.
As an upper bound on the gains-from-trade of every pair within the interval, it ensures that the leaf it selects is always worth at least the best fixed price; as a close estimate of the gains-from-trade of the reflected subsidy pair $(q,p)$ once the interval is narrow, it prices subsidy candidates without further exploration.

\paragraph{Hardness of learning.}
Global budget balance helps only through subsidies, so our hard instances make the subsidy region unprofitable.
Both of our lower bounds build on the construction of \citet{chen2025} and follow their proof (\Cref{sec:lower}).
With interest, however, their global price balance~\citep[Lemma~13]{chen2025} holds only with the price difference of round $\ti$ weighted by $(1+\intrate)^{-\ti}$ (\Cref{lem:corr-price-balance}, and \Cref{lem:adv-weighted-price-balance} for our adversarial instances), whereas the regret weights all rounds equally, so the weighted balance can no longer be combined with the regret as in their argument.
We overcome this difficulty differently in the two settings, so that the regret carries the same weights as the balance.
For adversarial valuations, we mix the hard instance in round $\ti$ with a two-point distribution (\Cref{fig:adv-hard-instance}), which scales the regret of that round by $(1+\intrate)^{-\ti}$ (\cref{eq:adv-lb-regret}).
Correlated i.i.d.\ valuations are stationary, so their instance cannot change over the rounds; we instead add points with negative gains-from-trade, which only subsidy pairs trade, and the loss they cause is subtracted from the gains-from-trade of every subsidy pair (\Cref{lem:corr-gaps}), discounting them approximately by the terminal weight $(1+\intrate)^{-\tiH}$ (\Cref{cor:corr-discount}).

\subsection{Related Work}
\paragraph{Bilateral trade.}
Bilateral trade has long been studied in economics as a basic model of exchange between privately informed traders.
A central question is whether such traders can trade efficiently.
\citet{chatterjee1983} show that strategic bidding in a sealed-bid double auction leads to inefficient trade, and \citet{myerson1983} further prove that this inefficiency is unavoidable for incentive-compatible and individually rational mechanisms that run without external subsidies.
Strengthening the incentive requirement of \citet{myerson1983} to dominant strategies, which work whatever the traders believe about each other, \citet{hagerty1987} show that posted prices are essentially the only such mechanisms that balance the budget.
Given these impossibility results, a large literature asks how well fixed prices and other simple mechanisms approximate the optimal gains-from-trade~\citep{blumrosen2016,colinibaldeschi2017,brustle2017,dutting2021,kang2022,deng2022approximately}.

\paragraph{Repeated bilateral trade.}
The repeated problem goes back to online market clearing~\citep{blum2006market}, and regret minimization for fixed-price bilateral trade was initiated by \citet{cesabianchi2021extended}.
Under strong or weak budget balance, the optimal regret has been characterized for stochastic, smoothed, and adversarial valuations under various feedback models~\citep{cesabianchi2021extended,cesabianchi2023colt,azar2022,bacchiocchi2025}, and it is linear for adversarial valuations~\citep{cesabianchi2021extended,azar2022}.
\citet{bernasconi2024} relax the budget constraint of each round to a nonnegative final profit, called global budget balance, which enables sublinear regret for adversarial valuations, and propose a stronger benchmark that compares the algorithm with the best feasible distribution over price pairs.
Against the best fixed price, \citet{chen2025} and \citet{jin2026} determine the tight rates under global budget balance with one- or two-bit feedback and with semi feedback, respectively.
From the perspective of external subsidies, \citet{lunghi2025} further characterize the trade-off between regret and budget violation.
Against the best feasible distribution, no algorithm attains sublinear regret for adversarial valuations~\citep{bernasconi2024}, so subsequent work restricts the valuations and attains sublinear regret for stochastic valuations with a bounded joint density~\citep{lunghi2026benchmark} and for stochastic valuations subject to adversarial corruption~\citep{lunghi2026perturbed}.
\citet{lunghi2025many} extend global budget balance to several buyers, running a second-price auction among them whose outcome is more informative than the feedback of bilateral trade.
Other extensions consider profit maximization~\citep{digregorio2025}, learning simple mechanisms from samples~\citep{babaioff2024}, two-sided markets with several buyers and sellers~\citep{feng2026}, contextual information~\citep{gaucher2025,feng2026}, fairness~\citep{bachoc2024fair}, and brokerage~\citep{bolic2024,bachoc2025parametric,bachoc2025tight}.

\paragraph{Interest and discounting.}
Interest and discounting are central to economic decisions over time, and the forward-looking saving and pricing decisions of households and firms underlie the work recognized by the 2026 Nobel Prize in Economics~\citep{nobel2026,woodford2003interest}.
The role of interest and discounting in bilateral trade has motivated a large body of work.
\citet{rubinstein1982} shows that in alternating-offer bargaining, the traders' discount factors endogenously determine the division of the surplus in the unique subgame-perfect equilibrium.
Beyond a single negotiation, when the same traders trade repeatedly, a high discount factor allows the budget to balance across periods, which can overcome the static impossibility of \citet{myerson1983} and realize efficient trade~\citep{athey2007efficiency,athey2013efficient,skrzypacz2015mechanisms}.
Discounting also appears throughout sequential decision making, including dynamic programming~\citep{bellman1957dynamic,howard1960dynamic,blackwell1965discounted}, Markov decision processes~\citep{puterman1994markov}, multi-armed bandits~\citep{gittins1979bandit,whittle1980multi}, reinforcement learning~\citep{sutton2018reinforcement,azar2013minimax}, and online pricing with strategic buyers~\citep{amin2013learning}.

\section{Preliminaries}
\label{sec:prelim}
We study repeated bilateral trade over a horizon of $\tiH$ rounds.  In each
round $\ti\in[\tiH]$, a new seller and a new buyer arrive with private
valuations $\sval{\ti}\in\tinterval{0,1}$ and $\bval{\ti}\in\tinterval{0,1}$
for an indivisible item.  We consider two classic models of valuations:
adversarial valuations and stochastic valuations.

For \emph{adversarial valuations}, we follow the convention of a randomized
oblivious adversary, which draws the whole sequence
$(\sval{\ti},\bval{\ti})_{\ti=1}^{\tiH}$ before the first round, with no restrictions on its joint distribution.

For \emph{stochastic valuations}, we consider \emph{correlated i.i.d.\
valuations}: the pairs $(\sval{\ti},\bval{\ti})$ are drawn i.i.d.\ from
an arbitrary unknown distribution $\valuelaw$ on
$\tinterval{0,1}^2$, under which the two valuations may be correlated.

\paragraph{Notation.}
For a positive integer $k$, let $[k]\deq\set{1,\ldots,k}$.  For a set
$A$, $\Unif(A)$ denotes the uniform distribution on $A$.  We write
$\plus{x}\deq\max\set{x,0}$, and $\softO{\cdot}$ hides polylogarithmic
factors in $\tiH$.

\paragraph{Learning protocol.}
In each round $\ti$, the platform posts a seller price
$\sprice{\ti}\in\tinterval{0,1}$ and a buyer price
$\bprice{\ti}\in\tinterval{0,1}$.  The seller accepts if and only if
$\sval{\ti}\le\sprice{\ti}$, the buyer accepts if and only if
$\bprice{\ti}\le\bval{\ti}$, and trade occurs if and only if both accept.
The platform observes only whether trade occurs,
$\tradeindicator{\ti}\deq\indicator{\sval{\ti}\le\sprice{\ti},\
\bprice{\ti}\le\bval{\ti}}$, which is known as \emph{one-bit feedback}.
The platform knows $\tiH$ and the interest rate defined below, and chooses
$(\sprice{\ti},\bprice{\ti})$ by an algorithm that uses the past prices and feedback.

When trade occurs, the platform pays the seller price to the seller and receives
the buyer price from the buyer.  For a price pair
$(p,q)\in\tinterval{0,1}^2$, the gains-from-trade and the profit of round
$\ti$ are therefore
\begin{gather*}
    \GFT_{\ti}(p,q)\deq(\bval{\ti}-\sval{\ti})\,\indicator{\sval{\ti}\le p,\ q\le\bval{\ti}},
    \\*
    \Profit_{\ti}(p,q)\deq(q-p)\,\indicator{\sval{\ti}\le p,\ q\le\bval{\ti}}.
\end{gather*}
We call $(p,q)$ an \emph{earning pair} if $p\le q$ and a \emph{subsidy pair}
if $p>q$.

\paragraph{Budget balance with interest.}
The platform keeps its profit in an account that earns interest at a per-round
rate $\intrate\ge0$.  Starting from $\account{1}=0$, the account evolves as
$\account{\ti+1}\deq(1+\intrate)\account{\ti}+\Profit_{\ti}(\sprice{\ti},\bprice{\ti})$,
so its final value is
\[
    \textstyle\account{\tiH+1}=\sum_{\ti=1}^{\tiH}(1+\intrate)^{\tiH-\ti}
      \Profit_{\ti}(\sprice{\ti},\bprice{\ti}).
\]
An algorithm is \emph{feasible} if it satisfies \emph{global budget balance with
interest}: $\account{\tiH+1}\ge0$ almost surely.  Without interest
($\intrate=0$), this is the global budget balance of \citet{bernasconi2024}:
subsidies are allowed as long as the total profit is nonnegative.  With
interest, a unit of profit earned in round $\ti$ pays for
$(1+\intrate)^{\tiH-\ti}$ units of subsidy at the end, so early profit pays
for more of the later subsidies.  We parametrize the rate as
$\intrate=\tiH^{-\rateexponent}$ with $\rateexponent\ge0$, and show in the
following sections how it affects learning.  An important threshold is
$\rateexponent=1$: the total growth $(1+\intrate)^{\tiH}$ of the account
tends to $1$ for $\rateexponent>1$, to $\ee$ for $\rateexponent=1$, and
to infinity for $0\le\rateexponent<1$.

\paragraph{Benchmarks and regret.}
We measure the performance of an algorithm by its regret against the best
fixed price.  It suffices to consider common prices: for $p\le z\le q$, every
trade at $(p,q)$ also occurs at $(z,z)$ and generates the same gains-from-trade.  For adversarial valuations, the benchmark is the best fixed price in
hindsight, whose gains-from-trade per round are
$\fixedpricebenchmark\deq\max_{z\in\tinterval{0,1}}\tfrac1{\tiH}\sum_{\ti=1}^{\tiH}\GFT_{\ti}(z,z)$,
and the regret of an algorithm is
\[
    \textstyle\advminimaxregret\deq\expect{\tiH\cdot\fixedpricebenchmark
      -\sum_{\ti=1}^{\tiH}\GFT_{\ti}(\sprice{\ti},\bprice{\ti})},
\]
where the expectation is over the valuations and the randomness of the
algorithm.  Given the valuation sequence, the best fixed price is easy to
compute: $\GFT_{\ti}(z,z)=(\bval{\ti}-\sval{\ti})\indicator{\sval{\ti}\le z\le\bval{\ti}}$,
so a sweep over $\tinterval{0,1}$ that adds $\bval{\ti}-\sval{\ti}$ at
$\sval{\ti}$ and subtracts it after $\bval{\ti}$ finds the maximizer.  For
correlated i.i.d.\ valuations, the benchmark is computed directly from
$\valuelaw$ as
$\fixedpricebenchmark\deq\max_{z\in\tinterval{0,1}}\expect{\GFT_{\ti}(z,z)}$,
and the regret of an algorithm is
\[
    \textstyle\corrminimaxregret\deq\tiH\cdot\fixedpricebenchmark
      -\expect{\sum_{\ti=1}^{\tiH}\GFT_{\ti}(\sprice{\ti},\bprice{\ti})}.
\]
Putting these together, the problem we study is
\[
    \begin{array}{@{}l@{}}
    \displaystyle\min\,\sup_{(\sval{\ti},\bval{\ti})_{\ti=1}^{\tiH}}\
    \textstyle\reg=\expect{\tiH\cdot\fixedpricebenchmark-\sum_{\ti=1}^{\tiH}\GFT_{\ti}(\sprice{\ti},\bprice{\ti})}
    \\[8pt]
    \text{s.t.}\ \
    \account{\tiH+1}=\sum_{\ti=1}^{\tiH}(1+\intrate)^{\tiH-\ti}\Profit_{\ti}(\sprice{\ti},\bprice{\ti})\ge0\ \text{a.s.},
    \end{array}
\]
where the minimum is over algorithms.  For correlated i.i.d.\ valuations,
$(\sval{\ti},\bval{\ti})\overset{\text{i.i.d.}}{\sim}\valuelaw$ and the
supremum is over $\valuelaw$; for
adversarial valuations, the sequence $(\sval{\ti},\bval{\ti})_{\ti=1}^{\tiH}$
is fixed before the algorithm runs.

\phantomsection\label{toc:adversarial-upper-bound}
\section{Warm-Up: Learning Algorithm for Adversarial Valuations}
\label{sec:adv-upper}
We begin with adversarial valuations, where interest helps in the most direct
way.  The following theorem gives the regret of \Cref{alg:adv-upper} at every
interest rate; together with the lower bound of \Cref{thm:adv-lower} in
\Cref{sec:adv-lower}, it determines the optimal regret up to polylogarithmic
factors.

\begin{adversarialtheorem}[Adversarial upper bound]
\label{thm:adv-upper}
\begin{flushleft}
For adversarial valuations under global budget balance with interest rate
$\intrate=\tiH^{-\rateexponent}$,
\Cref{alg:adv-upper} satisfies
\[
    \advminimaxregret
    =
    \begin{cases}
        \softO{\tiH^{\frac57}},
            & \rateexponent\le\frac{20}{21},\\[1mm]
        \softO{\tiH^{\frac{3\rateexponent}4}},
            & \frac{20}{21}<\rateexponent<1,\\[1mm]
        \softO{\tiH^{\frac34}},
            & \rateexponent\ge1.
    \end{cases}
\]
\end{flushleft}
\end{adversarialtheorem}

\paragraph{Overview.}
Interest lets a small profit grow, so the algorithm only needs to earn a little
at the start.  \Cref{alg:adv-upper} first posts random earning pairs of spread
$\frac1\tiH$ until one of them trades, which collects a profit of
$\frac1\tiH$ at a regret of $\bigO{}{1}$.  Over the next
$\ceil{4\log\tiH/\intrate}$ rounds, interest grows this profit to a balance of
at least $\tiH$, enough to fund an algorithm that may violate the budget.
\citet[Algorithm~5]{lunghi2025} give such an algorithm: with $\beta=\frac67$ and
$\delta=\frac1\tiH$, it attains regret $\softO{\tiH^{5/7}}$ at the price of a
budget violation of at most $\tiH^{6/7}$, which the balance covers.  We call it
$\subsidizedgft$ and run it on the remaining rounds.  The waiting rounds,
however, need their regret controlled: there are $\softTheta{\intrate^{-1}}$
of them, so without such control, the regret could become nearly linear as
$\rateexponent$ approaches $1$.  At the same time, they must protect the
balance, since a single subsidy could wipe out the initial profit before it has grown.  The
algorithm of \citet[Algorithm~1]{bernasconi2024}, configured for one-bit feedback
as in their Theorem~5.4, meets both requirements: it attains regret
$\softO{\tiH^{3/4}}$ under global budget balance, with nonnegative profit on
every prefix.  We use it as a subroutine, which we call $\gftmax$, so the
waiting rounds cost only $\softO{\intrate^{-3/4}}$ regret, and the bound of
\Cref{thm:adv-upper} follows for $\rateexponent<1$.  Below the threshold
$\intrate=\frac1\tiH$, interest cannot turn the initial profit into a balance of $\tiH$
within the horizon, and the algorithm simply runs $\gftmax$ on all rounds.  By
\Cref{thm:adv-lower}, no algorithm does better there up to polylogarithmic
factors, so the interest rate must reach this threshold before it changes the
market.

\begin{algorithm}[H]
\caption{\textsc{Interest-Financed Switching}}
\label{alg:adv-upper}
\begin{algorithmic}[1]
\linespread{0.9}\selectfont
    \If{$\intrate\le\tfrac{1}{\tiH}$}
        \State run $\gftmax$ on all rounds and stop
    \EndIf
    \State $\seedend\gets\tiH$
    \For{$\ti=1,\ldots,\tiH$}
        \State draw $\auxprice_{\ti}\sim\Unif\set{0,\ldots,\tiH-1}$ and post
            $(\tfrac{\auxprice_{\ti}}{\tiH},\tfrac{\auxprice_{\ti}+1}{\tiH})$
        \If{$\tradeindicator{\ti}=1$}
            \State $\seedend\gets\ti$
            \State \textbf{break}
        \EndIf
    \EndFor
    \State $\bridgeend\gets\seedend+\bigl\lceil\tfrac{4\log\tiH}{\intrate}\bigr\rceil$
    \If{$\bridgeend\ge\tiH$}
        \State run $\gftmax$ on rounds $\seedend+1,\ldots,\tiH$
    \Else
        \State run $\gftmax$ on rounds $\seedend+1,\ldots,\bridgeend$
        \State run $\subsidizedgft$ on rounds $\bridgeend+1,\ldots,\tiH$
    \EndIf
\end{algorithmic}
\end{algorithm}

\begin{proof}[Proof of \Cref{thm:adv-upper}]
If $\intrate\le1/\tiH$,
\Cref{alg:adv-upper} runs $\gftmax$ on all rounds,
and \citet[Theorem~5.4]{bernasconi2024} gives
$\advminimaxregret=\softO{\tiH^{3/4}}$.

We now analyze $\intrate>1/\tiH$, and first bound the regret over rounds
$\ti\le\seedend$ against the best fixed price on them.  A trade at round
$\ti$ collects the whole surplus $\plus{\bval{\ti}-\sval{\ti}}$, so this
regret is at most
\begin{align}
\expect{\textstyle\sum_{\ti\le\seedend}\plus{\bval{\ti}-\sval{\ti}}
    \parent{1-\tradeindicator{\ti}}}
&\textstyle=\sum_{\ti\in\tinterval{\tiH}}
    \prob{\seedend\ge\ti,\ \tradeindicator{\ti}=0}\plus{\bval{\ti}-\sval{\ti}}
\notag
\\
&\textstyle\le\sum_{\ti\in\tinterval{\tiH}}
    \prob{\seedend\ge\ti}\plus{\bval{\ti}-\sval{\ti}}
\notag
\\
&\textstyle\le\sum_{\ti\in\tinterval{\tiH}}
    \prob{\seedend\ge\ti}\bigl(
    \prob{\tradeindicator{\ti}=1\condition\seedend\ge\ti}+\tfrac2\tiH\bigr)
\notag
\\
&\textstyle=\sum_{\ti\in\tinterval{\tiH}}
    \prob{\seedend\ge\ti,\ \tradeindicator{\ti}=1}
    +\tfrac2\tiH\sum_{\ti\in\tinterval{\tiH}}\prob{\seedend\ge\ti}
\notag
\\
\mr{$\sum_{\ti\in\tinterval{\tiH}}\prob{\seedend\ge\ti,\ \tradeindicator{\ti}=1}\le1$}
&\textstyle\le3.
\label{eq:adv-seed-regret}
\end{align}

If $\bridgeend\ge\tiH$, which includes the case that the first phase never
trades, only $\gftmax$ runs after the first phase, and the bounds below hold
with the third phase empty; we may thus assume $\bridgeend<\tiH$.  Then the
first phase ends with a trade, so $\account{\seedend+1}=1/\tiH$.  \citet[Theorem~5.4 and its proof]{bernasconi2024}
show that $\gftmax$ has regret $\softO{(\bridgeend-\seedend)^{3/4}}$ on
rounds $\seedend+1,\ldots,\bridgeend$ and nonnegative profit on every prefix
of them.  The balance therefore compounds over the second phase to
$\account{\bridgeend+1}\ge(1+\intrate)^{\bridgeend-\seedend}/\tiH
\ge\ee^{\intrate(\bridgeend-\seedend)/2}/\tiH\ge\tiH$.
On the remaining rounds, \citet[Theorem~4.15]{lunghi2025} show that $\subsidizedgft$ violates the
budget by at most $\tiH^{6/7}$ and, except with probability $\tfrac3\tiH$, has
realized regret $\softO{\tiH^{5/7}}$, which on a fixed valuation sequence is
equivalent to the same bound on the expected regret.  The balance $\tiH$ covers the
violation, which guarantees global budget balance.

Splitting the best fixed price over the three phases and combining
\cref{eq:adv-seed-regret} with these two guarantees,
\begin{align*}
\advminimaxregret
&\textstyle\le3+\softO{(\bridgeend-\seedend)^{3/4}}+\softO{\tiH^{5/7}}
    +2\tiH\cdot\tfrac3\tiH
\\
&\textstyle=\softO{\tiH^{\max\set{3\rateexponent/4,\,5/7}}},
\end{align*}
which is $\softO{\tiH^{5/7}}$ for $\rateexponent\le20/21$ and
$\softO{\tiH^{3\rateexponent/4}}$ for $20/21<\rateexponent<1$.  Together
with the case $\intrate\le1/\tiH$, this proves \Cref{thm:adv-upper}.
\end{proof}

\section{Learning Algorithm for Correlated i.i.d.\ Valuations}
\label{sec:corr-upper}

We now turn to correlated i.i.d.\ valuations and see how stationarity makes
the difference in the optimal regret.  The following theorem gives the regret
of \Cref{alg:corr-upper} at every interest rate; together with the lower bound
of \Cref{thm:corr-lower} in \Cref{sec:corr-lower}, it determines the optimal
regret up to polylogarithmic factors.

\begin{correlatedtheorem}[Correlated i.i.d.\ upper bound]
\label{thm:corr-upper}
\begin{flushleft}
For correlated i.i.d.\ valuations under global budget balance with interest
rate $\intrate=\tiH^{-\rateexponent}$,
\Cref{alg:corr-upper} satisfies
\[
    \corrminimaxregret
    =
    \begin{cases}
        \softO{\tiH^{\frac57}},
            & \rateexponent\le1,\\[1mm]
        \softO{\tiH^{\frac{2\rateexponent+3}7}},
            & 1<\rateexponent<\frac98,\\[1mm]
        \softO{\tiH^{\frac34}},
            & \rateexponent\ge\frac98.
    \end{cases}
\]
\end{flushleft}
\end{correlatedtheorem}
We explain below how \Cref{alg:corr-upper} attains this bound, and then
analyze its regret.

\paragraph{Exploiting stationarity.}
Subsidy pairs let the algorithm approach the gains-from-trade of the best price
without locating it.  Since the regret is measured in expectation, it suffices
to post an earning pair $(p_E,q_E)$ in the first $\lambda\tiH$ rounds and a subsidy pair
$(p_S,q_S)$ in the remaining $(1-\lambda)\tiH$ rounds, where, for a target
accuracy $\accscale$ set below, $\lambda\in\tinterval{0,1}$ satisfies
\begin{gather*}
\textstyle\fixedpricebenchmark-\lambda\expect{\GFT_{\ti}(p_E,q_E)}
-(1-\lambda)\expect{\GFT_{\ti}(p_S,q_S)}\le\accscale,
\\
\textstyle\mathbb E\bigl[\sum_{\ti\le\lambda\tiH}(1+\intrate)^{-\ti}\Profit_{\ti}(p_E,q_E)
+\sum_{\ti>\lambda\tiH}(1+\intrate)^{-\ti}\Profit_{\ti}(p_S,q_S)\bigr]\ge0.
\end{gather*}
The first condition bounds the regret by $\accscale\tiH$, and the second
states that the expected discounted account is nonnegative; since the earning
rounds come first, their profit carries the larger weights.  Rather than first
maximizing the profit and then the gains-from-trade, as \citet{bernasconi2024}
do for $\intrate=0$, our algorithm chooses the earning pair, the subsidy pair,
and $\lambda$ jointly, maximizing the gains-from-trade subject to global budget
balance with interest.  Since profit earned early accrues interest before it
funds the subsidy, the gains-from-trade given up per unit of subsidy fall as the
rate grows, so the cheaper the subsidy, the earlier the algorithm can stop
learning and commit to its two pairs.

\paragraph{Overview of the algorithm.}
We measure the surplus available in the market by the first-best gains-from-trade
\[
\textstyle\firstbest\deq\expect[(s,b)\sim\valuelaw]{\plus{b-s}},
\]
which clearly satisfy $\fixedpricebenchmark\le\firstbest$.
\Cref{alg:corr-upper} begins by estimating
$\firstbest$ and, when the estimate exceeds $\accscale$, runs an adaptive
bisection over the grid of $K=\ceil{\widehat\firstbest^{-1}}$ intervals.  Each
iteration bisects the leaf $\tinterval{p^\star,q^\star}$ of $\mathcal A$ with the
largest $\widehat{\GFT}{}^{\scoremark}$, defined below, and records it in the
collection $\mathcal S_{q^\star-p^\star}$ of bisected intervals of the same
width.  If this leaf is already narrower than
$\accscale\widehat\firstbest$, the algorithm instead posts the subsidy price pair
$(q^\star,p^\star)$ from then on.  Once enough intervals of one width have been bisected,
\Cref{alg:corr-pair-intervals} pairs them into candidate earning pairs
$\mathcal Q_E$ and subsidy pairs $\mathcal Q_S$.  Taking these candidates as
the feasible set, the algorithm solves the joint problem above on the remaining
rounds, with estimated gains-from-trade and profits, to obtain the two pairs and
the round $t_1$ at which it switches from earning to subsidizing.  It then
posts $(p_E,q_E)$ before round $t_1$ and $(p_S,q_S)$ from round $t_1$ on.

\paragraph{A core quantity.}
The core quantity of the algorithm is, for an interval $I=\tinterval{p,q}$,%
\phantomsection\label{eq:corr-gft-ub}
\[
\textstyle\GFT^{\scoremark}(I)\deq\expect{\GFT_{\ti}(p,p)}+\expect{\GFT_{\ti}(q,q)}
-\expect{\GFT_{\ti}(p,q)}+(q-p)\prob{p<s\le q,\ p\le b<q},
\]
which $\gftplusest(I,\widehat\firstbest,k)$ estimates.  We first state three
properties of $\GFT^{\scoremark}$.

\begin{restatable}{correlatedlemma}{corrub}
\label{lem:corr-ub}
\begin{flushleft}
Every interval $\tinterval{p,q}$ satisfies the following:
\begin{lemmaitems}
\item\label{prop:corr-ub-upper}
\emph{Upper bound.}
$\sup_{u,v\in\tinterval{p,q}}\expect{\GFT_{\ti}(u,v)}
\le\GFT^{\scoremark}(\tinterval{p,q})$.
\item\label{prop:corr-ub-nested}
\emph{Monotonicity.}
$\GFT^{\scoremark}(\tinterval{u,v})\le\GFT^{\scoremark}(\tinterval{p,q})$
for every interval $\tinterval{u,v}\subseteq\tinterval{p,q}$.
\item\label{prop:corr-ub-cover}
\emph{Tightness.}
$0\le\GFT^{\scoremark}(\tinterval{p,q})-\expect{\GFT_{\ti}(q,p)}
\le2(q-p)\prob{p<s\le q,\ p\le b<q}$.
\end{lemmaitems}
\end{flushleft}
\end{restatable}
We prove \Cref{lem:corr-ub} in \Cref{app:corr-upper}.

Each estimate costs exploration rounds, so we ask why this design is
efficient.  Part of the answer
lies in the bisection, which learns along the diagonal: as a leaf narrows, its
subsidy pair $(q,p)$ approaches a common price.  Since $\GFT^{\scoremark}$
bounds the gains-from-trade of every pair in $I\times I$ (\Cref{prop:corr-ub-upper})
and the leaves cover $\tinterval{0,1}$, the leaf with the largest
$\GFT^{\scoremark}$ always carries at least $\fixedpricebenchmark$, and bisecting
it never raises its value (\Cref{prop:corr-ub-nested}).  Once this leaf is narrower
than $\accscale\widehat\firstbest$, nothing more needs to be learned, because its
subsidy pair gains at most $2(q-p)$ less than $\GFT^{\scoremark}$
(\Cref{prop:corr-ub-cover}) and each of its trades costs a subsidy of at most
$\accscale\widehat\firstbest$, which few saving rounds of $\safepost$ fund.

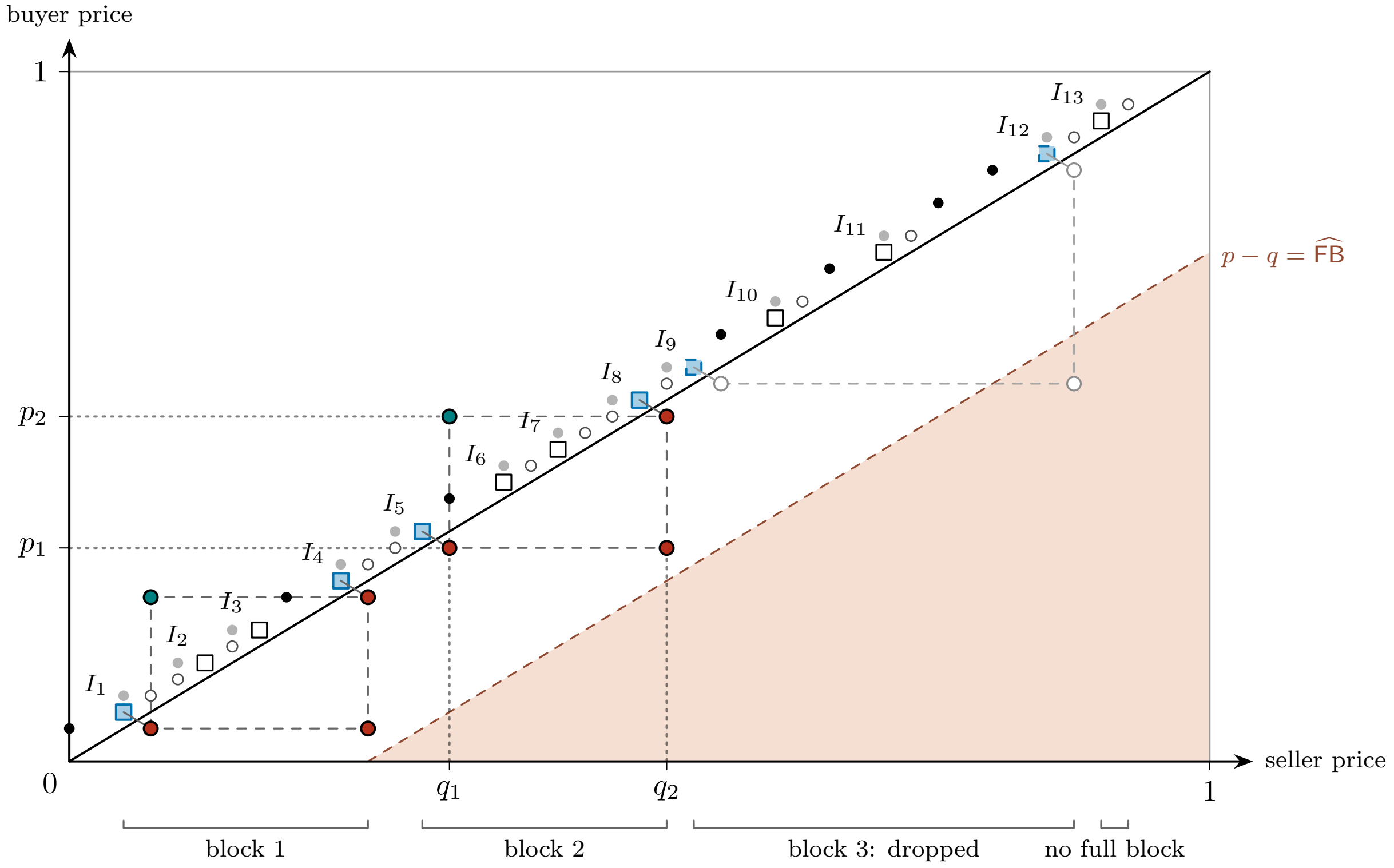
\begin{figure}[!ht]
\centering
\begin{tikzpicture}[cor picture,font=\small,x=.314cm,y=.19cm]
\tikzset{
    qe in pair/.style={draw=corblue,fill=corblue!35,line width=.75pt},
    qe in block/.style={draw=black,fill=white,line width=.6pt},
    qe pair dropped/.style={qe in pair,dashed},
    qe earn/.style={circle,draw=black,fill=corteal,line width=.75pt,minimum size=4.5pt,inner sep=0},
    qe subsidy/.style={circle,draw=black,fill=corbrickred,line width=.75pt,minimum size=4.5pt,inner sep=0},
    qe would be/.style={circle,draw=black!45,fill=white,line width=.6pt,minimum size=4.5pt,inner sep=0},
    qe bracket/.style={draw=black!60,line width=.6pt},
    qe point/.style={rectangle,minimum size=5pt,inner sep=0},
    qe bisected/.style={circle,fill=black!30,minimum size=3.5pt,inner sep=0},
    qe unbisected/.style={circle,fill=black,minimum size=3.5pt,inner sep=0},
    qe child/.style={circle,draw=black!70,fill=white,line width=.5pt,minimum size=3.5pt,inner sep=0},
    qe rectangle/.style={draw=black!60,line width=.6pt,dashed},
    qe reflect/.style={draw=black!60,line width=.6pt}
}

\fill[corbrickred!12] (11,0) -- (42,0) -- (42,31) -- cycle;
\draw[corbrickred!70!black,line width=.6pt,dashed] (11,0) -- (42,31)
    node[right,font=\scriptsize,text=corbrickred!70!black] {$p-q=\widehat\firstbest$};
\draw[black!40,line width=.5pt] (0,0) rectangle (42,42);
\draw[cor grid] (14,0) -- (14,13);
\draw[cor grid] (22,0) -- (22,13);
\draw[cor grid] (0,13) -- (14,13);
\draw[cor grid] (0,21) -- (14,21);
\draw[qe rectangle] (3,2) rectangle (11,10);
\draw[qe rectangle] (14,13) rectangle (22,21);
\draw[black!35,line width=.6pt,dashed] (24,23) -- (37,23) -- (37,36);
\draw[cor axis,-Stealth] (0,0) -- (43.6,0) node[right,font=\scriptsize] {seller price};
\draw[cor axis,-Stealth] (0,0) -- (0,44) node[above,font=\scriptsize] {buyer price};
\foreach \x/\lab in {14/q_1,22/q_2,42/1}{
    \draw (\x,0) -- (\x,-.5) node[below] {$\lab$};
}
\foreach \y/\lab in {13/p_1,21/p_2,42/1}{
    \draw (0,\y) -- (-.35,\y) node[left] {$\lab$};
}
\node[below left] at (0,0) {$0$};
\draw[cor diagonal] (0,0) -- (42,42);
\foreach \a/\b/\lab in {2/11/{block 1},13/22/{block 2},
                        23/37/{block 3: dropped},
                        38/39/{no full block}}{
    \draw[qe bracket] (\a,-3.6) -- ++(0,-.45) -- (\b,-4.05) -- ++(0,.45);
    \node[font=\scriptsize,below] at ({(\a+\b)/2},-4.05) {\lab};
}
\foreach \p in {2,4,6,10,12,16,18,20,22,26,30,36,38}{
    \node[qe bisected] at (\p,\p+2) {};
}
\foreach \p in {0,8,14,24,28,32,34}{
    \node[qe unbisected] at (\p,\p+2) {};
}
\foreach \p in {3,4,6,11,12,17,19,20,22,27,31,37,39}{
    \node[qe child] at (\p,\p+1) {};
}
\foreach \k/\x/\sty in {1/2/qe in pair,2/5/qe in block,3/7/qe in block,
                        4/10/qe in pair,5/13/qe in pair,6/16/qe in block,
                        7/18/qe in block,8/21/qe in pair,9/23/qe pair dropped,
                        10/26/qe in block,11/30/qe in block,12/36/qe pair dropped,
                        13/38/qe in block}{
    \node[\sty,qe point] at (\x,\x+1) {};
    \node[font=\scriptsize,anchor=south east] at (\x-.15,\x+1.35) {$I_{\k}$};
}
\draw[qe reflect] (2,3) -- (3,2);
\draw[qe reflect] (10,11) -- (11,10);
\draw[qe reflect] (13,14) -- (14,13);
\draw[qe reflect] (21,22) -- (22,21);
\draw[qe reflect,draw=black!40] (23,24) -- (24,23);
\draw[qe reflect,draw=black!40] (36,37) -- (37,36);
\foreach \lo/\hi in {2/10,13/21}{
    \node[qe earn]    at (\lo+1,\hi) {};
    \node[qe subsidy] at (\lo+1,\lo) {};
    \node[qe subsidy] at (\hi+1,\hi) {};
    \node[qe subsidy] at (\hi+1,\lo) {};
}
\foreach \x/\y in {24/23,37/36,37/23}{
    \node[qe would be] at (\x,\y) {};
}
\end{tikzpicture}
\caption{Pairing intervals into candidate price pairs.}
\label{fig:corr-pair-intervals}
\end{figure}

\paragraph{An economical feasible set.}
The heart of the efficiency, however, lies in the feasible set of the joint
problem.  Until the algorithm solves this problem, no phase is devoted to
collecting profit; which pair earns, which pair subsidizes, and when to switch
are all decided jointly in it, so its feasible set must be rich enough to
contain a good combination of the two pairs.  Yet
every candidate needs estimates of its gains-from-trade and its profit, so a
naive feasible set would cost many exploration rounds.  Ours is economical, as
\Cref{fig:corr-pair-intervals} shows.  For two kept intervals
$\tinterval{p_1,q_1}$ and $\tinterval{p_2,q_2}$, the candidates are the four
corners of the rectangle $\set{q_1,q_2}\times\set{p_1,p_2}$ of seller and buyer
prices: the earning pair $(q_1,p_2)$ at the upper left, the wide subsidy pair
$(q_2,p_1)$ at the lower right, and the narrow subsidy pairs $(q_1,p_1)$ and
$(q_2,p_2)$ next to the diagonal.  The narrow subsidy pairs come for free.
Since screening keeps the intervals with the smallest
$\prob{p<s\le q,\ p\le b<q}$, \Cref{prop:corr-ub-cover} lets the values of
$\GFT^{\scoremark}$ already computed during the bisection estimate their
gains-from-trade, which are close to $\fixedpricebenchmark$.  The earning pair needs one
new estimate, which $\gftest$ obtains without posting any subsidy.  The trade
probabilities at the four corners, which the profits require anyway, then give
the wide subsidy pair by a simple geometric relation: up to a correction term,
$\expect{\GFT_{\ti}(q_1,p_2)+\GFT_{\ti}(q_2,p_1)}\ge
\expect{\GFT_{\ti}(q_1,p_1)+\GFT_{\ti}(q_2,p_2)}\approx2\fixedpricebenchmark$.
One new estimate thus buys four candidates, and a simple combination of the
earning pair and the wide subsidy pair nearly attains the gains-from-trade of the
best common price.  Interest makes this combination affordable.  As the
paired intervals are $L\approx\frac{16}{\intrate\tiH}$ widths apart, the subsidy
spread $q_2-p_1$ exceeds the earning spread $p_2-q_1$ by $2(q_1-p_1)$, at most
a fraction $\frac{\intrate\tiH}8$ of it, which the interest accrued between the
earning and the subsidy rounds covers.

\paragraph{Guarantee of the two exits.}
Both exits of the bisection rest on \Cref{prop:corr-ub-cover}, by which the
gains-from-trade of the subsidy pair $(q,p)$ fall short of
$\GFT^{\scoremark}(\tinterval{p,q})$ by at most twice the width $q-p$ times
the mass $\prob{p<s\le q,\ p\le b<q}$.  In the narrow exit
(line~\ref{algline:corr-narrow} of \Cref{alg:corr-upper}), the width alone is
small.  In the pairing exit, no single interval need be narrow enough, but the
intervals are constrained as a group: those bisected at one width are disjoint,
so their widths and their masses each add up to at most $1$.  Once $2n$ of
them have accumulated (line~\ref{algline:corr-certify}), their common width is
at most $\frac1{2n}$, and at most $n$ of them can carry mass above
$\frac1n$, so screening (line~\ref{algline:corr-screen}) can always keep $n$
intervals whose product is small.  Both exits therefore certify the same
guarantee, and a single bisection can stop at whichever it reaches first.  In
short, the estimate of $\GFT^{\scoremark}$ is well worth its cost: the same
quantity steers the bisection, signals when to stop, and prices three of the
four corners.

We use \Cref{alg:corr-upper} only for $0<\intrate\le\frac1\tiH$, with target
accuracy and confidence parameter
$\accscale\deq\min\set{\tiH^{-1/4},(\intrate\tiH^2)^{-2/7}}$ and
$\delta\deq\accscale/\tiH$.  For $\intrate>\frac1\tiH$,
\Cref{obs:rate-monotone} gives
$\corrminimaxregret\le\corrregret{\frac1\tiH}$.  Combining the two cases
gives \Cref{thm:corr-upper}.

\begin{correlatedobservation}
\label{obs:rate-monotone}
\begin{flushleft}
For any two interest rates $0\le r\le r'$, we have
$\corrregret{r'}\le\corrregret{r}$.
\end{flushleft}
\end{correlatedobservation}

\begin{algorithm}[t]
\caption{\textsc{Interest-Financed Bisection}}
\label{alg:corr-upper}
\begin{algorithmic}[1]
\linespread{0.9}\selectfont
    \Statex \textbf{Input:} $\tiH$, $\accscale$, $\delta$.
    \State $k\gets\ceil{\accscale^{-2}\log(1/\delta)}$,\quad
        $N_0\gets\ceil{6\log(1/\delta)}$
    \State $\widehat\firstbest\gets{}$\Call{\hyperref[alg:corr-subroutines]{FB-Est}}{$N_0$}
    \If{$\widehat\firstbest\le\accscale$}
        \State post $(0,1)$ in every remaining round \Comment{$\Texploit$}
            \label{algline:corr-fallback}
    \EndIf
    \State $K\gets\ceil{\frac{1}{\widehat\firstbest}}$,\quad
        $\mathcal A\gets\set{[\frac iK,\frac{i+1}K]:i=0,\ldots,K-1}$
    \ForAll{$I\in\mathcal A$}
        \State $\widehat{\GFT}{}^{\scoremark}(I)\gets\hyperref[alg:corr-subroutines]{\gftplusest}(I,\widehat\firstbest,k)$ \Comment{$\Tscore$}
        \label{algline:corr-score-grid}
    \EndFor
    \While{$\ti\le\tiH$}
        \State $\tinterval{p^\star,q^\star}\gets\argmax_{I\in\mathcal A}
            \widehat{\GFT}{}^{\scoremark}(I)$
            \label{algline:corr-select}
        \If{$q^\star-p^\star\le\accscale\widehat\firstbest$}
            \label{algline:corr-narrow}
            \State $\hyperref[alg:corr-subroutines]{\safepost}((q^\star,p^\star),\widehat\firstbest)$ in every remaining round \Comment{$\Texploit$}
                \label{algline:corr-narrow-post}
        \EndIf
        \ForAll{$I\in\set{[p^\star,\frac{p^\star+q^\star}2],[\frac{p^\star+q^\star}2,q^\star]}$}
            \State $\widehat{\GFT}{}^{\scoremark}(I)\gets\hyperref[alg:corr-subroutines]{\gftplusest}(I,\widehat\firstbest,k)$ \Comment{$\Tscore$}
            \label{algline:corr-score}
        \EndFor
        \State $\mathcal A\gets(\mathcal A\setminus\set{\tinterval{p^\star,q^\star}})
            \cup\set{[p^\star,\frac{p^\star+q^\star}2],[\frac{p^\star+q^\star}2,q^\star]}$
        \State $\mathcal S_{q^\star-p^\star}\gets\mathcal S_{q^\star-p^\star}\cup\set{\tinterval{p^\star,q^\star}}$
            \label{algline:corr-record}
        \If{$\card{\mathcal S_{q^\star-p^\star}}
            =2\ceil{\frac{256}{\intrate\tiH\sqrt{\accscale\widehat\firstbest}}}$}
            \label{algline:corr-certify}
            \State $\mathcal S\gets\argmin\limits_{\substack{\mathcal S'\subseteq\mathcal S_{q^\star-p^\star}\\
                \card{\mathcal S'}=\card{\mathcal S_{q^\star-p^\star}}/2}}
                \sum_{\tinterval{p,q}\in\mathcal S'}\hprob{p<s\le q,\ p\le b<q}$
                \label{algline:corr-screen}
            \State $(\mathcal Q_E,\mathcal Q_S,\widehat{\GFT}(\cdot),\widehat{\Profit}(\cdot))\gets{}$\Call{\hyperref[alg:corr-pair-intervals]{Pair-Intervals}}{$\mathcal S,\widehat\firstbest,k$} \Comment{$\Tpair$}
                \label{algline:corr-pair}
            \State $t_0\gets\ti$
                \label{algline:corr-t0}
            \State $\bigl((p_E,q_E),(p_S,q_S),t_1\bigr)\gets\argmax\limits_{\substack{(p_E,q_E)\in\mathcal Q_E\cup\{(\auxprice,\auxprice+\frac13\widehat\firstbest)\}\\
                (p_S,q_S)\in\mathcal Q_S\\
                t_1\in\{t_0,\ldots,\tiH+1\}}}
                \sum_{\ti=t_0}^{t_1-1}\widehat{\GFT}(p_E,q_E)
                +\sum_{\ti=t_1}^{\tiH}\widehat{\GFT}(p_S,q_S)$
                \label{algline:corr-earn}\label{algline:corr-cross-test}
            \Statex \hspace{\algorithmicindent}\hspace{\algorithmicindent}s.t.\
                $\sum_{\ti=t_0}^{t_1-1}(1+\intrate)^{\tiH-\ti}\widehat{\Profit}(p_E,q_E)
                +\sum_{\ti=t_1}^{\tiH}(1+\intrate)^{\tiH-\ti}\widehat{\Profit}(p_S,q_S)\ge0$
            \For{$\ti=t_0,\ldots,t_1-1$} \Comment{$\Texploit$}
                \label{algline:corr-post-earn}
                \State draw $\auxprice\sim\Unif[0,1-\frac13\widehat\firstbest]$ and post $(p_E,q_E)$
            \EndFor
            \State $\hyperref[alg:corr-subroutines]{\safepost}((p_S,q_S),\widehat\firstbest)$ in every remaining round \Comment{$\Texploit$}
                \label{algline:corr-post-subsidy}
        \EndIf
    \EndWhile
\end{algorithmic}
\end{algorithm}

Our analysis assumes that \hyperref[alg:corr-subroutines]{\fbest} reaches
$N_0$ trades; otherwise it runs until the horizon, and the proof of
\Cref{thm:corr-upper} bounds the regret by $N_0$.  After $\fbest$, the
rounds of \Cref{alg:corr-upper} form an exploration and an exploitation phase.
The exploration rounds $\Texplore\deq\Tscore\cup\Tpair$ consist of the
scoring rounds $\Tscore$, in which \hyperref[alg:corr-subroutines]{\gftplusest}
estimates $\GFT^{\scoremark}$ on the discretization grid in
line~\ref{algline:corr-score-grid} and on the two halves of each selected leaf
in line~\ref{algline:corr-score}, and the pairing rounds $\Tpair$, in which
\hyperref[alg:corr-pair-intervals]{\textsc{Pair-Intervals}} estimates the gains-from-trade and profits of the
candidate pairs in line~\ref{algline:corr-pair}.  The exploitation rounds
$\Texploit$ commit to prices in one of three branches: the pair $(0,1)$ when
the first-best gains-from-trade are too small to be worth learning, in
line~\ref{algline:corr-fallback}; the subsidy pair $(q^\star,p^\star)$ when the
selected leaf is already narrow, in line~\ref{algline:corr-narrow-post}; and,
once enough intervals of one width have been bisected, the earning and subsidy
pairs of the joint problem, in lines~\ref{algline:corr-post-earn}
and~\ref{algline:corr-post-subsidy}.  In both phases, each subsidy pair is posted
through \hyperref[alg:corr-subroutines]{\safepost}, which first posts
$(\auxprice,\auxprice+\frac13\widehat\firstbest)$ in
line~\ref{algline:safepost-save} of \Cref{alg:corr-subroutines} until the
account covers the subsidy; we write $\Tsave$ for these saving rounds.  For
simplicity, we also assume $\tiH>256$, so that
$\accscale\le\tiH^{-1/4}<\frac14$; the proof of \Cref{thm:corr-upper} covers
smaller horizons.

\begin{correlatedlemma}
\label{lem:corr-confidence}
\begin{flushleft}
There is an event $\mathcal E$ with $\prob{\mathcal E}\ge1-4\tiH\delta$ on which
every completed estimate satisfies the following:
\begin{lemmaitems}
\item\label{lem:corr-confidence-fb}
$\bigl|\widehat\firstbest-\firstbest\bigr|\le\frac{\firstbest}{2}$;
\item\label{lem:corr-confidence-gft}
$\bigl|\widehat{\GFT}(p,q)-\expect{\GFT_{\ti}(p,q)}\bigr|\le\accscale$
for $\gftest$;
\item\label{lem:corr-confidence-prob}
$\bigl|\hprob{s\le p,\ b\ge q}-\prob{s\le p,\ b\ge q}\bigr|\le\accscale$
for every sampled price pair $(p,q)$.
\end{lemmaitems}
\end{flushleft}
\end{correlatedlemma}

\begin{proof}
$\fbest$ posts $(\auxprice,\auxprice)$ with
$\auxprice\sim\Unif\tinterval{0,1}$, which trades exactly when
$s\le\auxprice\le b$.  Integrating over $\auxprice$,
\[
\textstyle\prob{\tradeindicator{\ti}=1\condition s,b}
=\int_0^1\indicator{s\le\auxprice\le b}\,\dd\auxprice
=\plus{b-s},
\]
so $\prob{\tradeindicator{\ti}=1}=\expect{\plus{b-s}}=\firstbest$.
Chernoff's inequality therefore gives
\[
\textstyle\prob{\bigl|\widehat\firstbest-\firstbest\bigr|>\frac12\firstbest}
\le2\tiH\delta.
\]

For $p\le q$, $\gftest((p,q),\cdot)$ posts $(\auxprice,q)$, $(p,q)$ or
$(p,\auxprice)$ according to whether $\auxprice<p$,
$p\le\auxprice\le q$ or $\auxprice>q$.  Integrating over $\auxprice$,
\begin{align*}
\textstyle\prob{\tradeindicator{\ti}=1\condition s,b}
&\textstyle=\int_0^p\indicator{s\le\auxprice,\ b\ge q}\,\dd\auxprice
+(q-p)\indicator{s\le p,\ b\ge q}
+\int_q^1\indicator{s\le p,\ b\ge\auxprice}\,\dd\auxprice
\\*
&\textstyle=\bigl((p-s)+(q-p)+(b-q)\bigr)\indicator{s\le p,\ b\ge q}
\\*
&\textstyle=\GFT_{\ti}(p,q),
\end{align*}
so $\prob{\tradeindicator{\ti}=1}=\expect{\GFT_{\ti}(p,q)}$.
Conditional on its starting history, each estimate averages $k$ i.i.d.\
Bernoulli samples, so Hoeffding's inequality gives both
\begin{gather*}
\textstyle\prob{\bigl|\widehat{\GFT}(p,q)-\expect{\GFT_{\ti}(p,q)}\bigr|
>\accscale}\le2\delta
\\*
\shortintertext{and}
\textstyle\prob{\bigl|\hprob{s\le p,\ b\ge q}-\prob{s\le p,\ b\ge q}\bigr|
>\accscale}\le2\delta.
\end{gather*}
A union bound over $\fbest$ and the at most $\tiH$ completed estimates
gives $\prob{\mathcal E^c}\le4\tiH\delta$.
\end{proof}

$\gftplusest$ combines its estimates linearly, so \Cref{lem:corr-confidence}
controls its output.

\begin{correlatedcorollary}
\label{cor:corr-accuracy}
\begin{flushleft}
On $\mathcal E$, every interval $\tinterval{p,q}$ scored by $\gftplusest$
satisfies
\begin{gather*}
\textstyle\bigl|\hprob{p<s\le q,\ p\le b<q}-\prob{p<s\le q,\ p\le b<q}\bigr|
\le4\accscale
\\*
\shortintertext{and}
\textstyle\bigl|\widehat{\GFT}{}^{\scoremark}(\tinterval{p,q})
-\GFT^{\scoremark}(\tinterval{p,q})\bigr|\le7\accscale.
\end{gather*}
\end{flushleft}
\end{correlatedcorollary}

\begin{proof}
$\gftplusest$ forms the rectangle estimate from its four corner estimates by
inclusion--exclusion, as the probability is formed from the four corner
probabilities.  Hence
\begin{align*}
&\textstyle\bigl|\hprob{p<s\le q,\ p\le b<q}-\prob{p<s\le q,\ p\le b<q}\bigr|
\\*
&\textstyle=\bigl|\bigl(\hprob{s\le q,\ b\ge p}-\prob{s\le q,\ b\ge p}\bigr)
-\bigl(\hprob{s\le p,\ b\ge p}-\prob{s\le p,\ b\ge p}\bigr)
\\*
&\textstyle\qquad-\bigl(\hprob{s\le q,\ b\ge q}-\prob{s\le q,\ b\ge q}\bigr)
+\bigl(\hprob{s\le p,\ b\ge q}-\prob{s\le p,\ b\ge q}\bigr)\bigr|
\\*
&\textstyle\le4\accscale.
\tag*{\Cref{lem:corr-confidence}}
\end{align*}
By the definitions of $\GFT^{\scoremark}$ and of its estimate, and the first
bound,
\begin{align*}
&\textstyle\bigl|\widehat{\GFT}{}^{\scoremark}(\tinterval{p,q})
-\GFT^{\scoremark}(\tinterval{p,q})\bigr|
\\*
&\textstyle=\bigl|\bigl(\widehat{\GFT}(p,p)-\expect{\GFT_{\ti}(p,p)}\bigr)
+\bigl(\widehat{\GFT}(q,q)-\expect{\GFT_{\ti}(q,q)}\bigr)
-\bigl(\widehat{\GFT}(p,q)-\expect{\GFT_{\ti}(p,q)}\bigr)
\\*
&\textstyle\qquad+(q-p)\bigl(\hprob{p<s\le q,\ p\le b<q}
-\prob{p<s\le q,\ p\le b<q}\bigr)\bigr|
\\*
&\textstyle\le3\accscale+4\accscale(q-p)\le7\accscale.
\tag*{\Cref{lem:corr-confidence}}
\end{align*}
\end{proof}

Since the leaves of $\mathcal A$ cover $\tinterval{0,1}$,
\Cref{prop:corr-ub-upper} turns \Cref{cor:corr-accuracy} into the following.

\begin{correlatedcorollary}
\label{cor:corr-optimism}
\begin{flushleft}
On $\mathcal E$, the leaf $\tinterval{p^\star,q^\star}$ selected in any
iteration of \Cref{alg:corr-upper} satisfies
$\widehat{\GFT}{}^{\scoremark}(\tinterval{p^\star,q^\star})\ge\fixedpricebenchmark-7\accscale$.
\end{flushleft}
\end{correlatedcorollary}

\begin{proof}
Every price lies in some leaf of $\mathcal A$, so
\begin{align*}
\textstyle\fixedpricebenchmark
&\textstyle\le\max_{I\in\mathcal A}\GFT^{\scoremark}(I)
\tag*{\Cref{prop:corr-ub-upper}}
\\*
&\textstyle\le\max_{I\in\mathcal A}\widehat{\GFT}{}^{\scoremark}(I)+7\accscale
\tag*{\Cref{cor:corr-accuracy}}
\\*
\mr{\Cref{alg:corr-upper}, line~\ref{algline:corr-select}}
&\textstyle=\widehat{\GFT}{}^{\scoremark}(\tinterval{p^\star,q^\star})+7\accscale.
\end{align*}
\end{proof}

\begin{correlatedlemma}
\label{lem:corr-screening}
\begin{flushleft}
On $\mathcal E$, every $\tinterval{p,q}$ in the set $\mathcal S$ selected
in line~\ref{algline:corr-screen} of \Cref{alg:corr-upper} satisfies
\[
\textstyle\expect{\GFT_{\ti}(q,p)}+4\accscale
\ge\widehat{\GFT}{}^{\scoremark}(\tinterval{p,q})
\ge\widehat{\GFT}{}^{\scoremark}(\tinterval{p^\star,q^\star})-14\accscale.
\]
\end{flushleft}
\end{correlatedlemma}

\begin{proof}
Line~\ref{algline:corr-certify} of \Cref{alg:corr-upper} gives
\[
\textstyle\card{\mathcal S}
=\ceil{\frac{256}{\intrate\tiH\sqrt{\accscale\widehat\firstbest}}}
\ge\frac{256}{\sqrt\accscale},
\]
since $\intrate\tiH\le1$ and $\widehat\firstbest\le1$.

Now fix $\tinterval{p,q}\in\mathcal S$.  The $2\card{\mathcal S}$ intervals
of $\mathcal S_{q^\star-p^\star}$ have width $q-p$ and pairwise disjoint
interiors, so
$\sum_{\tinterval{u,v}\in\mathcal S_{q^\star-p^\star}}(v-u)=2\card{\mathcal S}(q-p)\le1$
and their rectangle probabilities sum to at most $1$.  At least $\card{\mathcal S}$ of these probabilities are
therefore at most $1/\card{\mathcal S}$, while
line~\ref{algline:corr-screen} keeps the $\card{\mathcal S}$ intervals with
the least estimates.  Hence
\begin{align}
&\textstyle(q-p)\bigl(\prob{p<s\le q,\ p\le b<q}+\hprob{p<s\le q,\ p\le b<q}\bigr)
\notag\\*
&\textstyle\le(q-p)\bigl(2\,\hprob{p<s\le q,\ p\le b<q}+4\accscale\bigr)
\tag*{\Cref{cor:corr-accuracy}}
\\*
\mr{\Cref{alg:corr-upper}, line~\ref{algline:corr-screen}}
&\textstyle\le(q-p)\bigl(\frac2{\card{\mathcal S}}+12\accscale\bigr)
\tag*{\Cref{cor:corr-accuracy}}
\\*
\mr{$2\card{\mathcal S}(q-p)\le1$}
&\textstyle\le\frac1{\card{\mathcal S}^2}+\frac{6\accscale}{\card{\mathcal S}}
\notag\\*
\mr{$\card{\mathcal S}\ge256/\sqrt\accscale$}
&\textstyle\le\accscale.
\label{eq:corr-screen-rect}
\end{align}
For the left inequality, we compare $\GFT^{\scoremark}(\tinterval{p,q})$ with
its estimate term by term:
\begin{align*}
&\textstyle\expect{\GFT_{\ti}(q,p)}
\\*
&\textstyle\ge\GFT^{\scoremark}(\tinterval{p,q})-2(q-p)\prob{p<s\le q,\ p\le b<q}
\tag*{\Cref{prop:corr-ub-cover}}
\\*
&\textstyle=\widehat{\GFT}{}^{\scoremark}(\tinterval{p,q})
+\bigl(\expect{\GFT_{\ti}(p,p)}-\widehat{\GFT}(p,p)\bigr)
+\bigl(\expect{\GFT_{\ti}(q,q)}-\widehat{\GFT}(q,q)\bigr)
\\*
&\textstyle\quad-\bigl(\expect{\GFT_{\ti}(p,q)}-\widehat{\GFT}(p,q)\bigr)
-(q-p)\bigl(\prob{p<s\le q,\ p\le b<q}+\hprob{p<s\le q,\ p\le b<q}\bigr)
\\*
&\textstyle\ge\widehat{\GFT}{}^{\scoremark}(\tinterval{p,q})-3\accscale
-(q-p)\bigl(\prob{p<s\le q,\ p\le b<q}+\hprob{p<s\le q,\ p\le b<q}\bigr)
\tag*{\Cref{lem:corr-confidence-gft}}
\\*
&\textstyle\ge\widehat{\GFT}{}^{\scoremark}(\tinterval{p,q})-4\accscale.
\tag*{\cref{eq:corr-screen-rect}}
\end{align*}

For the right inequality, recall that every $\tinterval{p,q}\in\mathcal S$ is
a leaf that line~\ref{algline:corr-select} selected in this or an earlier
iteration; the bound is trivial if
$\tinterval{p,q}=\tinterval{p^\star,q^\star}$.  Otherwise, let
$I\supseteq\tinterval{p^\star,q^\star}$ be the ancestor of the current leaf
that was a leaf of $\mathcal A$ when $\tinterval{p,q}$ was selected.  Then
\begin{align*}
\textstyle\widehat{\GFT}{}^{\scoremark}(\tinterval{p^\star,q^\star})
&\textstyle\le\GFT^{\scoremark}(\tinterval{p^\star,q^\star})+7\accscale
\tag*{\Cref{cor:corr-accuracy}}
\\*
\mr{$I\supseteq\tinterval{p^\star,q^\star}$}
&\textstyle\le\GFT^{\scoremark}(I)+7\accscale
\tag*{\Cref{prop:corr-ub-nested}}
\\*
&\textstyle\le\widehat{\GFT}{}^{\scoremark}(I)+14\accscale
\tag*{\Cref{cor:corr-accuracy}}
\\*
\mr{\Cref{alg:corr-upper}, line~\ref{algline:corr-select}}
&\textstyle\le\widehat{\GFT}{}^{\scoremark}(\tinterval{p,q})+14\accscale.
\end{align*}
\end{proof}

\begin{correlatedlemma}
\label{lem:corr-funding}
\begin{flushleft}
If $\fbest$ returns $\widehat\firstbest$ with
$\bigl|\widehat\firstbest-\firstbest\bigr|\le\frac{\firstbest}{2}$, then
every $\mathcal T\in\set{\Texplore,\Texploit}$ satisfies
\[
\textstyle\expect{\card{\mathcal T\cap\Tsave}}
\le\frac6{\firstbest\cdot\widehat\firstbest}\cdot
\expect{\sum_{i\in\mathcal T\setminus\Tsave}\plus{-\Profit_i}}+\frac8{\firstbest}.
\]
\end{flushleft}
\end{correlatedlemma}

\begin{proof}
While $(1+\intrate)\account{\ti}<p-q$, $\safepost$ posts
$(\auxprice,\auxprice+\frac13\widehat\firstbest)$ with
$\auxprice\sim\Unif[0,1-\frac13\widehat\firstbest]$, which trades exactly when
$s\le\auxprice\le b-\frac13\widehat\firstbest$.  Integrating over
$\auxprice$,
\begin{align}
\textstyle\prob{\tradeindicator{\ti}=1}
&\textstyle=\frac{1}{1-\widehat\firstbest/3}
\expect{\int_0^{1-\widehat\firstbest/3}
\indicator{s\le\auxprice\le b-\widehat\firstbest/3}\,\dd\auxprice}
\notag\\*
&\textstyle=\frac{\expect{\plus{b-s-\widehat\firstbest/3}}}{1-\widehat\firstbest/3}
\notag\\*
\mr{$\plus{b-s-\frac13\widehat\firstbest}\ge\plus{b-s}-\frac13\widehat\firstbest$}
&\textstyle\ge\firstbest-\frac13\widehat\firstbest
\notag\\*
\mr{$\widehat\firstbest\le\frac32\firstbest$}
&\textstyle\ge\frac12\firstbest,
\label{eq:corr-bank-quality}
\end{align}
and we have
\begin{equation}
\textstyle\expect{\Profit_{\ti}(\auxprice,\auxprice+\frac13\widehat\firstbest)}
=\frac13\widehat\firstbest\,\prob{\tradeindicator{\ti}=1}
\ge\frac19\widehat\firstbest^2
\deq\widehat{\Profit}(\auxprice,\auxprice+\frac13\widehat\firstbest).
\label{eq:corr-bank-profit}
\end{equation}
The same holds in every earning round of line~\ref{algline:corr-post-earn}
of \Cref{alg:corr-upper} that posts this pair, as $\auxprice$ is drawn
afresh there too.  We also set $\widehat{\GFT}(\auxprice,\auxprice+\frac13\widehat\firstbest)\deq0$.

Every pair requested from $\safepost$ has $p-q\le\widehat\firstbest$, by
the choice of $K$ and line~\ref{algline:corr-blocks} of \Cref{alg:corr-pair-intervals}.
Fix $\mathcal T$, and let $t'$ be its last round in which
$(\auxprice,\auxprice+\frac13\widehat\firstbest)$ trades, during a call
requesting $(p,q)$.  The account at the start of $\mathcal T$ and its
interest are nonnegative, and $\Profit_i\ge-\plus{-\Profit_i}$, so summing the
cash flows of $\mathcal T$ up to $t'$ gives
\begin{align*}
&\textstyle\sum_{i\in\mathcal T\cap\Tsave}\Profit_i
-\sum_{i\in\mathcal T\setminus\Tsave}\plus{-\Profit_i}
\\*
&\textstyle\le\account{t'+1}
=(1+\intrate)\account{t'}+\frac13\widehat\firstbest
<p-q+\frac13\widehat\firstbest
\le\frac43\widehat\firstbest.
\end{align*}
If the pair never trades, the left-hand side is nonpositive.

Whether a round enters each sum is decided before its valuations and random
price are drawn, so the tower property and the account bound above give
\begin{align*}
\textstyle\expect{\card{\mathcal T\cap\Tsave}}
&\textstyle\le\frac2{\firstbest}\cdot
    \expect{\sum_{i\in\mathcal T\cap\Tsave}\tradeindicator{i}}
\tag*{\cref{eq:corr-bank-quality}}
\\*
\mr{$\Profit_i=\frac13\widehat\firstbest\,\tradeindicator{i}$ in $\Tsave$}
&\textstyle=\frac6{\firstbest\cdot\widehat\firstbest}\cdot
    \expect{\sum_{i\in\mathcal T\cap\Tsave}\Profit_i}
\\*
&\textstyle\le\frac6{\firstbest\cdot\widehat\firstbest}\cdot
    \expect{\sum_{i\in\mathcal T\setminus\Tsave}\plus{-\Profit_i}}+\frac8{\firstbest}.
\end{align*}
\end{proof}

\begin{correlatedlemma}
\label{lem:corr-batch}
\begin{flushleft}
On $\mathcal E$, the prices $(p_E,q_E)$, $(p_S,q_S)$ and the round
$t_1$ selected in line~\ref{algline:corr-earn} of \Cref{alg:corr-upper}
satisfy
\[
\textstyle(t_1-t_0)\widehat{\GFT}(p_E,q_E)
+(\tiH-t_1+1)\widehat{\GFT}(p_S,q_S)
\ge(\tiH-t_0+1)(\fixedpricebenchmark-40\accscale)
-(t_0-1)\fixedpricebenchmark.
\]
\end{flushleft}
\end{correlatedlemma}

\begin{proof}
By the maximality in line~\ref{algline:corr-earn}, it suffices to exhibit
candidate pairs $(p_E,q_E)$, $(p_S,q_S)$ and a round $t_1$ that satisfy
its constraint with objective at least the right side of the claim.

If $\tiH-t_0+1\le t_0-1$, the right side of the claim is nonpositive, and
the earning pair $(\auxprice,\auxprice+\frac13\widehat\firstbest)$ with
$t_1=\tiH+1$ suffices, as it satisfies the constraint and has objective
$0$.  Otherwise $\tiH-t_0+1>\tfrac{\tiH}{2}$ and
$(\tiH-t_0+1)\accscale>\tfrac12\tiH^{5/7}>16$ as
$\accscale\ge\tiH^{-2/7}$ and $\tiH>256$, and we show objective at least
$(\tiH-t_0+1)(\fixedpricebenchmark-40\accscale)$ in two cases.

\medskip\noindent\textbf{Case 1: there is a candidate earning pair $(p,q)$ with
$\widehat{\GFT}(p,q)\ge\widehat{\GFT}{}^{\scoremark}(\tinterval{p^\star,q^\star})-33\accscale$.}
\par\nopagebreak\noindent Taking it as the earning pair with $t_1=\tiH+1$
satisfies the constraint, as $\widehat{\Profit}(p,q)\ge0$ for $p\le q$, and
gives objective at least $(\tiH-t_0+1)(\fixedpricebenchmark-40\accscale)$, by
\Cref{cor:corr-optimism}.

\medskip\noindent\textbf{Case 2: every candidate earning pair $(p,q)$ has
$\widehat{\GFT}(p,q)<\widehat{\GFT}{}^{\scoremark}(\tinterval{p^\star,q^\star})-33\accscale$.}
\par\nopagebreak\noindent We set
\begin{align*}
(p_E,q_E)
&\textstyle\deq\argmin_{(p,q)\in\mathcal Q_E\cup\{(\auxprice,\auxprice+\frac13\widehat\firstbest)\},\
\widehat{\Profit}(p,q)>0}
\frac{\widehat{\GFT}{}^{\scoremark}(\tinterval{p^\star,q^\star})-\widehat{\GFT}(p,q)}
{\widehat{\Profit}(p,q)},
\\
(p_S,q_S)
&\textstyle\deq(q_2,p_1)
\quad\text{for}\quad
(\tinterval{p_1,q_1},\tinterval{p_2,q_2})\deq\argmin_{\mathcal P}\,\hyperlink{corr-batch-dagger}{(\dagger)},
\\
t_1
&\textstyle\deq\min\bigl\{t\ge t_0:
\sum_{i=t_0}^{t-1}(1+\intrate)^{\tiH-i}\widehat{\Profit}(p_E,q_E)
+\sum_{i=t}^{\tiH}(1+\intrate)^{\tiH-i}\widehat{\Profit}(p_S,q_S)\ge0\bigr\},
\end{align*}
where $(\dagger)$ is the term marked in \cref{eq:corr-batch-subsidy}.  Since
$\widehat{\GFT}(\auxprice,\auxprice+\frac13\widehat\firstbest)=0$, the case
assumption gives
$\widehat{\GFT}{}^{\scoremark}(\tinterval{p^\star,q^\star})>33\accscale$,
while \Cref{lem:corr-screening} gives
$\widehat{\GFT}{}^{\scoremark}(\tinterval{p^\star,q^\star})\le
\firstbest+18\accscale$, so $15\accscale<\firstbest$.  Comparing with
$(\auxprice,\auxprice+\frac13\widehat\firstbest)$,
\begin{align*}
\textstyle\varphi\deq\frac{\widehat{\GFT}{}^{\scoremark}(\tinterval{p^\star,q^\star})
-\widehat{\GFT}(p_E,q_E)}{\widehat{\Profit}(p_E,q_E)}
&\textstyle\le\frac{9}{\widehat\firstbest^2}\,
\widehat{\GFT}{}^{\scoremark}(\tinterval{p^\star,q^\star})
\le\frac{9}{\widehat\firstbest^2}(\firstbest+18\accscale)
\\*
\mathllap{\mr{$15\accscale<\firstbest$}}
&\textstyle<\frac{20\firstbest}{\widehat\firstbest^2}
\le\frac{40}{\widehat\firstbest}.
\tag*{\Cref{lem:corr-confidence}}
\end{align*}
Fewer than $\tfrac{1}{\widehat\firstbest}$ blocks have span greater than
$\widehat\firstbest$, and $L\le\tfrac{19}{\intrate\tiH}$ as
$\intrate\tiH\le1$, so
\begin{align}
\textstyle\card{\mathcal P}
&\textstyle>\frac{n}{L}-1-\frac{1}{\widehat\firstbest}
\notag\\*
\mathllap{\mr{\Cref{alg:corr-upper}, line~\ref{algline:corr-certify}; $\accscale<\widehat\firstbest$}}
&\textstyle\ge\frac{256}{19\sqrt{\accscale\widehat\firstbest}}-1
-\frac{1}{\widehat\firstbest}
>\frac{11}{\sqrt{\accscale\widehat\firstbest}}.
\label{eq:corr-batch-count}
\end{align}
By construction, the intervals $\tinterval{p_1,q_2}$ spanned by the elements
$(\tinterval{p_1,q_1},\tinterval{p_2,q_2})$ of $\mathcal P$ have pairwise
disjoint interiors, so
\[
\left\{
\begin{aligned}
&\textstyle\sum_{\mathcal P}\prob{q_1<s\le q_2,\ p_1\le b<p_2}\le1,
\\
&\textstyle\sum_{\mathcal P}\bigl(\prob{s\le q_2,\ b\ge p_1}
-\prob{s\le q_1,\ b\ge p_2}\bigr)\le2,
\\
&\textstyle\sum_{\mathcal P}(q_2-p_1)\le1.
\end{aligned}
\right.
\]
For the chosen element,
\begin{align}
&\textstyle\widehat{\GFT}(p_S,q_S)
\notag\\
&\textstyle=\widehat{\GFT}{}^{\scoremark}(\tinterval{p_1,q_1})
+\widehat{\GFT}{}^{\scoremark}(\tinterval{p_2,q_2})-\widehat{\GFT}(q_1,p_2)
-(q_2-p_1)\bigl(\hprob{q_1<s\le q_2,\ p_1\le b<p_2}+4\accscale\bigr)
\notag\\
&\textstyle\ge2\,\widehat{\GFT}{}^{\scoremark}(\tinterval{p^\star,q^\star})
-28\accscale-\widehat{\GFT}(q_1,p_2)
-(q_2-p_1)\bigl(\prob{q_1<s\le q_2,\ p_1\le b<p_2}+8\accscale\bigr)
\tag*{\Cref{lem:corr-confidence,lem:corr-screening}}
\\
&\textstyle\stepmark{a}{\ge}\widehat{\GFT}{}^{\scoremark}(\tinterval{p^\star,q^\star})-28\accscale
+\varphi(p_2-q_1)\,\hprob{s\le q_1,\ b\ge p_2}
\notag\\*
&\textstyle\quad-(q_2-p_1)\bigl(\prob{q_1<s\le q_2,\ p_1\le b<p_2}+8\accscale\bigr)
\notag\\
&\textstyle=\widehat{\GFT}{}^{\scoremark}(\tinterval{p^\star,q^\star})-28\accscale
+\varphi\frac{p_2-q_1}{q_2-p_1}\bigl(-\widehat{\Profit}(p_S,q_S)\bigr)
\notag\\*
&\textstyle\quad-\frac{1}{\widehat\firstbest}(q_2-p_1)\Bigl(\widehat\firstbest\,\prob{q_1<s\le q_2,\ p_1\le b<p_2}
\notag\\*
&\textstyle\qquad+\varphi\widehat\firstbest\,\frac{p_2-q_1}{q_2-p_1}
\bigl(\hprob{s\le q_2,\ b\ge p_1}-\hprob{s\le q_1,\ b\ge p_2}\bigr)
+8\widehat\firstbest\accscale\Bigr)
\notag\\
&\textstyle\stepmark{b}{\ge}\widehat{\GFT}{}^{\scoremark}(\tinterval{p^\star,q^\star})-28\accscale
+\varphi\frac{p_2-q_1}{q_2-p_1}\bigl(-\widehat{\Profit}(p_S,q_S)\bigr)
\notag\\*
&\textstyle-\frac{1}{\widehat\firstbest}\,\underbrace{\textstyle(q_2-p_1)
\bigl(\prob{q_1<s\le q_2,\ p_1\le b<p_2}
+40\bigl(\prob{s\le q_2,\ b\ge p_1}-\prob{s\le q_1,\ b\ge p_2}\bigr)
+88\accscale\bigr)}_{\hypertarget{corr-batch-dagger}{(\dagger)}}
\tag*{\Cref{lem:corr-confidence}}
\\
&\textstyle\stepmark{c}{\ge}\widehat{\GFT}{}^{\scoremark}(\tinterval{p^\star,q^\star})-28\accscale
+\varphi\frac{p_2-q_1}{q_2-p_1}\bigl(-\widehat{\Profit}(p_S,q_S)\bigr)
-\frac{1}{\widehat\firstbest}\Bigl(\frac{1}{\card{\mathcal P}}\sum_{\mathcal P}\sqrt{(\dagger)}\Bigr)^2
\notag\\
&\textstyle\stepmark{d}{\ge}\widehat{\GFT}{}^{\scoremark}(\tinterval{p^\star,q^\star})-28\accscale
+\varphi\frac{p_2-q_1}{q_2-p_1}\bigl(-\widehat{\Profit}(p_S,q_S)\bigr)
-\frac{81+88\accscale\card{\mathcal P}}{\widehat\firstbest\card{\mathcal P}^2}
\notag\\
&\textstyle>\widehat{\GFT}{}^{\scoremark}(\tinterval{p^\star,q^\star})-28\accscale
+\varphi\frac{p_2-q_1}{q_2-p_1}\bigl(-\widehat{\Profit}(p_S,q_S)\bigr)
-\frac{81\accscale}{121}
-8\accscale\sqrt{\frac{\accscale}{\widehat\firstbest}}
\tag*{\cref{eq:corr-batch-count}}
\\
&\textstyle\stepmark{e}{>}\widehat{\GFT}{}^{\scoremark}(\tinterval{p^\star,q^\star})-32\accscale
+\varphi\frac{p_2-q_1}{q_2-p_1}\bigl(-\widehat{\Profit}(p_S,q_S)\bigr),
\label{eq:corr-batch-subsidy}
\end{align}
where (a) uses the minimality of $\varphi$ for the candidate earning pair
$(q_1,p_2)$, which also holds if $\widehat{\Profit}(q_1,p_2)=0$, as the case
assumption gives $\widehat{\GFT}(q_1,p_2)<\widehat{\GFT}{}^{\scoremark}(\tinterval{p^\star,q^\star})$,
(b) uses $p_2-q_1\le q_2-p_1$, $\varphi\le\tfrac{40}{\widehat\firstbest}$
and $\widehat\firstbest\le1$, (c) uses that the element minimizes $(\dagger)$,
(d) uses Cauchy--Schwarz and the three sums, and (e) uses
$15\accscale<\firstbest\le2\widehat\firstbest$.

If $t_1=t_0$, \cref{eq:corr-batch-subsidy} and \Cref{cor:corr-optimism} give objective at
least $(\tiH-t_0+1)(\fixedpricebenchmark-39\accscale)$.  Otherwise, we relax
$t_1$ to a real switching time $\tilde t_1$, writing the two sums of the
constraint through their geometric-series formulas.  The constraint then binds
at a unique $\tilde t_1\in(t_0,\tiH+1)$, and $t_1=\ceil{\tilde t_1}$.
There,
\begin{align}
\textstyle\frac{(\tiH-\tilde t_1+1)\bigl(-\widehat{\Profit}(p_S,q_S)\bigr)}
{(\tilde t_1-t_0)\,\widehat{\Profit}(p_E,q_E)}
&\textstyle=\frac{\tiH-\tilde t_1+1}{\tilde t_1-t_0}\cdot
\frac{(1+\intrate)^{\tilde t_1-t_0}-1}{1-(1+\intrate)^{-(\tiH-\tilde t_1+1)}}
\stepmark{a}{\ge}1+\frac12(\tiH-t_0+1)\log(1+\intrate)
\notag\\*
&\textstyle\stepmark{b}{\ge}1+\frac{\intrate\tiH}{8}
\stepmark{c}{\ge}1+\frac{2(q_1-p_1)}{p_2-q_1}
=\frac{q_2-p_1}{p_2-q_1},
\label{eq:corr-batch-switch}
\end{align}
where (a) uses $\tfrac{\ee^x-1}{x}\ge1+\tfrac{x}{2}$ and $\tfrac{x}{1-\ee^{-x}}\ge1+\tfrac{x}{2}$ for
$x>0$, (b) uses $\log(1+\intrate)\ge\tfrac{\intrate}{2}$, which holds for
$0\le\intrate\le1$, and
$\tiH-t_0+1>\tfrac{\tiH}{2}$, and (c) uses
$p_2-q_1\ge(L-2)(q_1-p_1)\ge\tfrac{16(q_1-p_1)}{\intrate\tiH}$, as the $L-2$
intervals between $\tinterval{p_1,q_1}$ and $\tinterval{p_2,q_2}$ have
width $q_1-p_1$ and disjoint interiors.  Hence, as
$(\tiH-t_0+1)\accscale>16$,
\begin{align*}
&\textstyle(t_1-t_0)\widehat{\GFT}(p_E,q_E)+(\tiH-t_1+1)\widehat{\GFT}(p_S,q_S)
\\*
&\textstyle\stepmark{a}{\ge}(\tilde t_1-t_0)\widehat{\GFT}(p_E,q_E)
+(\tiH-\tilde t_1+1)\widehat{\GFT}(p_S,q_S)-2-8\accscale
\tag*{\Cref{lem:corr-screening}}
\\*
&\textstyle\ge(\tilde t_1-t_0)\widehat{\GFT}(p_E,q_E)
+(\tiH-\tilde t_1+1)\bigl(\widehat{\GFT}{}^{\scoremark}(\tinterval{p^\star,q^\star})
-32\accscale\bigr)-2-8\accscale
\\*
&\textstyle\quad+\varphi\frac{p_2-q_1}{q_2-p_1}(\tiH-\tilde t_1+1)\bigl(-\widehat{\Profit}(p_S,q_S)\bigr)
\tag*{\cref{eq:corr-batch-subsidy}}
\\*
&\textstyle\ge(\tilde t_1-t_0)\widehat{\GFT}(p_E,q_E)
+(\tiH-\tilde t_1+1)\bigl(\widehat{\GFT}{}^{\scoremark}(\tinterval{p^\star,q^\star})
-32\accscale\bigr)-2-8\accscale
\\*
&\textstyle\quad+(\tilde t_1-t_0)\bigl(\widehat{\GFT}{}^{\scoremark}(\tinterval{p^\star,q^\star})
-\widehat{\GFT}(p_E,q_E)\bigr)
\tag*{\cref{eq:corr-batch-switch}}
\\*
&\textstyle=(\tiH-t_0+1)\,\widehat{\GFT}{}^{\scoremark}(\tinterval{p^\star,q^\star})
-32(\tiH-\tilde t_1+1)\accscale-2-8\accscale
\\*
&\textstyle\ge(\tiH-t_0+1)\bigl(\widehat{\GFT}{}^{\scoremark}(\tinterval{p^\star,q^\star})
-33\accscale\bigr)
\ge(\tiH-t_0+1)(\fixedpricebenchmark-40\accscale),
\tag*{\Cref{cor:corr-optimism}}
\end{align*}
where (a) uses $0\le t_1-\tilde t_1<1$, $\widehat{\GFT}(p_E,q_E)\ge0$, as
$\gftest$ averages trade indicators, and
$\widehat{\GFT}(p_S,q_S)\le\widehat{\GFT}{}^{\scoremark}(\tinterval{p_1,q_1})
+\widehat{\GFT}{}^{\scoremark}(\tinterval{p_2,q_2})$.
\end{proof}

\begin{correlatedlemma}
\label{lem:corr-execution}
\begin{flushleft}
On $\mathcal E$, conditional on the history before round $t_0$, the prices
posted by \Cref{alg:corr-upper} in rounds $t_0,\ldots,\tiH$ satisfy
\[
\textstyle\expect{\sum_{\ti=t_0}^{\tiH}\GFT_{\ti}(\sprice{\ti},\bprice{\ti})}
\ge(\tiH-t_0+1)(\fixedpricebenchmark-80\accscale)
-(t_0-1)\fixedpricebenchmark.
\]
\end{flushleft}
\end{correlatedlemma}

\begin{proof}
For an element $(\tinterval{p_1,q_1},\tinterval{p_2,q_2})$ of
$\mathcal P$, the trade event of $(q_2,p_1)$ splits by
inclusion--exclusion, and $b-s\ge p_1-q_2$ on
$\set{q_1<s\le q_2,\ p_1\le b<p_2}$, so
\begin{align}
&\textstyle\expect{\GFT_{\ti}(q_2,p_1)}
\notag\\*
&\textstyle=\expect{\GFT_{\ti}(q_1,p_1)}+\expect{\GFT_{\ti}(q_2,p_2)}
-\expect{\GFT_{\ti}(q_1,p_2)}
+\expect{(b-s)\indicator{q_1<s\le q_2,\ p_1\le b<p_2}}
\notag\\*
&\textstyle\ge\widehat{\GFT}{}^{\scoremark}(\tinterval{p_1,q_1})
+\widehat{\GFT}{}^{\scoremark}(\tinterval{p_2,q_2})
-\widehat{\GFT}(q_1,p_2)-9\accscale
\notag\\*
&\textstyle\quad-(q_2-p_1)\bigl(\hprob{q_1<s\le q_2,\ p_1\le b<p_2}
+4\accscale\bigr)
\tag*{\Cref{lem:corr-confidence,lem:corr-screening}}
\\*
&\textstyle=\widehat{\GFT}(q_2,p_1)-9\accscale.
\label{eq:corr-exec-welfare}
\end{align}
If $\tiH-t_0+1\le t_0-1$, the right side of the claim is at most
$-9(\tiH-t_0+1)\accscale$, and the left side is always at least this bound.  Every posted pair other than
$(p_S,q_S)$ has its seller price below its buyer price, hence nonnegative
gains-from-trade, and if $t_1\le\tiH$, maximality gives
$\widehat{\GFT}(p_S,q_S)\ge\widehat{\GFT}(p_E,q_E)\ge0$, as $t_1=\tiH+1$
also satisfies the constraint of line~\ref{algline:corr-earn}, so
$\expect{\GFT_{\ti}(p_S,q_S)}\ge-9\accscale$, by \Cref{lem:corr-screening}
and \cref{eq:corr-exec-welfare}.

Otherwise, as in the proof of \Cref{lem:corr-batch},
$\tiH-t_0+1>\tfrac{\tiH}{2}$ and $(\tiH-t_0+1)\accscale>16$.

Rounds $t_0,\ldots,t_1-1$ post $(p_E,q_E)$, and the rounds
$i\ge t_1$ outside $\Tsave$ post $(p_S,q_S)$.  The saving rounds from
$t_1$ on refill what these subsidies take from the account, so their trades
are bounded by the final account and by the shortfall of the cash flow of the
two pairs.  If there are none, the left side of \cref{eq:corr-exec-account} is
zero.  Otherwise, let $t'$ be the last one; $\safepost$ saves in round
$t'$ only if $(1+\intrate)\account{t'}<p_S-q_S$, so
$\account{t'+1}<p_S-q_S+\frac13\widehat\firstbest\le\frac43\widehat\firstbest$,
by line~\ref{algline:corr-blocks} of \Cref{alg:corr-pair-intervals}.  Then
\begin{align}
&\textstyle\expect{\sum_{i\in\Tsave,\ i\ge t_1}\tradeindicator{i}}
\notag\\*
&\textstyle\stepmark{a}{\le}\frac3{\widehat\firstbest}\,
\expect{\sum_{i\in\Tsave,\ t_1\le i\le t'}
(1+\intrate)^{\tiH-i}\Profit_i}
\notag\\*
&\textstyle=\frac3{\widehat\firstbest}\,
\mathbb E\Bigl[(1+\intrate)^{\tiH-t'}\account{t'+1}
-(1+\intrate)^{\tiH-t_0+1}\account{t_0}
-\sum_{i=t_0}^{t_1-1}(1+\intrate)^{\tiH-i}\Profit_i(p_E,q_E)
\notag\\*
&\textstyle\qquad-\sum_{i\notin\Tsave,\ t_1\le i\le t'}
(1+\intrate)^{\tiH-i}\Profit_i(p_S,q_S)\Bigr]
\notag\\*
&\textstyle\stepmark{b}{\le}12+\frac3{\widehat\firstbest}\,\expect{\Bigl(
-\sum_{i=t_0}^{t_1-1}(1+\intrate)^{\tiH-i}\Profit_i(p_E,q_E)
-\sum_{i=t_1}^{\tiH}(1+\intrate)^{\tiH-i}\Profit_i(p_S,q_S)\Bigr)^+}
\notag\\*
&\textstyle\stepmark{c}{\le}12+3\ee(\tiH-t_0+1)\accscale
+3\ee\sqrt{\tiH-t_0+1}
\tag*{\Cref{lem:corr-confidence}; \cref{eq:corr-bank-profit}}
\\*
&\textstyle\stepmark{d}{\le}12+\tfrac{9\ee}{2}(\tiH-t_0+1)\accscale,
\label{eq:corr-exec-account}
\end{align}
where (a) uses $\Profit_i=\frac13\widehat\firstbest\tradeindicator{i}$ in
$\Tsave$; (b) uses $\account{t'+1}<\frac43\widehat\firstbest$,
$\account{t_0}\ge0$, $(1+\intrate)^{\tiH}\le\ee$ and
$\Profit_i(p_S,q_S)\le0$; (c) bounds the mean of the cash flow by
$\expect{\Profit_{\ti}(p,q)}\ge\widehat{\Profit}(p,q)-\accscale\widehat\firstbest$
for both pairs and the constraint of line~\ref{algline:corr-earn}, and its
deviation by Cauchy--Schwarz, as its terms are independent, the random price
being redrawn in each round, and have absolute value at most
$\ee\,\widehat\firstbest$; and (d)
uses $\accscale\sqrt{\tiH-t_0+1}>\tiH^{-2/7}\sqrt{\tiH/2}>2$.

In each round $i\ge t_1$ of $\Tsave$, a pair
with nonnegative gains-from-trade replaces $(p_S,q_S)$, whose expected
gains-from-trade are at most $\firstbest$, and whether a round is one of them is decided before its
valuations and price are drawn.  Hence, as $(\tiH-t_0+1)\accscale>16$,
\begin{align*}
&\textstyle\expect{\sum_{\ti=t_0}^{\tiH}\GFT_{\ti}(\sprice{\ti},\bprice{\ti})}
\\*
&\textstyle\ge(t_1-t_0)\expect{\GFT_{\ti}(p_E,q_E)}
+(\tiH-t_1+1)\expect{\GFT_{\ti}(p_S,q_S)}
-\firstbest\,\expect{
\sum_{i\in\Tsave,\ i\ge t_1}1}
\\*
&\textstyle\ge(t_1-t_0)\expect{\GFT_{\ti}(p_E,q_E)}
+(\tiH-t_1+1)\expect{\GFT_{\ti}(p_S,q_S)}
-2\,\expect{\sum_{i\in\Tsave,\ i\ge t_1}\tradeindicator{i}}
\tag*{\cref{eq:corr-bank-quality}}
\\*
&\textstyle\ge(t_1-t_0)\widehat{\GFT}(p_E,q_E)
+(\tiH-t_1+1)\widehat{\GFT}(p_S,q_S)
-(9+9\ee)(\tiH-t_0+1)\accscale-24
\tag*{\Cref{lem:corr-confidence,lem:corr-screening}; \cref{eq:corr-exec-welfare,eq:corr-exec-account}}
\\*
&\textstyle\ge(\tiH-t_0+1)(\fixedpricebenchmark-40\accscale)
-(t_0-1)\fixedpricebenchmark
-(9+9\ee)(\tiH-t_0+1)\accscale-24
\tag*{\Cref{lem:corr-batch}}
\\*
&\textstyle\ge(\tiH-t_0+1)(\fixedpricebenchmark-80\accscale)
-(t_0-1)\fixedpricebenchmark.
\end{align*}
\end{proof}

\begin{proof}[Proof of \Cref{thm:corr-upper}]
The algorithms use only trade indicators.  Every subsidized pair $(p,q)$
is posted by $\safepost$ only once $(1+\intrate)\account{\ti}\ge p-q$, and
all other pairs have nonnegative cash flow.  This gives pathwise budget
balance.
Horizons $\tiH\le256$ cost at most $2\tiH$.  Otherwise
$N_0=\softO{1}$ and $k=\softO{\accscale^{-2}}$, and $\tiH\accscale^3\ge1$
as $\accscale\ge\tiH^{-2/7}$.

Write $\history{\ti-1}$ for the history before round $\ti$.  Round $\ti$
draws its prices given $\history{\ti-1}$, before its valuations, and loses
at most $2$, so the rounds off $\mathcal E$ cost at most
$2\tiH\prob{\mathcal E^c}$.  We work on $\mathcal E$ from now on and split
its rounds into those of $\fbest$, $\Texplore$ and $\Texploit$; by the
tower property,
\begin{align}
&\textstyle\tiH\fixedpricebenchmark
-\expect{\sum_{\ti\in\tinterval{\tiH}}\GFT_{\ti}(\sprice{\ti},\bprice{\ti})}
\notag\\*
&\textstyle\le\expect{\sum_{\ti\,\in\,\textnormal{\fbest{} rounds}}
\bigl(\fixedpricebenchmark
-\expect{\GFT_{\ti}(\sprice{\ti},\bprice{\ti})\condition\history{\ti-1}}\bigr)
\cdot\indicator{\mathcal E}}
\label{eq:corr-ub-calibration}
\\*
&\textstyle\quad+\expect{\sum_{\ti\in\Texplore}
\bigl(\fixedpricebenchmark
-\expect{\GFT_{\ti}(\sprice{\ti},\bprice{\ti})\condition\history{\ti-1}}\bigr)
\cdot\indicator{\mathcal E}}
\label{eq:corr-ub-learning}
\\*
&\textstyle\quad+\expect{\sum_{\ti\in\Texploit}
\bigl(\fixedpricebenchmark
-\expect{\GFT_{\ti}(\sprice{\ti},\bprice{\ti})\condition\history{\ti-1}}\bigr)
\cdot\indicator{\mathcal E}}
\label{eq:corr-ub-final}
\\*
&\textstyle\quad+2\tiH\prob{\mathcal E^c}.
\notag
\end{align}
We bound the three terms in order.

$\fbest$ stops at its $N_0$-th trade or at the horizon, and each of its
rounds posts a random common price, which trades with probability
$\firstbest$ and loses between $0$ and
$\fixedpricebenchmark\le\firstbest$ in expectation, so
\[
\textstyle\textup{\cref{eq:corr-ub-calibration}}
\le\firstbest\cdot\frac{N_0}{\firstbest}=N_0=\softO{1}.
\]

For \cref{eq:corr-ub-learning}, recall that on $\mathcal E$ the estimate of
$\fbest$ satisfies $\bigl|\widehat\firstbest-\firstbest\bigr|\le\frac{\firstbest}{2}$,
by \Cref{lem:corr-confidence-fb}.  This gives
$\firstbest\le2\widehat\firstbest$ and is the premise of
\Cref{lem:corr-funding}, decided when $\fbest$ returns.  Moreover,
$\widehat\firstbest>\accscale$ unless $\Texplore=\emptyset$, by
line~\ref{algline:corr-fallback} of \Cref{alg:corr-upper}.  Then
\begin{align}
\textstyle\textup{\cref{eq:corr-ub-learning}}
&\textstyle=\expect{\sum_{\ti\in\Texplore\setminus\Tsave}
\bigl(\fixedpricebenchmark
-\expect{\GFT_{\ti}(\sprice{\ti},\bprice{\ti})\condition\history{\ti-1}}\bigr)
\cdot\indicator{\mathcal E}}
\notag\\*
&\textstyle\quad+\expect{\sum_{\ti\in\Texplore\cap\Tsave}
\bigl(\fixedpricebenchmark
-\expect{\GFT_{\ti}(\sprice{\ti},\bprice{\ti})\condition\history{\ti-1}}\bigr)
\cdot\indicator{\mathcal E}}
\notag\\
&\textstyle\stepmark{a}{\le}\mathbb E\Bigl[2\widehat\firstbest\cdot
\card{\Texplore\setminus\Tsave}\cdot\indicator{\mathcal E}
\label{eq:corr-ub-requests}
\\*
&\textstyle\qquad+\sum_{\ti\in\Texplore\setminus\Tsave}
\expect{\plus{-\Profit_{\ti}}\condition\history{\ti-1}}\cdot\indicator{\mathcal E}\Bigr]
\label{eq:corr-ub-debits}
\\*
&\textstyle\quad+\expect{\firstbest\cdot
\card{\Texplore\cap\Tsave}\cdot\indicator{\mathcal E}},
\label{eq:corr-ub-bank}
\end{align}
where (a) holds as $\fixedpricebenchmark\le\firstbest$,
$-\GFT_{\ti}(\sprice{\ti},\bprice{\ti})\le\plus{-\Profit_{\ti}}$, and
$\GFT_{\ti}(\sprice{\ti},\bprice{\ti})\ge0$ in $\Tsave$.  Then
\begin{align*}
\textstyle\textup{\cref{eq:corr-ub-requests}}
&\textstyle=\expect{2\widehat\firstbest\bigl(\card{\Tscore\setminus\Tsave}
+\card{\Tpair\setminus\Tsave}\bigr)\cdot\indicator{\mathcal E}}
\\*
&\textstyle\stepmark{b}{\le}\expect{2\widehat\firstbest\Bigl(7k\bigl(K+2+2\sum_w\card{\mathcal S_w}\bigr)
+5k\cdot\card{\mathcal P}\Bigr)\cdot\indicator{\mathcal E}}
\\*
\mr{$\card{\mathcal P}\le\card{\mathcal S}\le\sum_w\card{\mathcal S_w}$}
&\textstyle\le\expect{20k\widehat\firstbest\bigl(K+2+2\sum_w\card{\mathcal S_w}\bigr)
\cdot\indicator{\mathcal E}}
\\*
\mr{\Cref{alg:corr-upper}, lines~\ref{algline:corr-narrow} and~\ref{algline:corr-certify}}
&\textstyle\le\expect{20k\widehat\firstbest\Bigl(K+2
+2\sum_{w>\accscale\widehat\firstbest}
\min\set{\frac1{\accscale\widehat\firstbest},
2\ceil{\frac{256}{\intrate\tiH\sqrt{\accscale\widehat\firstbest}}}}\Bigr)\cdot\indicator{\mathcal E}}
\\*
\mr{$\widehat\firstbest(K+2)\le4$}
&\textstyle\le\softO{k\min\set{\frac1\accscale,\frac1{\intrate\tiH\sqrt\accscale}}+k}
\\*
\mr{$\accscale=\min\set{\tiH^{-1/4},(\intrate\tiH^2)^{-2/7}}$}
&\textstyle=\softO{k\tiH\accscale^3+k}
\\*
\mr{$\tiH\accscale^3\ge1$}
&\textstyle\le\softO{\tiH\accscale},
\end{align*}
where (b) holds as \hyperref[alg:corr-subroutines]{\gftplusest}
requests $7k$ pairs per call and is called $K$ times in
line~\ref{algline:corr-score-grid} of \Cref{alg:corr-upper} and twice in
line~\ref{algline:corr-score} for each leaf, all recorded in
line~\ref{algline:corr-record} but the last, while
\hyperref[alg:corr-pair-intervals]{\textsc{Pair-Intervals}}
requests $5k$ pairs per element of $\mathcal P$.
Next,
\begin{align*}
\textstyle\textup{\cref{eq:corr-ub-debits}}
&\textstyle=\expect{\sum_{\ti\in\Tscore\setminus\Tsave}
\expect{\plus{-\Profit_{\ti}}\condition\history{\ti-1}}\cdot\indicator{\mathcal E}}
\\*
&\textstyle\quad+\expect{\sum_{\ti\in\Tpair\setminus\Tsave}
\expect{\plus{-\Profit_{\ti}}\condition\history{\ti-1}}\cdot\indicator{\mathcal E}}
\\*
&\textstyle\stepmark{c}{\le}\mathbb E\Bigl[k\Bigl(\sum_{i=0}^{K-1}\frac1K\cdot\prob{s\le\frac{i+1}K,\ b\ge\frac iK}+\frac1K
\\*
&\textstyle\qquad+\sum_w\sum_{[p,q]\in\mathcal S_w}(q-p)\cdot\prob{s\le q,\ b\ge p}\Bigr)\cdot\indicator{\mathcal E}\Bigr]
\\*
&\textstyle\quad+\expect{k\sum_{(u,v)\in\mathcal Q_S}(u-v)\cdot\prob{s\le u,\ b\ge v}\cdot\indicator{\mathcal E}}
\\*
&\textstyle\stepmark{d}{\le}\expect{k\Bigl(\firstbest+\frac3K
+\sum_w(\firstbest+2w)+3(\firstbest+2\widehat\firstbest)\Bigr)\cdot\indicator{\mathcal E}}
\\*
\mr{$1/K\le\widehat\firstbest$}
&\textstyle\le\expect{\softO{k\widehat\firstbest}\cdot\indicator{\mathcal E}},
\end{align*}
where (c) uses that a round posting $(u,v)$ has expected subsidy
$\plus{u-v}\prob{s\le u,\ b\ge v}$, which is zero unless $u>v$.  Since the
pairs of \hyperref[alg:corr-subroutines]{\gftest} have $u\le v$, all subsidies come from the corners
that \hyperref[alg:corr-subroutines]{\safepost} posts, at most $k$ times each.  For an interval
$\tinterval{p,q}$ passed to \hyperref[alg:corr-subroutines]{\gftplusest}, the only corner with $u>v$
is $(q,p)$.  Each grid interval therefore gives a term of the first sum.  The
two halves of a leaf $\tinterval{p,q}$ together give at most
$(q-p)\prob{s\le q,\ b\ge p}$.  This is a term of the second sum, since
$\tinterval{p,q}\in\mathcal S_{q-p}$, except for the last leaf, which adds at
most $\frac1K$.
For an element of $\mathcal P$, the corners with $u>v$ are the pairs of
$\mathcal Q_S$, as $q_1\le p_2$, and give the third sum.  And (d) bounds
each sum of the form $\sum_{[p,q]}(q-p)\prob{s\le q,\ b\ge p}$ by
$\firstbest+2\max_{[p,q]}(q-p)$, which holds for intervals with disjoint
interiors; the third sum splits into three such families,
$\tinterval{p_1,q_1}$, $\tinterval{p_2,q_2}$ and $\tinterval{p_1,q_2}$
over $\mathcal P$, of width at most $\widehat\firstbest$ by
line~\ref{algline:corr-blocks} of \Cref{alg:corr-pair-intervals}.
Finally,
\begin{align*}
\textstyle\textup{\cref{eq:corr-ub-bank}}
&\textstyle=\expect{\firstbest\cdot
\card{\Texplore\cap\Tsave}\cdot\indicator{\mathcal E}}
\\*
&\textstyle\le\expect{\firstbest\cdot
\card{\Texplore\cap\Tsave}\cdot
\indicator{\bigl|\widehat\firstbest-\firstbest\bigr|\le\frac{\firstbest}2}}
\\*
&\textstyle\le\expect{\frac6{\widehat\firstbest}\cdot\softO{k\widehat\firstbest}\cdot
\indicator{\bigl|\widehat\firstbest-\firstbest\bigr|\le\frac{\firstbest}2}}+8
\tag*{\Cref{lem:corr-funding}}
\\*
&\textstyle=\softO{k}.
\end{align*}
By nonnegativity, the second step keeps only the part of $\mathcal E$ decided
when $\fbest$ returns, on which \Cref{lem:corr-funding} applies; the bound
$\softO{k\widehat\firstbest}$ on \cref{eq:corr-ub-debits} holds there as
well, since it uses only the algorithm and $\firstbest\le2\widehat\firstbest$.
As $k=\softO{\accscale^{-2}}\le\softO{\tiH\accscale}$,
$\textup{\cref{eq:corr-ub-learning}}\le\softO{\tiH\accscale}$.

For \cref{eq:corr-ub-final}, every estimate is completed before the first round
of $\Texploit$, so the history before it decides $\mathcal E$ and the line
of \Cref{alg:corr-upper} at which $\Texploit$ starts.

\medskip\noindent\textbf{Case 1: $\Texploit$ starts at
line~\ref{algline:corr-fallback}.}
\par\nopagebreak\noindent Then $\widehat\firstbest\le\accscale$, and
each round loses at most $\firstbest\le2\widehat\firstbest\le2\accscale$, so
\[
\textstyle\expect{\sum_{\ti\in\Texploit}
\bigl(\fixedpricebenchmark-\GFT_{\ti}(\sprice{\ti},\bprice{\ti})\bigr)}
\le2\tiH\accscale.
\]

\medskip\noindent\textbf{Case 2: $\Texploit$ starts at
line~\ref{algline:corr-narrow-post}.}
\par\nopagebreak\noindent Then
$q^\star-p^\star\le\accscale\widehat\firstbest$, the rounds of
$\Texploit\setminus\Tsave$ post $(q^\star,p^\star)$, and
\begin{align*}
\textstyle\expect{\GFT_{\ti}(q^\star,p^\star)}
&\textstyle\ge\GFT^{\scoremark}(\tinterval{p^\star,q^\star})
-2(q^\star-p^\star)\prob{p^\star<s\le q^\star,\ p^\star\le b<q^\star}
\tag*{\Cref{prop:corr-ub-cover}}
\\*
&\textstyle\ge\widehat{\GFT}{}^{\scoremark}(\tinterval{p^\star,q^\star})
-7\accscale-2(q^\star-p^\star)
\tag*{\Cref{cor:corr-accuracy}}
\\*
&\textstyle\ge\fixedpricebenchmark-14\accscale-2\accscale\widehat\firstbest
\tag*{\Cref{cor:corr-optimism}}
\\*
\mr{$\widehat\firstbest\le1$}
&\textstyle\ge\fixedpricebenchmark-16\accscale.
\end{align*}
Hence
\begin{align*}
\textstyle\expect{\sum_{\ti\in\Texploit}
\bigl(\fixedpricebenchmark-\GFT_{\ti}(\sprice{\ti},\bprice{\ti})\bigr)}
&\textstyle=\expect{\sum_{\ti\in\Texploit\setminus\Tsave}
\bigl(\fixedpricebenchmark-\GFT_{\ti}(q^\star,p^\star)\bigr)}
\\*
&\textstyle\quad+\expect{\sum_{\ti\in\Texploit\cap\Tsave}
\bigl(\fixedpricebenchmark-\GFT_{\ti}(\sprice{\ti},\bprice{\ti})\bigr)}
\\*
\mr{the bound above, $\fixedpricebenchmark\le\firstbest$}
&\textstyle\le16\accscale\cdot\expect{\card{\Texploit\setminus\Tsave}}
+\firstbest\cdot\expect{\card{\Texploit\cap\Tsave}}
\\*
&\textstyle\le16\tiH\accscale
+\frac6{\widehat\firstbest}\cdot\expect{\sum_{\ti\in\Texploit\setminus\Tsave}\plus{-\Profit_{\ti}}}+8
\tag*{\Cref{lem:corr-funding}}
\\*
\mr{$\plus{-\Profit_{\ti}}\le q^\star-p^\star\le\accscale\widehat\firstbest$}
&\textstyle\le22\tiH\accscale+8.
\end{align*}

\medskip\noindent\textbf{Case 3: $\Texploit$ starts at
line~\ref{algline:corr-post-earn}.}
\par\nopagebreak\noindent Then
\begin{align*}
&\textstyle\expect{\sum_{\ti\in\Texploit}
\bigl(\fixedpricebenchmark-\GFT_{\ti}(\sprice{\ti},\bprice{\ti})\bigr)}
\\*
&\textstyle\le80\tiH\accscale+(t_0-1)\fixedpricebenchmark
\tag*{\Cref{lem:corr-execution}}
\\*
&\textstyle\stepmark{a}{=}80\tiH\accscale
+\bigl(\frac{N_0}{\widehat\firstbest}+\card{\Texplore\setminus\Tsave}
+\card{\Texplore\cap\Tsave}\bigr)\cdot\fixedpricebenchmark
\\*
\mr{$\fixedpricebenchmark\le\firstbest\le2\widehat\firstbest$}
&\textstyle\le80\tiH\accscale+2N_0+2\widehat\firstbest\cdot\card{\Texplore\setminus\Tsave}
+\firstbest\cdot\card{\Texplore\cap\Tsave},
\end{align*}
where (a) holds as $\fbest$ returns $N_0$ divided by its number of rounds.

In every case,
\[
\textstyle\textup{\cref{eq:corr-ub-final}}
\le80\tiH\accscale+8+2N_0
+\textup{\cref{eq:corr-ub-requests}}+\textup{\cref{eq:corr-ub-bank}}
\le\softO{\tiH\accscale}.
\]
Adding the three bounds and $2\tiH\prob{\mathcal E^c}\le8\tiH\accscale$,
by \Cref{lem:corr-confidence}, gives
\[
\textstyle\corrminimaxregret\le\softO{\tiH\accscale}.
\]
Substituting the prescribed value of $\accscale$ completes the proof.
\end{proof}

\section{Hardness of Learning}
\label{sec:lower}

Both of our lower bounds build on the construction of
\citet[Theorem~3]{chen2025}.  As in their setting, they hold even if the
algorithm observes the two-bit feedback
$(\indicator{\sval{\ti}\le\sprice{\ti}},\indicator{\bprice{\ti}\le\bval{\ti}})$,
which determines whether trade occurs, so they also hold for the one-bit
feedback of our model.  On their instances, every feasible algorithm
posting on their grid satisfies the global price balance
$\expect{\sum_{\ti}(\bprice{\ti}-\sprice{\ti})}\ge0$ (their Lemma~13), so
every subsidy must be paid for by profit, which costs at least as much regret
as the subsidy saves.  With interest, the same argument yields only the
weighted balance
$\expect{\sum_{\ti}(1+\intrate)^{-\ti}(\bprice{\ti}-\sprice{\ti})}\ge0$,
whereas the regret weights all rounds equally, so the two can no longer be
combined as in their analysis.  We give the regret the same weights as the
balance, differently in the two settings.  For adversarial valuations
(\Cref{sec:adv-lower}), we mix the instance of round $\ti$ with a two-point
distribution, which scales the regret of that round by $(1+\intrate)^{-\ti}$.
For correlated i.i.d.\ valuations (\Cref{sec:corr-lower}), we add points that
only subsidizing actions trade, which discount the gain of every subsidy by the
terminal weight $(1+\intrate)^{-\tiH}$, up to a small error.

\subsection{Adversarial Valuations}
\label{sec:adv-lower}

\begin{adversarialtheorem}[Adversarial lower bound]
\label{thm:adv-lower}
\begin{flushleft}
Suppose that $\intrate=\tiH^{-\rateexponent}$.  Then
\[
    \advminimaxregret
    =
    \begin{cases}
        \bomega{\tiH^{\frac57}},
            & \rateexponent\le\frac{20}{21},\\[1mm]
        \bomega{\tiH^{\frac{3\rateexponent}4}},
            & \frac{20}{21}<\rateexponent<1,\\[1mm]
        \bomega{\tiH^{\frac34}},
            & \rateexponent\ge1.
    \end{cases}
\]
\end{flushleft}
\end{adversarialtheorem}

The bound $\bomega{\tiH^{5/7}}$ holds for every $\rateexponent$ by
\citet[Theorem~5.5]{bernasconi2024}: their hard instances add points just below
the diagonal to the weakly budget-balanced construction of
\citet{cesabianchi2023colt}, so that every action with $q<p$ is dominated and
the budget constraint, with or without interest, never binds.

For $\rateexponent>\frac{20}{21}$, we modify the $\bomega{\tiH^{3/4}}$
construction of \citet[Theorem~3]{chen2025}.  In their instances, only actions
in a bad region $\badset$ reveal which instance is in force, and each costs
constant regret.  Subsidizing actions in a good region $\goodset$ gain more
than the best fixed price, but they cannot offset this cost: no action on their
grid with $q>p$ trades more often than the fixed price $\frac12$, and
none with $q<p$ trades less often, which gives the global price balance.  The
trade-off between the cost of exploration and the loss from not identifying
the instance then gives $\bomega{\tiH^{3/4}}$.

With interest, profit earned early pays for later subsidies at a discount, the
same effect our algorithms exploit, so subsidies can offset the cost of
exploration again.  We draw the valuations of round $\ti$ from each of their
instances with probability $(1+\intrate)^{-\ti}$ and otherwise from a
two-point distribution on $(0,1)$ and $(1,0)$ (\Cref{fig:adv-hard-instance}).
Under the two-point
distribution, every action on their grid and every fixed price trade exactly
at $(0,1)$, so the comparison of trade probabilities with the fixed price
$\frac12$ is unchanged, and the budget constraint still relaxes to the
weighted balance (\Cref{lem:adv-weighted-price-balance}).  The mixture scales
the one-round regret of every action on their grid by $(1+\intrate)^{-\ti}$
(\cref{eq:adv-lb-regret}) and the information it reveals by at most this
factor (\cref{eq:adv-lb-change-of-measure}).  Their argument then goes through with the horizon $\tiH$
replaced by
$\sum_{\ti}(1+\intrate)^{-\ti}=\btheta{\min\set{\tiH,\intrate^{-1}}}$, which
gives $\bomega{\tiH^{3\rateexponent/4}}$ for $\rateexponent<1$ and
$\bomega{\tiH^{3/4}}$ for $\rateexponent\ge1$.  We give the proof in
\Cref{app:adv-lower}.

\subsection{Correlated i.i.d.\ Valuations}
\label{sec:corr-lower}

\begin{correlatedtheorem}[Correlated i.i.d.\ lower bound]
\label{thm:corr-lower}
\begin{flushleft}
Suppose that $\intrate=\tiH^{-\rateexponent}$.  Then
\[
    \corrminimaxregret
    =
    \begin{cases}
        \bomega{\tiH^{\frac57}},
            & \rateexponent\le1,\\[1mm]
        \bomega{\tiH^{\frac{2\rateexponent+3}7}},
            & 1<\rateexponent<\frac98,\\[1mm]
        \bomega{\tiH^{\frac34}},
            & \rateexponent\ge\frac98.
    \end{cases}
\]
\end{flushleft}
\end{correlatedtheorem}

Together with \Cref{thm:corr-upper}, this determines the optimal regret up to
polylogarithmic factors.  We add points to the hard instances of
\citet[Theorem~3]{chen2025} and keep each instance fixed over the rounds.  As
in \Cref{sec:adv-lower}, interest lets early earning pay for later subsidies,
and against a fixed instance a unit of subsidy may need only $(1+\intrate)^{-\tiH}$ units of profit, so a subsidy
of spread $x$ saves more regret than the earning that pays for it costs, by an
amount linear in $x$.  The added points have $s>b$, so only subsidizing
actions trade them, at a loss that cancels this gain.  This loss must grow with
the interest accrued over the horizon, so the added points vanish without
interest, where the construction reduces to theirs.  With interest, penalizing
every subsidy takes added mass proportional to this interest.  Since the
available mass is limited, the perturbation must shrink as the interest grows,
which is why the exponent falls below $\frac34$ for
$\rateexponent<\frac98$.  We present the construction and the proof below,
deferring the proofs of several lemmas to \Cref{app:corr-iid}.

For all sufficiently large $\tiH$, choose
\begin{equation}
\label{eq:corr-parameterization}
    \chenperturb
    \deq\frac1{10}
    \begin{cases}
        \tiH^{-2/7}, & 0\le\rateexponent\le1,\\[1mm]
        \tiH^{-2(2-\rateexponent)/7}, & 1<\rateexponent<\frac98,\\[1mm]
        \tiH^{-1/4}, & \rateexponent\ge\frac98,
    \end{cases}
    \qquad
    \chenfamilysize\deq2\bceil{40\chenperturb^3\tiH}.
\end{equation}
The support consists of five parts.  The hard instances differ on the two
chains
\[
    \textstyle\uppersupport
    \deq\set{\bigl(\frac{j}{5\chenfamilysize},\frac25+\frac{j}{5\chenfamilysize}\bigr)}_{j=0}^{3\chenfamilysize},
    \qquad
    \lowersupport
    \deq\set{\bigl(\frac15+\frac{j}{5\chenfamilysize},\frac25+\frac{j}{5\chenfamilysize}\bigr)}_{j=0}^{2\chenfamilysize},
\]
whose $j$-th points we denote by $\uppersupport[j]$ and
$\lowersupport[j]$.  To them we add the chain
\[
    \textstyle\revsupport
    \deq\set{\bigl(\frac25+\frac{j+\revspanindex}{5\chenfamilysize},
    \frac25+\frac{j}{5\chenfamilysize}\bigr)}_{j=0}^{\chenfamilysize-\revspanindex},
    \qquad
    \revspanindex\deq\bceil{30\bigl(1-(1+\intrate)^{-\tiH}\bigr)\chenperturb\chenfamilysize^2},
\]
whose points have $s-b=\frac{\revspanindex}{5\chenfamilysize}>0$, so no fixed
price trades them.  To complete the distribution, the corners
$\cornersupport\deq\set{0,1}^2$ make every feedback outcome likely, and the
majority point $\majorityatom\deq(\frac25,\frac35)$, which coincides with
$\lowersupport[\chenfamilysize]$, makes the prices in
$\tinterval{\frac25,\frac35}$ the best and carries the remaining mass.  With
$\revmass\deq3\bigl(1-(1+\intrate)^{-\tiH}\bigr)\chenperturb\chenfamilysize$,
the base instance is
\[
    \mathcal D^0(v)
    \deq
    \begin{cases}
        \chenperturb,
            & v\in(\uppersupport\cup\lowersupport)\setminus\set{\majorityatom},\\[1mm]
        \frac{\revmass}{\revspanindex},
            & v\in\revsupport,\\[1mm]
        \frac1{10},
            & v\in\cornersupport,\\[1mm]
        \frac35-(5\chenfamilysize+1)\chenperturb
        -\frac{\chenfamilysize-\revspanindex+1}{\revspanindex}\revmass,
            & v=\majorityatom.
    \end{cases}
\]
Since $\chenperturb\chenfamilysize\le\frac1{100}$ for all sufficiently large
$\tiH$, the choice of $\revspanindex$ ensures
$1\le\revspanindex\le\frac{\chenfamilysize}2$ and puts total mass at most
$\frac1{10}$ on $\revsupport$, so $\mathcal D^0(\majorityatom)\ge\frac25$.
For $k\in[\chenfamilysize]$,
the hard instance $\mathcal D^k$ moves mass $\chenperturb$ from
$\uppersupport[k-1]$ and $\lowersupport[k]$ to $\uppersupport[k]$ and
$\lowersupport[k-1]$:
\[
    \mathcal D^k(v)
    \deq
    \begin{cases}
        \mathcal D^0(v)+\chenperturb,
            & v\in\set{\uppersupport[k],\ \lowersupport[k-1]},\\
        \mathcal D^0(v)-\chenperturb,
            & v\in\set{\uppersupport[k-1],\ \lowersupport[k]},\\
        \mathcal D^0(v),
            & \text{otherwise}.
    \end{cases}
\]
In possibility $k\in\set{0,\ldots,\chenfamilysize}$, the valuations of all
rounds are drawn i.i.d.\ from $\mathcal D^k$, and $\mathbb E^k$ denotes the
expectation in this possibility.  Since an algorithm chooses
$(\sprice{\ti},\bprice{\ti})$ before observing $(\sval{\ti},\bval{\ti})$, the
valuations of round $\ti$ are independent of the history and of
$(\sprice{\ti},\bprice{\ti})$.

\begin{figure}[!t]
\centering
\begin{tikzpicture}[cor picture,font=\small]
\pgfmathtruncatemacro{\K}{8}
\pgfmathtruncatemacro{\J}{2}
\pgfmathtruncatemacro{\Umax}{3*\K}
\pgfmathtruncatemacro{\Lmax}{2*\K}
\pgfmathtruncatemacro{\Rmax}{\K-\J}
\pgfmathsetmacro{\gridstep}{1/(5*\K)}
\pgfmathsetmacro{\revoff}{\J/(5*\K)}

\begin{scope}[x=12cm,y=12cm]
    \path[cor bad region] (0,.4) rectangle (.4,1);
    \path[cor bad region] (.4,.6) rectangle (.6,1);
    \path[cor good region] (.4,.4) rectangle (.6,.6);
    \draw[cor grid,step=\gridstep] (0,.4) grid (.6,1);
    \draw[cor diagonal] (.4,.4) -- (.6,.6);
    \draw[cor axis] (0,.4) rectangle (.6,1);

    \pgfmathtruncatemacro{\kind}{3}
    \pgfmathsetmacro{\st}{\gridstep}
    \pgfmathsetmacro{\qk}{.4+\kind*\st}
    \pgfmathsetmacro{\qkm}{.4+(\kind-1)*\st}
    \pgfmathsetmacro{\pleft}{(\kind-1)*\st}
    \pgfmathsetmacro{\pright}{.2+(\kind-1)*\st}
    \begin{scope}
        \clip (0,.4) rectangle (.6,1);
        \draw[cor informative,line width=3pt] ({(\kind-.5)*\st},\qk) -- (\pright,\qk);
        \draw[cor informative,line width=3pt] (\pleft,.39) -- (\pleft,1.01);
        \draw[cor informative,line width=3pt] (\pright,.39) -- (\pright,{\qkm+.5*\st});
        \draw[cor informative,line width=3pt] (\pright,{\qk+.5*\st}) -- (\pright,1.01);
    \end{scope}
    \draw[cor good line,line width=3pt] (.4,\qk) -- (.6,\qk);

    \draw[cor reversal line] (.4+\revoff,.4) -- (.6,.6-\revoff);
    \foreach \j in {0,...,\Rmax} {
        \pgfmathsetmacro{\sx}{.4+(\j+\J)/(5*\K)}
        \pgfmathsetmacro{\by}{.4+\j/(5*\K)}
        \node[cor reversal large] at (\sx,\by) {};
    }
    \draw[cor span arrow] (.622,.6-\revoff) -- (.622,.6);
    \node[cor legend,anchor=west,text=corviolet] at (.638,.6-.5*\revoff)
        {$\dfrac{\revspanindex}{5\chenfamilysize}$};

    \foreach \j in {0,...,\Umax} {
        \pgfmathsetmacro{\sx}{\j/(5*\K)}
        \pgfmathsetmacro{\by}{.4+\j/(5*\K)}
        \node[cor upper large] at (\sx,\by) {};
    }
    \foreach \j in {0,...,\Lmax} {
        \ifnum\j=\K\relax
        \else
            \pgfmathsetmacro{\sx}{.2+\j/(5*\K)}
            \pgfmathsetmacro{\by}{.4+\j/(5*\K)}
            \node[cor lower large] at (\sx,\by) {};
        \fi
    }
    \node[cor corner large] at (0,1) {};
    \node[cor majority large] at (.4,.6) {};
    \draw[black,line width=.65pt]
        (.4-.0028,.6-.0028) -- (.4+.0028,.6+.0028);
    \node[cor plus] at ({\kind*\st},\qk) {};
    \node[cor minus] at (\pleft,\qkm) {};
    \node[cor plus] at (\pright,\qkm) {};
    \node[cor minus] at ({.2+\kind*\st},\qk) {};

    \node[anchor=north east] at (0,.4) {$(0,\tfrac25)$};
    \node[anchor=south east] at (0,1) {$(0,1)$};
    \node[anchor=north west] at (.6,.4) {$(\tfrac35,\tfrac25)$};
    \node[anchor=south west] at (.6,1) {$(\tfrac35,1)$};
    \node[below=8pt] at (.3,.4) {seller value $s$};
    \node[above=8pt,rotate=90] at (0,.7) {buyer value $b$};
\end{scope}

\begin{scope}[x=12cm,y=12cm]
    \foreach \sty/\lab [count=\i] in {
        cor upper large/\uppersupport,
        cor lower large/\lowersupport,
        cor reversal large/\revsupport,
        cor corner large/\cornersupport,
        cor majority large/\majorityatom,
        cor plus/+\chenperturb,
        cor minus/-\chenperturb}{
        \node[\sty] at (.75,{1-.05*\i}) {};
        \node[cor legend,anchor=west] at (.77,{1-.05*\i}) {$\lab$};
    }
    \draw[black,line width=.55pt] (.75-.0031,.75-.0031) -- (.75+.0031,.75+.0031);
    \draw[cor informative,line width=3pt] (.735,.6) -- (.765,.6);
    \node[cor legend,anchor=west] at (.77,.6) {$\infoset{k}$};
    \draw[cor good line,line width=3pt] (.735,.55) -- (.765,.55);
    \node[cor legend,anchor=west] at (.77,.55) {$\goodrow{k}$};
    \path[cor bad region] (.735,.488) rectangle (.765,.512);
    \draw[black,line width=.5pt] (.735,.488) rectangle (.765,.512);
    \node[cor legend,anchor=west] at (.77,.5) {$\badset$};
    \path[cor good region] (.735,.438) rectangle (.765,.462);
    \draw[black,line width=.5pt] (.735,.438) rectangle (.765,.462);
    \node[cor legend,anchor=west] at (.77,.45) {$\goodset$};
\end{scope}
\end{tikzpicture}
\caption{A diagram of the support
$\uppersupport\cup\lowersupport\cup\revsupport\cup\cornersupport\cup\set{\majorityatom}$
and of the actions $\goodrow{k}$ and $\infoset{k}$ for $k=3$, drawn for
$\chenfamilysize=8$ and $\revspanindex=2$; the corners other than $(0,1)$ lie
outside the drawn region.}
\label{fig:corr-hard-instance-geometry}
\end{figure}

The hard instances live on a grid, which lets us discretize the actions: apart
from the corners, all support points lie on the multiples of
$\frac1{5\chenfamilysize}$ inside
$\tinterval{0,\frac35}\times\tinterval{\frac25,1}$.  Rounding the seller price
down and the buyer price up onto this grid, within the same square, thus loses
only trades at corners with $s\ge b$, so neither the gains-from-trade nor the
profit of any round decreases.  Hence, as the following lemma shows, it
suffices to prove the lower bound for algorithms that post actions in
\[
    \textstyle\actiongrid
    \deq\set{\frac{i}{5\chenfamilysize}}_{i=0}^{3\chenfamilysize}
    \times\set{\frac25+\frac{j}{5\chenfamilysize}}_{j=0}^{3\chenfamilysize}.
\]

\begin{restatable}{correlatedlemma}{corrprojection}
\label{lem:corr-projection}
\begin{flushleft}
For every algorithm $\learner$, there is an algorithm $\learner'$ that,
whenever $\learner$ posts $(p,q)$, posts
\[
    (p',q')\deq
    \bigl(\min\set{\tfrac{\floor{5\chenfamilysize p}}{5\chenfamilysize},\tfrac35},\
    \max\set{\tfrac{\ceil{5\chenfamilysize q}}{5\chenfamilysize},\tfrac25}\bigr)\in\actiongrid,
\]
and on every valuation sequence in the support satisfies, in every round $\ti$,
\[
    \GFT_{\ti}(p',q')\ge\GFT_{\ti}(p,q),
    \qquad
    \Profit_{\ti}(p',q')\ge\Profit_{\ti}(p,q).
\]
In particular, $\learner'$ is feasible whenever $\learner$ is, and its regret
is no larger.
\end{flushleft}
\end{restatable}

The good actions are those with both prices in $\tinterval{\frac25,\frac35}$,
grouped by buyer price into the rows
\[
    \textstyle\goodrow{k}
    \deq\set{\bigl(\frac25+\frac{i}{5\chenfamilysize},\frac25+\frac{k}{5\chenfamilysize}\bigr)}_{i=0}^{\chenfamilysize},
    \qquad
    \goodset\deq\bigcup_{k=0}^{\chenfamilysize}\goodrow{k},
    \qquad
    \badset\deq\actiongrid\setminus\goodset.
\]
The possibilities differ only in the four masses of the perturbation, and these
change the law of the feedback $(\indicator{s\le p},\indicator{q\le b})$
only for the actions in one row and two columns of the grid
(\Cref{fig:corr-hard-instance-geometry}):
\[
    \textstyle\infoset{k}\deq\bigl\{\bigl(\frac{i}{5\chenfamilysize},\frac25+\frac{j}{5\chenfamilysize}\bigr)\in\actiongrid:\
    i\in\set{k-1,\chenfamilysize+k-1}
    \text{ or }j=k,\ k-1\le i\le\chenfamilysize+k-1\bigr\}.
\]
All of them have $p<\frac25$, so an algorithm can tell the possibilities apart
only by posting bad actions.  A bad action, however, misses the majority point,
which every good action trades, so each such round costs a constant regret.
Among the good actions, the regret is determined by the spread, except that in
possibility $k\ge1$ the row $\goodrow{k}$ is better than the others by
$\frac{\chenperturb}5$.  An algorithm must therefore either pay a constant per
round to explore with bad actions or stay in $\goodset$ without knowing which
row is best.  A simple calculation gives the one-round regrets.

\begin{restatable}{correlatedlemma}{corrgaps}
\label{lem:corr-gaps}
\begin{flushleft}
For every $k\in\set{0,\ldots,\chenfamilysize}$ and $(p,q)\in\actiongrid$,
\[
    \reg_{\mathcal D^k}(p,q)
    \ge
    \begin{cases}
        \frac2{25}, & (p,q)\in\badset,\\[1mm]
        3\chenperturb\chenfamilysize(q-p)
        +\smash[b]{\underbrace{\revmass\plus{p-q-\tfrac{\revspanindex-1}{5\chenfamilysize}}}_{\text{penalty of }\revsupport}}
        +\frac{\chenperturb}5\indicator{k\ge1,\ (p,q)\notin\goodrow{k}},
        & (p,q)\in\goodset,
    \end{cases}
    \rule[-4ex]{0pt}{0pt}
\]
with equality on $\goodset$.
\end{flushleft}
\end{restatable}

Apart from the penalty, these are the one-round regrets of their instances.
For a subsidy in $\goodset$, the first term is negative, so without the
penalty, the subsidy would beat the best fixed price.  In their analysis, this
gain is harmless, as the global price balance (their Lemma~13) makes earning pay
for it in full.  With interest, however, the budget constraint relaxes only to
the weighted balance of \Cref{lem:corr-price-balance} below, under which early
earning pays for more later subsidy, and the first term is no longer under
control.  This is why we add $\revsupport$, whose penalty offsets the
compounding: since
$\revmass=3\bigl(1-(1+\intrate)^{-\tiH}\bigr)\chenperturb\chenfamilysize$,
a subsidy of spread $x=p-q\ge\frac{\revspanindex-1}{5\chenfamilysize}$ has
$-3\chenperturb\chenfamilysize x
+\revmass\bigl(x-\frac{\revspanindex-1}{5\chenfamilysize}\bigr)
=-3(1+\intrate)^{-\tiH}\chenperturb\chenfamilysize x
-\frac{\revmass(\revspanindex-1)}{5\chenfamilysize}$, so its gain is discounted
by the terminal weight $(1+\intrate)^{-\tiH}$.  Subsidies of smaller spread
keep their gain, but by the choice of $\revspanindex$ this costs at most
$\frac{\chenperturb}{50}$ per round, so \Cref{lem:corr-gaps} gives the
following.

\begin{restatable}{correlatedcorollary}{corrdiscount}
\label{cor:corr-discount}
\begin{flushleft}
For every $k\in\set{0,\ldots,\chenfamilysize}$ and $(p,q)\in\goodset$,
\[
    \textstyle\reg_{\mathcal D^k}(p,q)
    \ge3\chenperturb\chenfamilysize\bigl(\plus{q-p}
    -(1+\intrate)^{-\tiH}\plus{p-q}\bigr)
    +\frac{\chenperturb}5\indicator{k\ge1,\ (p,q)\notin\goodrow{k}}
    -\frac{\chenperturb}{50}.
\]
\end{flushleft}
\end{restatable}

We now derive the weighted balance.  Their Lemma~13 needs only a reference
action whose trade probability no earning action exceeds and no subsidy falls
below; in their instances, this is the common price
$(\frac12,\frac12)\in\goodset$.  Let
$\refprob\deq\prob[\mathcal D^0]{s\le\frac12\le b}$ be its trade probability
under $\mathcal D^0$.  Our instances keep this property.  In them,
$\revsupport$ takes its mass from the majority point, which every
subsidy in $\actiongrid$ trades, as $(\frac12,\frac12)$ does; both thus lose
the same mass, and only the subsidy may trade points of $\revsupport$ in
return, which only raises its trade probability.  An earning action trades no
point of $\revsupport$.  If it trades the majority point, it is a good action
and compares with $(\frac12,\frac12)$ as in their instances; otherwise it
trades with probability at most $\frac12$, whereas the masses of the majority
point and the corner $(0,1)$ keep $\refprob\ge\frac12$.

\begin{restatable}{correlatedlemma}{corrreferenceprobability}
\label{lem:corr-reference-probability}
\begin{flushleft}
For every $k\in\set{0,\ldots,\chenfamilysize}$ and $(p,q)\in\actiongrid$,
$(q-p)\bigl(\prob[\mathcal D^k]{s\le p,\ q\le b}-\refprob\bigr)\le0$.
\end{flushleft}
\end{restatable}

The argument of their Lemma~13 then gives a global price balance, weighted by
the interest.

\begin{correlatedlemma}
\label{lem:corr-price-balance}
\begin{flushleft}
Every feasible algorithm posting actions in $\actiongrid$ satisfies
$\mathbb E^k\bigl[\sum_{\ti}(1+\intrate)^{-\ti}(\bprice{\ti}-\sprice{\ti})\bigr]\ge0$
in every possibility $k$.
\end{flushleft}
\end{correlatedlemma}

\begin{proof}
Since
$\account{\tiH+1}=\sum_{\ti}(1+\intrate)^{\tiH-\ti}(\bprice{\ti}-\sprice{\ti})\tradeindicator{\ti}$,
feasibility gives
$\sum_{\ti}(1+\intrate)^{-\ti}(\bprice{\ti}-\sprice{\ti})\tradeindicator{\ti}\ge0$,
so
\begin{align*}
0
&\textstyle\le\mathbb E^k\bigl[\sum_{\ti}(1+\intrate)^{-\ti}
  (\bprice{\ti}-\sprice{\ti})\tradeindicator{\ti}\bigr]
\\*
&\textstyle=\mathbb E^k\bigl[\sum_{\ti}(1+\intrate)^{-\ti}
  (\bprice{\ti}-\sprice{\ti})
  \prob[\mathcal D^k]{s\le \sprice{\ti},\ \bprice{\ti}\le b}\bigr]
\\*
&\textstyle\le\refprob\,\mathbb E^k\bigl[\sum_{\ti}(1+\intrate)^{-\ti}
  (\bprice{\ti}-\sprice{\ti})\bigr],
\tag*{\Cref{lem:corr-reference-probability}}
\end{align*}
and $\refprob>0$.
\end{proof}

Let $T_S$ be the number of rounds in which an algorithm posts an action in
$S\subseteq\actiongrid$.  Bounding the weights of positive spreads by $1$
and those of negative spreads by the terminal weight $(1+\intrate)^{-\tiH}$
can only increase the sum in \Cref{lem:corr-price-balance}; since
$0<q-p\le1$ on $\badset$, we have the following.

\begin{correlatedcorollary}
\label{cor:corr-good-balance}
\begin{flushleft}
Every feasible algorithm posting actions in $\actiongrid$ satisfies, in every
possibility $k$,
\[
    \textstyle\mathbb E^k\bigl[\sum_{\ti}\plus{\bprice{\ti}-\sprice{\ti}}
    \indicator{(\sprice{\ti},\bprice{\ti})\in\goodset}\bigr]
    -(1+\intrate)^{-\tiH}\,
    \mathbb E^k\bigl[\sum_{\ti}\plus{\sprice{\ti}-\bprice{\ti}}
    \indicator{(\sprice{\ti},\bprice{\ti})\in\goodset}\bigr]
    \ge-\mathbb E^k\bigl[T_{\badset}\bigr].
\]
\end{flushleft}
\end{correlatedcorollary}

The added points do not change how the possibilities can be told apart: they
are the same in all possibilities, the perturbation still affects only the
actions of $\infoset{k}$, and the corners keep every feedback outcome at
probability at least $\frac1{10}$.  Their Lemma~14 therefore carries over.

\begin{restatable}{correlatedlemma}{corrchangeofmeasure}
\label{lem:corr-change-of-measure}
\begin{flushleft}
For every algorithm posting actions in $\actiongrid$ and every
$k\in[\chenfamilysize]$,
\[
    \mathbb E^k\bigl[T_{\goodrow{k}}\bigr]-\mathbb E^0\bigl[T_{\goodrow{k}}\bigr]
    \le4\chenperturb\tiH\sqrt{\mathbb E^0\bigl[T_{\infoset{k}}\bigr]}.
\]
\end{flushleft}
\end{restatable}

The rows $\goodrow{k}$ are disjoint, and every action lies in at most two of
the sets $\infoset{k}$, one through its row and one through its column, so
pathwise
\begin{equation}
\label{eq:corr-occupation}
    \textstyle\sum_{k=1}^{\chenfamilysize}T_{\goodrow{k}}\le\tiH,
    \qquad
    \sum_{k=1}^{\chenfamilysize}T_{\infoset{k}}\le2T_{\badset}.
\end{equation}
We can now prove \Cref{thm:corr-lower}.  The proof also shows why the chain
$\revsupport$ is effective: \Cref{cor:corr-good-balance} charges subsidies at
the same terminal weight at which \Cref{cor:corr-discount} discounts their
gain, so this gain is paid for by earning, up to the profit of the bad rounds.

\begin{proof}[Proof of \Cref{thm:corr-lower}]
By \Cref{lem:corr-projection}, we may consider feasible algorithms that post
actions in $\actiongrid$.  We first bound the regret in possibility $0$.
Since $\actiongrid=\goodset\cup\badset$,
\begin{align*}
\reg_{\mathcal D^0}
&\textstyle=\mathbb E^0\Bigl[\sum_{\ti}\reg_{\mathcal D^0}(\sprice{\ti},\bprice{\ti})
\bigl(\indicator{(\sprice{\ti},\bprice{\ti})\in\badset}
+\indicator{(\sprice{\ti},\bprice{\ti})\in\goodset}\bigr)\Bigr]
\\*
&\textstyle\ge\frac2{25}\mathbb E^0[T_{\badset}]
+\mathbb E^0\Bigl[\sum_{\ti}\reg_{\mathcal D^0}(\sprice{\ti},\bprice{\ti})
\indicator{(\sprice{\ti},\bprice{\ti})\in\goodset}\Bigr]
\tag*{\Cref{lem:corr-gaps}}
\\*
&\textstyle\ge\frac2{25}\mathbb E^0[T_{\badset}]-\frac{\chenperturb\tiH}{50}
\\*
&\textstyle\quad+3\chenperturb\chenfamilysize\,\mathbb E^0\Bigl[\sum_{\ti}
\bigl(\plus{\bprice{\ti}-\sprice{\ti}}
-(1+\intrate)^{-\tiH}\plus{\sprice{\ti}-\bprice{\ti}}\bigr)
\indicator{(\sprice{\ti},\bprice{\ti})\in\goodset}\Bigr]
\tag*{\Cref{cor:corr-discount}}
\\*
&\textstyle\ge\frac2{25}\mathbb E^0[T_{\badset}]
-3\chenperturb\chenfamilysize\,\mathbb E^0[T_{\badset}]
-\frac{\chenperturb\tiH}{50}
\tag*{\Cref{cor:corr-good-balance}}
\\*
\mr{$\chenperturb\chenfamilysize\le\frac1{100}$}
&\textstyle\ge\frac1{20}\mathbb E^0[T_{\badset}]-\frac{\chenperturb\tiH}{50}.
\end{align*}
In possibility $k\in[\chenfamilysize]$, the same steps apply, and
\Cref{cor:corr-discount} adds $\frac{\chenperturb}5$ in every round with an
action in $\goodset\setminus\goodrow{k}$.  Hence
\begin{align*}
\reg_{\mathcal D^k}
&\textstyle\ge\frac1{20}\mathbb E^k[T_{\badset}]
+\frac{\chenperturb}5\mathbb E^k\bigl[T_{\goodset\setminus\goodrow{k}}\bigr]
-\frac{\chenperturb\tiH}{50}
\\*
\mr{$T_{\goodset\setminus\goodrow{k}}=\tiH-T_{\badset}-T_{\goodrow{k}}$}
&\textstyle=\frac1{20}\mathbb E^k[T_{\badset}]
+\frac{\chenperturb}5\bigl(\tiH-\mathbb E^k[T_{\badset}]
-\mathbb E^k[T_{\goodrow{k}}]\bigr)
-\frac{\chenperturb\tiH}{50}
\\*
\mr{$\frac1{20}\ge\frac{\chenperturb}5$}
&\textstyle\ge\frac{\chenperturb}5\bigl(\tiH-\mathbb E^k[T_{\goodrow{k}}]\bigr)
-\frac{\chenperturb\tiH}{50}
\\*
&\textstyle\ge\frac{\chenperturb}5\Bigl(\tiH-\mathbb E^0[T_{\goodrow{k}}]
-4\chenperturb\tiH\sqrt{\mathbb E^0[T_{\infoset{k}}]}\Bigr)
-\frac{\chenperturb\tiH}{50}.
\tag*{\Cref{lem:corr-change-of-measure}}
\end{align*}
Averaging over $k\in[\chenfamilysize]$ gives
\begin{align*}
\textstyle\frac1{\chenfamilysize}\sum_k\reg_{\mathcal D^k}
&\textstyle\ge\frac{\chenperturb}5\Bigl(\tiH
-\frac1{\chenfamilysize}\sum_k\mathbb E^0[T_{\goodrow{k}}]
-\frac{4\chenperturb\tiH}{\chenfamilysize}\sum_k
\sqrt{\mathbb E^0[T_{\infoset{k}}]}\Bigr)
-\frac{\chenperturb\tiH}{50}
\\*
\mr{$\sum_kT_{\goodrow{k}}\le\tiH$}
&\textstyle\ge\frac{\chenperturb}5\Bigl(\tiH-\frac{\tiH}{\chenfamilysize}
-\frac{4\chenperturb\tiH}{\chenfamilysize}\sum_k
\sqrt{\mathbb E^0[T_{\infoset{k}}]}\Bigr)
-\frac{\chenperturb\tiH}{50}
\\*
\mr{Cauchy--Schwarz inequality}
&\textstyle\ge\frac{\chenperturb}5\Bigl(\tiH-\frac{\tiH}{\chenfamilysize}
-4\chenperturb\tiH\sqrt{\tfrac1{\chenfamilysize}\sum_k\mathbb E^0[T_{\infoset{k}}]}\Bigr)
-\frac{\chenperturb\tiH}{50}
\\*
\mr{$\sum_kT_{\infoset{k}}\le2T_{\badset}$}
&\textstyle\ge\frac{\chenperturb}5\Bigl(\tiH-\frac{\tiH}{\chenfamilysize}
-4\chenperturb\tiH\sqrt{\tfrac2{\chenfamilysize}\mathbb E^0[T_{\badset}]}\Bigr)
-\frac{\chenperturb\tiH}{50},
\end{align*}
where the second and the last steps use \cref{eq:corr-occupation}.  Combining
the two bounds with $\chenfamilysize\ge80\chenperturb^3\tiH$ and
$\chenfamilysize\ge8$, it is easy to verify through elementary algebra that
\begin{align*}
\textstyle\max\set{\reg_{\mathcal D^0},
\frac1{\chenfamilysize}\sum_k\reg_{\mathcal D^k}}
&\textstyle\ge\max\set{\frac1{20}\mathbb E^0[T_{\badset}],\
\frac{\chenperturb\tiH}5\Bigl(1-\frac1{\chenfamilysize}
-4\chenperturb\sqrt{\tfrac2{\chenfamilysize}\mathbb E^0[T_{\badset}]}\Bigr)}
-\frac{\chenperturb\tiH}{50}
\\*
&\textstyle\ge\frac{\chenperturb\tiH}{25}-\frac{\chenperturb\tiH}{50}
=\frac{\chenperturb\tiH}{50}.
\end{align*}
Substituting \cref{eq:corr-parameterization} yields the three claimed rates,
with the constant $\frac1{500}$.
\end{proof}

\section{\texorpdfstring{$\alpha$}{Alpha}-Regret}
\label{sec:alpha}

Under global budget balance, \citet{bernasconi2024} attain sublinear regret
against the best fixed price for adversarial valuations.  Since a feasible
algorithm may subsidize trade, they further propose a stronger benchmark, the
best feasible distribution over price pairs, and show that, in the same
setting, no algorithm attains sublinear regret against it.  With interest, its
gains-from-trade on a valuation sequence are
\[
    \begin{array}{@{}l@{}}
    \displaystyle\tiH\cdot\distributionbenchmark\deq\sup_{\comparatordist}\
    \textstyle\sum_{\ti=1}^{\tiH}\expect[a\sim\comparatordist]{\GFT_{\ti}(a)}
    \\[8pt]
    \text{s.t.}\ \
    \textstyle\sum_{\ti=1}^{\tiH}(1+\intrate)^{\tiH-\ti}
      \expect[a\sim\comparatordist]{\Profit_{\ti}(a)}\ge0,
    \end{array}
\]
where $\comparatordist$ ranges over distributions on $\tinterval{0,1}^2$.
Common prices make no profit, so $\distributionbenchmark\ge\fixedpricebenchmark$.
To measure the difficulty of this benchmark, \citet{bernasconi2024} adopt the
classical $\regretfactor$-regret, which compares the benchmark with
$\regretfactor\ge1$ times the gains-from-trade of the algorithm.  We define the
minimax $\regretfactor$-regret and the \emph{optimal competitive ratio}, the
threshold below which no feasible algorithm attains sublinear
$\regretfactor$-regret, by
\begin{gather*}
    \alpharegret{\regretfactor}{\intrate}
    \deq\inf_{\learner}\ \sup
        \biggl\{\tiH\distributionbenchmark
          -\regretfactor\expect[\learner]{\sum_{\ti=1}^{\tiH}
            \GFT_{\ti}(\sprice{\ti},\bprice{\ti})}\biggr\},\\
    \critfactor{\rateexponent}
    \deq\sup\biggl\{\regretfactor\ge1:
       \liminf_{\tiH\to\infty}
       \frac{\alpharegret{\regretfactor}{\tiH^{-\rateexponent}}}{\tiH}>0
       \biggr\}.
\end{gather*}
Here the valuations $(\sval{\ti},\bval{\ti})_{\ti=1}^{\tiH}$ are
adversarial and the algorithm receives one-bit feedback.

\begin{alphatheorem}[Optimal competitive ratios]
\label{thm:alpha-main}
\begin{flushleft}
For every fixed $\rateexponent$ and
$\intrate=\tiH^{-\rateexponent}$, the optimal competitive ratio equals
\[
    \critfactor{\rateexponent}=
    \begin{cases}
        2,&\rateexponent>1,\\
        \criticalfactor=2.1831\ldots,&\rateexponent=1,\\
        3,&\rateexponent<1.
    \end{cases}
\]
There exist algorithms with one-bit feedback satisfying $\account{\ti+1}\ge0$ at every
round that achieve
\[
    \alpharegret{\critfactor{\rateexponent}}{\tiH^{-\rateexponent}}
    =
    \begin{cases}
      \softO{\tiH^{\max\{3/4,\,2-\rateexponent\}}},&\rateexponent>1,\\
      \softO{\tiH^{4/5}},&\rateexponent=1,\\
      \softO{\tiH^{\max\{4/5,\,\rateexponent\}}},&\rateexponent<1.
    \end{cases}
\]
For every fixed $1\le\regretfactor<\critfactor{\rateexponent}$,
$\alpharegret{\regretfactor}{\tiH^{-\rateexponent}}=\bomega{\tiH}$.
\end{flushleft}
\end{alphatheorem}
In particular, this tight characterization closes the gap between $\tfrac{36}{35}$ and
$2$ that \citet{bernasconi2024} left open without interest.

\subsection{Upper Bound}
\label{sec:alpha-upper}

We prove the upper bounds separately for three ranges of $\rateexponent$.
All three start from the same comparison of the benchmark with fixed price
pairs $(p,q)\in\tinterval{0,1}^2$, which, unlike common prices, may subsidize
trade.  Since a feasible distribution has a nonnegative expected account,
adding $\lambda\ge0$ times this account to its gains-from-trade and maximizing
over price pairs gives
\[
\textstyle\tiH\distributionbenchmark
\le\max_{(p,q)\in\tinterval{0,1}^2}\sum_{\ti=1}^{\tiH}
\bigl(\GFT_{\ti}(p,q)+\lambda(1+\intrate)^{\tiH-\ti}\Profit_{\ti}(p,q)\bigr);
\]
with $\lambda=0$, the benchmark is at most the gains-from-trade of the best price pair.
For $\rateexponent>1$, the interest rate is too small to affect the
comparison, so taking $\lambda=1$ extends the factor~$2$ between the
benchmark and the best common price of \citet[Theorem~6.3]{bernasconi2024},
and the $2$-regret follows from $\gftmax$ as in their Corollary~6.4.  At $\rateexponent=1$, the best
common price over the whole horizon no longer suffices, so we split the
horizon into blocks and compare the benchmark with a weighted sum of the best
common prices of the blocks.  For $0\le\rateexponent<1$, the interest rate
can no longer be ignored; \Cref{alg:alpha-upper} modifies
\Cref{alg:adv-upper} by replacing $\subsidizedgft$ with Exp3 over price
pairs, whose regret is measured against the best price pair
($\lambda=0$).  Since $\distributionbenchmark\ge0$, these bounds also
give sublinear regret for every larger multiplicative factor.

\paragraph{Weak interest ($\rateexponent>1$).}
We first compare the benchmark with the best fixed price.  Fix a
deterministic valuation sequence and a price pair $(p,q)$.

\medskip\noindent\textbf{Case 1: earning pair, $p\le q$.}
\par\nopagebreak\noindent Every profit is nonnegative, and
every trade also occurs at the common price $p$, so
$\Profit_{\ti}(p,q)\le\GFT_{\ti}(p,q)\le\GFT_{\ti}(p,p)$.  Hence
\begin{align*}
\textstyle\sum_{\ti=1}^{\tiH}\bigl(\GFT_{\ti}(p,q)+(1+\intrate)^{\tiH-\ti}\Profit_{\ti}(p,q)\bigr)
&\textstyle\le\sum_{\ti=1}^{\tiH}\GFT_{\ti}(p,q)
  +(1+\intrate)^{\tiH-1}\sum_{\ti=1}^{\tiH}\Profit_{\ti}(p,q)
\\*
&\textstyle\le\bigl(1+(1+\intrate)^{\tiH-1}\bigr)\tiH\fixedpricebenchmark.
\end{align*}

\medskip\noindent\textbf{Case 2: subsidy pair, $q<p$.}
\par\nopagebreak\noindent Every profit is nonpositive, and
checking whether $p$ or $q$ lies in $\tinterval{\sval{\ti},\bval{\ti}}$
gives $\GFT_{\ti}(p,q)+\Profit_{\ti}(p,q)\le\GFT_{\ti}(p,p)+\GFT_{\ti}(q,q)$.
Hence
\begin{align*}
\textstyle\sum_{\ti=1}^{\tiH}\bigl(\GFT_{\ti}(p,q)+(1+\intrate)^{\tiH-\ti}\Profit_{\ti}(p,q)\bigr)
&\textstyle\le\sum_{\ti=1}^{\tiH}
  \bigl(\GFT_{\ti}(p,q)+\Profit_{\ti}(p,q)\bigr)
\\*
&\textstyle\le2\tiH\fixedpricebenchmark.
\end{align*}

Since $(1+\intrate)^{\tiH-1}\ge1$, the bound of Case~1 holds for every
pair.  Taking expectations under any feasible $\comparatordist$, and using
$\sum_{\ti=1}^{\tiH}(1+\intrate)^{\tiH-\ti}\expect[a\sim\comparatordist]{\Profit_{\ti}(a)}\ge0$, gives
\begin{align*}
\textstyle\tiH\distributionbenchmark
&\textstyle\le
    \bigl(1+(1+\intrate)^{\tiH-1}\bigr)\tiH\fixedpricebenchmark
\\*
&\textstyle\le2\tiH\fixedpricebenchmark+\bO{\tiH^{2-\rateexponent}},
\end{align*}
where $\fixedpricebenchmark\le1$ and
$(1+\intrate)^{\tiH-1}-1=\bO{\tiH^{1-\rateexponent}}$.

Running $\gftmax$ of \citet[Theorem~5.4]{bernasconi2024} over
$\tinterval{\tiH}$ is feasible, since its cumulative profit is nonnegative
on every prefix, and has regret $\softO{\tiH^{3/4}}$ against the best
common price on $\tinterval{\tiH}$.
Together with the benchmark bound, this gives
$\alpharegret{2}{\intrate}\le\softO{\tiH^{\max\{3/4,\,2-\rateexponent\}}}$.

\paragraph{Interest rate $1/\tiH$ ($\rateexponent=1$).}
\Cref{alg:alpha-critical} splits the horizon into blocks of about
$\tiH^{4/5}$ rounds and runs $\gftmax$ on each block.

\begin{algorithm}[H]
\caption{\textsc{Blockwise GFT-Max}}
\label{alg:alpha-critical}
\begin{algorithmic}[1]
\linespread{0.9}\selectfont
    \Statex \textbf{Input:} $\tiH$.
    \For{$j=1,\ldots,\ceil{\tiH^{1/5}}$}
        \State $\suffixrounds_j\gets
            \set{\ti\in\tinterval{\tiH}:(j-1)\tiH^{4/5}<\ti\le j\tiH^{4/5}}$
        \State run $\gftmax$ on rounds $\suffixrounds_j$
    \EndFor
\end{algorithmic}
\end{algorithm}

Fix a deterministic valuation sequence and a factor $\regretfactor$, and
let $\blockgft{j}$ be the benchmark $\fixedpricebenchmark$ of the
valuations in block $\suffixrounds_j$, so that
$\card{\suffixrounds_j}\blockgft{j}
=\max_{v\in\tinterval{0,1}}\sum_{\ti\in\suffixrounds_j}\GFT_{\ti}(v,v)$.

The $\regretfactor$-regret of \Cref{alg:alpha-critical} splits as
\begin{align*}
\textstyle\tiH\distributionbenchmark
 -\regretfactor\expect{\sum_{\ti=1}^{\tiH}\GFT_{\ti}(\sprice{\ti},\bprice{\ti})}
&\textstyle=\Bigl(\tiH\distributionbenchmark
   -\regretfactor\sum_j\card{\suffixrounds_j}\blockgft{j}\Bigr)
\\*
&\textstyle\quad+\regretfactor\Bigl(\sum_j\card{\suffixrounds_j}\blockgft{j}
   -\expect{\sum_{\ti=1}^{\tiH}\GFT_{\ti}(\sprice{\ti},\bprice{\ti})}\Bigr).
\end{align*}
Each run of $\gftmax$ is feasible, since its cumulative profit is
nonnegative on every prefix of its block, and by
\citet[Theorem~5.4]{bernasconi2024} has regret
$\softO{\card{\suffixrounds_j}^{3/4}}$ against the best common price
on $\suffixrounds_j$.  As there are
$\ceil{\tiH^{1/5}}$ blocks of at most $\ceil{\tiH^{4/5}}$ rounds, the
second term is $\softO{\tiH^{4/5}}$.

We bound the first term through a Lagrangian relaxation of the budget
constraint of the benchmark.  Since every feasible $\comparatordist$
satisfies
$\sum_{\ti=1}^{\tiH}(1+\intrate)^{\tiH-\ti}\expect[a\sim\comparatordist]{\Profit_{\ti}(a)}\ge0$,
for every $\lambda\ge0$,
\begin{align*}
\textstyle\expect[a\sim\comparatordist]{\sum_{\ti=1}^{\tiH}\GFT_{\ti}(a)}
&\textstyle\le\expect[a\sim\comparatordist]{\sum_{\ti=1}^{\tiH}\bigl(\GFT_{\ti}(a)
   +\lambda(1+\intrate)^{\tiH-\ti}\Profit_{\ti}(a)\bigr)}
\\*
&\textstyle\le\sup_{(p,q)\in\tinterval{0,1}^2}
   \sum_{\ti=1}^{\tiH}\bigl(\GFT_{\ti}(p,q)+\lambda(1+\intrate)^{\tiH-\ti}\Profit_{\ti}(p,q)\bigr).
\end{align*}
If every $\blockgft{j}$ is zero, no round has positive surplus and the
first term is nonpositive.  Otherwise, let $\lambda_\star$ satisfy
\[
\textstyle\lambda_\star\sum_j(1+\intrate)^{\tiH-\max\suffixrounds_j}
 \card{\suffixrounds_j}\blockgft{j}
=(\regretfactor-1)\sum_j\card{\suffixrounds_j}\blockgft{j}.
\]
Since $\intrate=1/\tiH$, we have
$1\le(1+\intrate)^{\tiH-\max\suffixrounds_j}<\ee$, and hence
$\tfrac{\regretfactor-1}{\ee}<\lambda_\star\le\regretfactor-1$.  It remains
to bound the right side of the relaxation at $\lambda=\lambda_\star$ for
every pair $(p,q)$.

\medskip\noindent\textbf{Case 1: earning pair, $p\le q$.}
\par\nopagebreak\noindent Since
$\sum_j\card{\suffixrounds_j}\blockgft{j}\le\tiH$, splitting the sum into
blocks gives
\begin{align*}
\textstyle\sum_{\ti=1}^{\tiH}\bigl(\GFT_{\ti}(p,q)+\lambda_\star(1+\intrate)^{\tiH-\ti}\Profit_{\ti}(p,q)\bigr)
&\textstyle\stepmark{a}{\le}\sum_j\bigl(1+\lambda_\star
   (1+\intrate)^{\tiH-\max\suffixrounds_j+\tiH^{4/5}}\bigr)\card{\suffixrounds_j}\blockgft{j}
\\*
\mathllap{\mr{choice of $\lambda_\star$}}
&\textstyle=\bigl(1+(\regretfactor-1)(1+\intrate)^{\tiH^{4/5}}\bigr)
   \sum_j\card{\suffixrounds_j}\blockgft{j}
\\*
\mathllap{\mr{$(1+\tfrac1\tiH)^{\tiH^{4/5}}=1+\bO{\tiH^{-1/5}}$}}
&\textstyle\le\regretfactor\sum_j\card{\suffixrounds_j}\blockgft{j}
  +\bO{\tiH^{4/5}},
\end{align*}
where (a) uses
$0\le\Profit_{\ti}(p,q)\le\GFT_{\ti}(p,q)\le\GFT_{\ti}(p,p)$, since every
trade also occurs at the common price $p$, and
$\tiH-\ti\le\tiH-\max\suffixrounds_j+\tiH^{4/5}$ for
$\ti\in\suffixrounds_j$.

\medskip\noindent\textbf{Case 2: subsidy pair, $q<p$.}
\par\nopagebreak\noindent Every profit is nonpositive, so
$(1+\intrate)^{\tiH-\ti}\Profit_{\ti}(p,q)
\le(1+\intrate)^{\tiH-\max\suffixrounds_j}\Profit_{\ti}(p,q)\le0$ for
$\ti\in\suffixrounds_j$.  Using this bound on the weights and splitting the
rounds of a block by whether $p$ or $q$ lies in
$\tinterval{\sval{\ti},\bval{\ti}}$, we obtain
\begin{align*}
&\textstyle\sum_{\ti\in\suffixrounds_j}\bigl(\GFT_{\ti}(p,q)
 +\lambda_\star(1+\intrate)^{\tiH-\ti}\Profit_{\ti}(p,q)\bigr)
\\*
&\textstyle\qquad=\sum_{\substack{\ti\in\suffixrounds_j\\
   \set{p,q}\cap\tinterval{\sval{\ti},\bval{\ti}}\ne\emptyset}}
 \bigl(\GFT_{\ti}(p,q)+\lambda_\star(1+\intrate)^{\tiH-\ti}\Profit_{\ti}(p,q)\bigr)
\\*
&\textstyle\qquad+\sum_{\substack{\ti\in\suffixrounds_j\\
   \set{p,q}\cap\tinterval{\sval{\ti},\bval{\ti}}=\emptyset}}
 \bigl(\GFT_{\ti}(p,q)+\lambda_\star(1+\intrate)^{\tiH-\ti}\Profit_{\ti}(p,q)\bigr)
\\*
&\textstyle\qquad\stepmark{a}{\le}2\card{\suffixrounds_j}\blockgft{j}
 +\sum_{\substack{\ti\in\suffixrounds_j\\q<\sval{\ti}<\bval{\ti}<p}}
  \bigl(\bval{\ti}-\sval{\ti}
  -\lambda_\star(1+\intrate)^{\tiH-\max\suffixrounds_j}(p-q)\bigr),
\end{align*}
where (a) uses $\GFT_{\ti}(p,q)\le\GFT_{\ti}(p,p)+\GFT_{\ti}(q,q)$ in the
first sum; every trade in the second sum has $q<\sval{\ti}$ and
$\bval{\ti}<p$, and those with $\bval{\ti}\le\sval{\ti}$ are dropped,
as their gains-from-trade and profit are both nonpositive.
For each round of the last sum, let
$\rho_{\ti}\deq\tfrac{p-q}{\bval{\ti}-\sval{\ti}}\ge1$.  Then
\begin{align*}
\textstyle\bval{\ti}-\sval{\ti}
 -\lambda_\star(1+\intrate)^{\tiH-\max\suffixrounds_j}(p-q)
&\textstyle=\frac{(\bval{\ti}-\sval{\ti})^2}{p-q}
 \bigl(\rho_{\ti}-\lambda_\star(1+\intrate)^{\tiH-\max\suffixrounds_j}\rho_{\ti}^2\bigr)
\\*
&\textstyle\stepmark{b}{\le}\frac{(\bval{\ti}-\sval{\ti})^2}{p-q}\cdot
 \frac{\ee-(1+\intrate)^{\tiH-\max\suffixrounds_j}}{\ee-1}
 \max_{\rho\ge1}\plus{\rho-\lambda_\star\rho^2},
\end{align*}
where (b) holds as $\rho_{\ti}-\lambda_\star w\rho_{\ti}^2$ is linear in
$w\in\tinterval{1,\ee}$ and negative at $w=\ee$, since
$\lambda_\star\ee\rho_{\ti}>\regretfactor-1\ge1$ for $\regretfactor\ge2$.  Moreover,
\begin{align*}
\textstyle\sum_{\substack{\ti\in\suffixrounds_j\\q<\sval{\ti}<\bval{\ti}<p}}(\bval{\ti}-\sval{\ti})^2
&\textstyle=\sum_{\substack{\ti\in\suffixrounds_j\\q<\sval{\ti}<\bval{\ti}<p}}
  \int_{\sval{\ti}}^{\bval{\ti}}\GFT_{\ti}(v,v)\,\dd v
\\*
\mathllap{\mr{$\tinterval{\sval{\ti},\bval{\ti}}\subseteq\tinterval{q,p}$}}
&\textstyle\le\int_q^p\sum_{\ti\in\suffixrounds_j}\GFT_{\ti}(v,v)\,\dd v
\\*
&\textstyle\le(p-q)\card{\suffixrounds_j}\blockgft{j}.
\end{align*}
Combining the three displays and summing over blocks, where the choice of
$\lambda_\star$ gives
$\sum_j\bigl(\ee-(1+\intrate)^{\tiH-\max\suffixrounds_j}\bigr)\card{\suffixrounds_j}\blockgft{j}
=\bigl(\ee-\tfrac{\regretfactor-1}{\lambda_\star}\bigr)\sum_j\card{\suffixrounds_j}\blockgft{j}$,
yields
\begin{align*}
&\textstyle\sum_{\ti=1}^{\tiH}\bigl(\GFT_{\ti}(p,q)+\lambda_\star(1+\intrate)^{\tiH-\ti}\Profit_{\ti}(p,q)\bigr)
\\*
&\textstyle\qquad\le\Bigl(2+\frac{\ee-\frac{\regretfactor-1}{\lambda_\star}}{\ee-1}
 \max_{\rho\ge1}\plus{\rho-\lambda_\star\rho^2}\Bigr)
 \sum_j\card{\suffixrounds_j}\blockgft{j}
\\*
&\textstyle\qquad\stepmark{c}{\le}\Bigl(2+\frac{(\sqrt\ee-\sqrt{\regretfactor-1})^2}{\ee-1}\Bigr)
 \sum_j\card{\suffixrounds_j}\blockgft{j},
\end{align*}
where (c) maximizes
$\bigl(\ee-\tfrac{\regretfactor-1}{\lambda}\bigr)\plus{\rho-\lambda\rho^2}$
over $\lambda\ge\tfrac{\regretfactor-1}{\ee}$ and $\rho\ge1$ by
differentiation.

Combining the two cases with the relaxation at $\lambda=\lambda_\star$
bounds the first term by
\[
\textstyle\Bigl(\max\Bigl\{\regretfactor,\,
 2+\frac{(\sqrt\ee-\sqrt{\regretfactor-1})^2}{\ee-1}\Bigr\}-\regretfactor\Bigr)
 \sum_j\card{\suffixrounds_j}\blockgft{j}+\bO{\tiH^{4/5}}.
\]
The two entries of the maximum coincide at
$\criticalfactor\deq2+\tfrac{\ee-1}{(\sqrt\ee+\sqrt2)^2}=2.1831\ldots$, so for
$\regretfactor=\criticalfactor$ the first term is $\bO{\tiH^{4/5}}$ and
$\alpharegret{\criticalfactor}{1/\tiH}\le\softO{\tiH^{4/5}}$.

\paragraph{Strong interest ($0\le\rateexponent<1$).}
For $\rateexponent<1$, the interest rate changes what the benchmark can
achieve.  Comparing it with the best common price, as for
$\rateexponent\ge1$, only gives the factor $1+(1+\intrate)^{\tiH-1}$,
which is unbounded, so \Cref{alg:alpha-upper} competes with the benchmark
directly.  Like \Cref{alg:adv-upper}, it first collects a profit of
$1/\tiH$ and runs $\gftmax$ while this profit compounds.  Once the balance reaches
$\tiH$, it spends the balance on Exp3 over a grid of price pairs instead of
$\subsidizedgft$, and thereby competes with every price pair rather than
with the best common price.

\begin{algorithm}[tp]
\caption{\textsc{Interest-Financed Exp3}}
\label{alg:alpha-upper}
\begin{algorithmic}[1]
\linespread{0.9}\selectfont
    \Statex \textbf{Input:} $\tiH$.
    \State $\seedend\gets\tiH$
    \For{$\ti=1,\ldots,\tiH$}
        \State draw $\auxprice_{\ti}\sim\Unif\set{0,\ldots,\tiH-1}$ and post
            $(\tfrac{\auxprice_{\ti}}{\tiH},\tfrac{\auxprice_{\ti}+1}{\tiH})$
        \If{$\tradeindicator{\ti}=1$}
            \State $\seedend\gets\ti$
            \State \textbf{break}
        \EndIf
    \EndFor
    \State $\bridgeend\gets\min\bigl\{\tiH,\,
        \seedend+\bigl\lceil\tfrac{4\log\tiH}{\intrate}\bigr\rceil\bigr\}$
        \label{algline:alpha-wait}
    \If{$\seedend<\bridgeend$}
        \State run $\gftmax$ on rounds $\seedend+1,\ldots,\bridgeend$
    \EndIf
    \If{$\bridgeend<\tiH$}
        \State $\suffixrounds\gets\set{\bridgeend+1,\ldots,\tiH}$, \quad
            $\chenfamilysize\gets\ceil{\card{\suffixrounds}^{1/5}}$, \quad
            $\actiongrid\gets\{0,\tfrac{1}{\chenfamilysize},\ldots,1\}^2$
        \State $\exploreprob\gets\tfrac{1}{\chenfamilysize}$, \quad
            $\learnrate\gets\tfrac{1}{12\chenfamilysize^4}$
        \State $P_{\bridgeend+1}(a)\gets\tfrac{1}{\card{\actiongrid}}$
            for all $a\in\actiongrid$
        \For{$\ti\in\suffixrounds$}
            \State $\widehat{\GFT}_{\ti}(a)\gets0$ for all $a\in\actiongrid$
            \State draw $D_{\ti}\sim\Bern(\exploreprob)$
            \If{$D_{\ti}=0$}
                \State draw $a_{\ti}\sim P_{\ti}$ and post it
            \Else
                \State draw $a_{\ti}\sim\Unif(\actiongrid)$
                \State $\widehat{\GFT}_{\ti}(a_{\ti})\gets
                    \frac{\card{\actiongrid}}{\exploreprob}\,
                    \hyperref[alg:alpha-gft-est]{\gftest}(a_{\ti})$
                \label{algline:alpha-explore}
            \EndIf
            \State $P_{\ti+1}(a)\gets
                \frac{P_{\ti}(a)\ee^{\learnrate\widehat{\GFT}_{\ti}(a)}}
                    {\sum_{a'\in\actiongrid}P_{\ti}(a')\ee^{\learnrate\widehat{\GFT}_{\ti}(a')}}$
                for all $a\in\actiongrid$
        \EndFor
    \EndIf
\end{algorithmic}
\end{algorithm}

\begin{alphalemma}
\label{lem:alpha-grid}
\begin{flushleft}
Let $\chenfamilysize\ge1$ be an integer and
$\actiongrid=\{0,\tfrac1\chenfamilysize,\ldots,1\}^2$.  For every set of
rounds $\suffixrounds$ and every price pair $(p,q)$,
\[
    \sum_{\ti\in\suffixrounds}\GFT_{\ti}(p,q)
    \le3\max_{a\in\actiongrid}\sum_{\ti\in\suffixrounds}\GFT_{\ti}(a)
       +\tfrac{2\card{\suffixrounds}}{\chenfamilysize}.
\]
\end{flushleft}
\end{alphalemma}

\begin{proof}
Write $\floor{x}_\chenfamilysize\deq\tfrac{\floor{\chenfamilysize x}}{\chenfamilysize}$
and $\ceil{x}_\chenfamilysize\deq\tfrac{\ceil{\chenfamilysize x}}{\chenfamilysize}$.
Every trade at $(\floor{p}_\chenfamilysize,\ceil{q}_\chenfamilysize)$ also
occurs at $(p,q)$, but a trade at $(p,q)$ with
$\floor{p}_\chenfamilysize<\sval{\ti}\le p$ or
$q\le\bval{\ti}<\ceil{q}_\chenfamilysize$ may not occur there.  We bound
the gains-from-trade of these trades conservatively by their part above the
diagonal,
$\plus{\bval{\ti}-\sval{\ti}}$.  Hence
\begin{align*}
\textstyle\sum_{\ti\in\suffixrounds}\GFT_{\ti}(p,q)
&\textstyle\le\sum_{\ti\in\suffixrounds}\bigl(
  \GFT_{\ti}(\floor{p}_\chenfamilysize,\ceil{q}_\chenfamilysize)
  +\plus{\bval{\ti}-\sval{\ti}}\indicator{\floor{p}_\chenfamilysize<\sval{\ti}\le p}
\\*
&\textstyle\quad
  +\plus{\bval{\ti}-\sval{\ti}}\indicator{q\le\bval{\ti}<\ceil{q}_\chenfamilysize}\bigr)
\\
&\textstyle\stepmark{a}{\le}\sum_{\ti\in\suffixrounds}\bigl(
  \GFT_{\ti}(\floor{p}_\chenfamilysize,\ceil{q}_\chenfamilysize)
  +\GFT_{\ti}(\ceil{p}_\chenfamilysize,\floor{p}_\chenfamilysize)
  +\GFT_{\ti}(\ceil{q}_\chenfamilysize,\floor{q}_\chenfamilysize)\bigr)
  +\tfrac{2\card{\suffixrounds}}{\chenfamilysize}
\\
\mathllap{\mr{pairs in $\actiongrid$}}
&\textstyle\le3\max_{a\in\actiongrid}\sum_{\ti\in\suffixrounds}\GFT_{\ti}(a)
  +\tfrac{2\card{\suffixrounds}}{\chenfamilysize},
\end{align*}
where (a) holds as a positive-surplus trade with
$\floor{p}_\chenfamilysize<\sval{\ti}\le p$ occurs at
$(\ceil{p}_\chenfamilysize,\floor{p}_\chenfamilysize)$, every trade of which
has gains-from-trade at least
$\floor{p}_\chenfamilysize-\ceil{p}_\chenfamilysize\ge-\tfrac1\chenfamilysize$,
and symmetrically for $q$.
\end{proof}

Fix a deterministic valuation sequence.  The first two phases of
\Cref{alg:alpha-upper} coincide with those of \Cref{alg:adv-upper}, so by the
proof of \Cref{thm:adv-upper} the account stays nonnegative through them and,
if $\bridgeend<\tiH$, reaches $\tiH$, which covers the loss of at most one
in each later round.

Fix $\bridgeend<\tiH$.  In each round
$\ti\in\suffixrounds$, $\gftest$ is called on each $a\in\actiongrid$
with probability $\tfrac{\exploreprob}{\card{\actiongrid}}$, so
$\widehat{\GFT}_{\ti}(a)$ is an unbiased estimate of $\GFT_{\ti}(a)$
given the history before round $\ti$ \citep[Lemma~4.2]{lunghi2025}.  Since
$\card{\gftest}\le3$, it also satisfies
$\learnrate\widehat{\GFT}_{\ti}(a)
\le\tfrac{3\learnrate\card{\actiongrid}}{\exploreprob}\le1$ and
$\expect{\sum_aP_{\ti}(a)\widehat{\GFT}_{\ti}(a)^2}
\le\tfrac{9\card{\actiongrid}}{\exploreprob}$.

The regret bound of Exp3 in \citet[Chapter~11]{lattimore2020bandit} uses
only these properties of the estimates, and an exploration round loses at
most one more, since the gains-from-trade of every action at round $\ti$
are $0$ or $\bval{\ti}-\sval{\ti}$.  Hence,
\begin{align}
&\textstyle\max_{a\in\tinterval{0,1}^2}\sum_{\ti\in\suffixrounds}\GFT_{\ti}(a)
  -3\expect{\sum_{\ti\in\suffixrounds}\GFT_{\ti}(\sprice{\ti},\bprice{\ti})}
\notag\\*
&\textstyle\qquad\le3\Bigl(\max_{a\in\actiongrid}\sum_{\ti\in\suffixrounds}\GFT_{\ti}(a)
  -\expect{\sum_{\ti\in\suffixrounds}\GFT_{\ti}(\sprice{\ti},\bprice{\ti})}\Bigr)
  +\tfrac{2\card{\suffixrounds}}{\chenfamilysize}
\tag*{\Cref{lem:alpha-grid}}
\\*
&\textstyle\qquad\le3\Bigl(\tfrac{\log\card{\actiongrid}}{\learnrate}
  +\tfrac{9\learnrate\card{\actiongrid}\card{\suffixrounds}}{\exploreprob}
  +\exploreprob\card{\suffixrounds}\Bigr)
  +\tfrac{2\card{\suffixrounds}}{\chenfamilysize}
\notag\\*
&\textstyle\qquad\mathllap{\mr{$\chenfamilysize\approx\card{\suffixrounds}^{1/5}$}}
  =\softO{\card{\suffixrounds}^{4/5}}.
\label{eq:alpha-suffix}
\end{align}

Since
$\tiH\distributionbenchmark\le\max_a\sum_{\ti=1}^{\tiH}\GFT_{\ti}(a)$, with
$\suffixrounds=\emptyset$ when $\bridgeend=\tiH$,
\begin{align*}
\textstyle\tiH\distributionbenchmark
    -3\expect{\sum_{\ti=1}^{\tiH}\GFT_{\ti}(\sprice{\ti},\bprice{\ti})}
&\textstyle\stepmark{a}{\le}
    \expect{\sum_{\ti=1}^{\seedend}\plus{\bval{\ti}-\sval{\ti}}
      \parent{1-\tradeindicator{\ti}}}
    +\expect{\bridgeend-\seedend}\\*
&\textstyle\quad+
    \expect{\max_{a\in\tinterval{0,1}^2}\sum_{\ti\in\suffixrounds}\GFT_{\ti}(a)
        -3\sum_{\ti\in\suffixrounds}\GFT_{\ti}(\sprice{\ti},\bprice{\ti})}\\*
\mathllap{\mr{line~\ref{algline:alpha-wait} of \Cref{alg:alpha-upper}}}
&\textstyle\le3+\ceil{\tfrac{4\log\tiH}{\intrate}}+\softO{\tiH^{4/5}}
\tag*{\Cref{eq:adv-seed-regret,eq:alpha-suffix}}
\\*
&\textstyle=\softO{\tiH^{\max\{4/5,\,\rateexponent\}}},
\end{align*}
where (a) holds because a first-phase trade collects its whole surplus and
the completed $\gftmax$ call has nonnegative total gains-from-trade, as
$\GFT_{\ti}\ge\Profit_{\ti}$ on every trade.

\subsection{Lower Bound}
\label{sec:alpha-lower}

We now prove the lower bounds of \Cref{thm:alpha-main}.  In each hard
instance, a hidden subsidy pair trades every point of positive surplus, while
every action outside its good place, two thin strips around it, earns
substantially less.  We construct $4^{\tiH}$ such instances whose vertical
strips are pairwise disjoint, as are their horizontal strips, so that every
action lies in the good places of at most two instances, and an algorithm
cannot locate the good place of the hidden pair from its feedback.

The benchmark distribution, in contrast, can place part of its mass on the
hidden pair and thus inflict regret on the algorithm.  To obtain a feasible
distribution, we follow the idea of \Cref{alg:adv-upper}, where profit is
earned first and grows with interest before it pays for subsidies.  The
benchmark mixes the hidden pair with two earning pairs, and the valuations
pass through three phases.  The earning pairs collect profit during the first
$\floor{d\tiH}$ rounds.  This profit then accrues interest over a long middle
phase, in which every valuation is $(1,0)$ and none of these pairs trades.  In
the last $\floor{d\tiH}$ rounds, the compounded profit finances the subsidies
of the hidden pair at a central point that neither earning pair trades.  The
interest accrued in between thus decides how much weight the hidden pair can
receive, and hence the optimal competitive ratio.

\begin{figure}[!t]
\centering
\begin{tikzpicture}[alr picture,font=\small]
\begin{scope}[x=7.4cm,y=7.4cm]
    \path[alr bad region] (0,0) rectangle (1,1);
    \path[alr good region] (.62,0) rectangle (.68,1);
    \path[alr good region] (0,.38) rectangle (1,.44);
    \path[alr good region j] (.80,0) rectangle (.86,1);
    \path[alr good region j] (0,.24) rectangle (1,.30);
    \draw[alr diagonal] (0,0) -- (1,1);
    \draw[alr axis] (0,0) rectangle (1,1);
    \draw[alr guide] (.40,0) -- (.40,.60);
    \draw[alr guide] (0,.60) -- (.86,.60);

    \draw[alr dim] (.62,.90) -- (.68,.90);
    \node[alr legend,anchor=west] at (.69,.90) {$\needlewidth$};
    \draw[alr dim] (.93,.38) -- (.93,.44);
    \node[alr legend,anchor=west] at (.94,.41) {$\needlewidth$};
    \draw[alr dim] (.10,.30) -- (.10,.44);
    \node[alr legend,anchor=west] at (.11,.34) {$(j-i)\needlewidth$};

    \node[alr earning] at (.62,1) {};
    \node[alr earning] at (0,.44) {};
    \node[alr trap] at (.68,.60) {};
    \node[alr trap] at (.40,.38) {};
    \node[alr earning j] at (.80,1) {};
    \node[alr earning j] at (0,.30) {};
    \node[alr trap j] at (.86,.60) {};
    \node[alr trap j] at (.40,.24) {};
    \node[alr central] at (.40,.60) {};
    \node[alr residual] at (1,0) {};

    \foreach \x/\lab in {.40/\tfrac12,.62/\sprice{i}^{\star},.80/\sprice{j}^{\star}}
        \draw[alr axis,line width=.5pt] (\x,0) -- (\x,-.014)
            node[below=1pt,font=\footnotesize] {$\lab$};
    \foreach \y/\lab in {.30/\bprice{j}^{\star},.44/\bprice{i}^{\star},.60/\tfrac12+d}
        \draw[alr axis,line width=.5pt] (0,\y) -- (-.014,\y)
            node[left=1pt,font=\footnotesize] {$\lab$};
    \node[anchor=north] at (.5,-.085) {seller value $\sval{}$};
    \node[rotate=90,anchor=south] at (-.20,.5) {buyer value $\bval{}$};

    \node[alr earning] at (1.075,.90) {};
    \node[alr earning j] at (1.115,.90) {};
    \node[alr legend,anchor=west] at (1.145,.90) {mass $\tfrac{d^2}2$};
    \node[alr central] at (1.075,.82) {};
    \node[alr legend,anchor=west] at (1.145,.82) {mass $\tfrac d4$};
    \node[alr trap] at (1.075,.74) {};
    \node[alr trap j] at (1.115,.74) {};
    \node[alr legend,anchor=west] at (1.145,.74) {mass $\tfrac14$};
    \node[alr residual] at (1.075,.66) {};
    \node[alr legend,anchor=west] at (1.145,.66) {remaining mass};
    \path[alr good region] (1.055,.565) rectangle (1.095,.595);
    \draw[black,line width=.5pt] (1.055,.565) rectangle (1.095,.595);
    \node[alr legend,anchor=west] at (1.145,.58) {$\goodset_i$};
    \path[alr good region j] (1.055,.485) rectangle (1.095,.515);
    \draw[black,line width=.5pt] (1.055,.485) rectangle (1.095,.515);
    \node[alr legend,anchor=west] at (1.145,.50) {$\goodset_j$};
\end{scope}
\end{tikzpicture}
\caption{Supports of $\mathcal D^i$ (filled) and $\mathcal D^j$ (hollow)
for two possibilities $i<j$, and their good places; all other actions are
bad.  The figure is not to scale.}
\label{fig:alpha-hard-instance}
\end{figure}
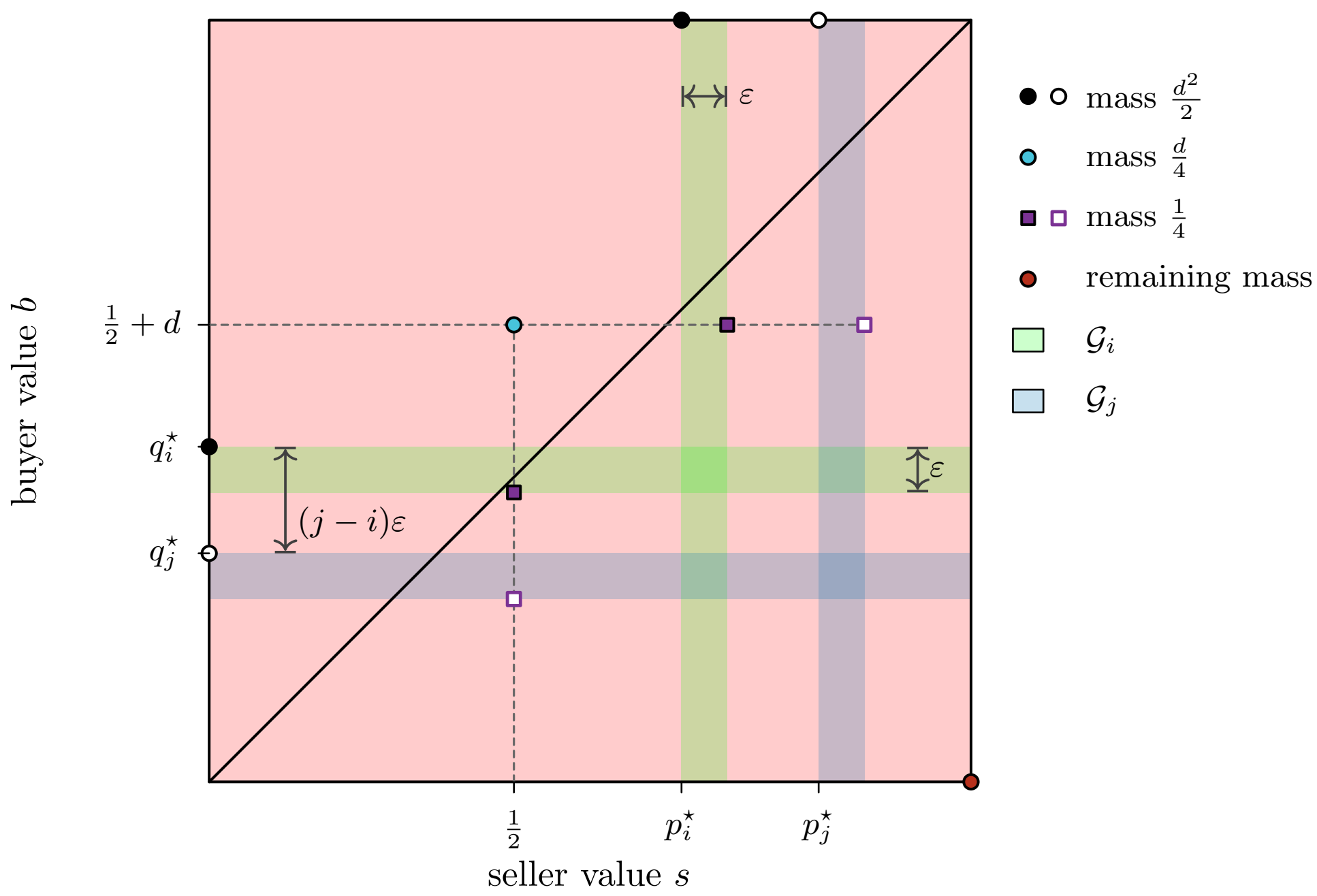

Fix $1\le\regretfactor<\critfactor{\rateexponent}$.  For all sufficiently
large $\tiH$, choose
\[
    d\deq\tfrac{\critfactor{\rateexponent}-\regretfactor}{100},
    \qquad
    \zeta\deq\begin{cases}
        d,&0\le\rateexponent<1,\\
        \sqrt{\tfrac{2}{\ee}},&\rateexponent\ge1,
    \end{cases}
    \qquad
    \chenfamilysize\deq4^{\tiH},
    \qquad
    \needlewidth\deq\tfrac{d^2}{\chenfamilysize},
\]
so that $0<d\le\tfrac1{50}$ and $1\le\floor{d\tiH}\le\tfrac{\tiH}2$.
For each possibility $i\in\set{\chenfamilysize,\ldots,2\chenfamilysize-1}$,
the hidden pair is $(\sprice{i}^{\star},\bprice{i}^{\star})$, where
$\sprice{i}^{\star}\deq\tfrac12+d+i\needlewidth$ and
$\bprice{i}^{\star}\deq\tfrac12-i\needlewidth$.  In possibility $i$, the
valuations are independent across rounds but not identically distributed:
\[
    (\sval{\ti},\bval{\ti})
    \begin{cases}
        \sim\widetilde{\mathcal D}^i, & 1\le\ti\le\floor{d\tiH},\\[1mm]
        =(1,0), & \floor{d\tiH}<\ti\le\tiH-\floor{d\tiH},\\[1mm]
        \sim\mathcal D^i, & \tiH-\floor{d\tiH}<\ti\le\tiH,
    \end{cases}
\]
where
\[
    \widetilde{\mathcal D}^i(\sval{},\bval{})
    \deq
    \begin{cases}
        \tfrac{\zeta d^2}2,
            &(\sval{},\bval{})\in\set{(\sprice{i}^{\star},1),(0,\bprice{i}^{\star})},\\[1mm]
        \tfrac \zeta4,
            &(\sval{},\bval{})\in\set{(\sprice{i}^{\star}+\needlewidth,\tfrac12+d),
             (\tfrac12,\bprice{i}^{\star}-\needlewidth)},\\[1mm]
        1-\zeta d^2-\tfrac \zeta2,
            &(\sval{},\bval{})=(1,0),
    \end{cases}
\]
and
\[
    \mathcal D^i(\sval{},\bval{})
    \deq
    \begin{cases}
        \tfrac{d^2}2,
            &(\sval{},\bval{})\in\set{(\sprice{i}^{\star},1),(0,\bprice{i}^{\star})},\\[1mm]
        \tfrac d4,
            &(\sval{},\bval{})=(\tfrac12,\tfrac12+d),\\[1mm]
        \tfrac14,
            &(\sval{},\bval{})\in\set{(\sprice{i}^{\star}+\needlewidth,\tfrac12+d),
             (\tfrac12,\bprice{i}^{\star}-\needlewidth)},\\[1mm]
        \tfrac12-\tfrac d4-d^2,
            &(\sval{},\bval{})=(1,0).
    \end{cases}
\]
The hidden pair trades the two earning points $(\sprice{i}^{\star},1)$ and
$(0,\bprice{i}^{\star})$ and the central point $(\tfrac12,\tfrac12+d)$,
which have positive surplus; the two points of mass $\tfrac14$ have
$\sval{}>\bval{}$ and lie just beyond its prices.  See
\Cref{fig:alpha-hard-instance}.

Algorithms do not know $i$, which we draw uniformly from
$\set{\chenfamilysize,\ldots,2\chenfamilysize-1}$, independently of their
internal randomization.

The actions split into a good place, around the hidden pair, and a bad place:
\[
    \goodset_i\deq\set{(p,q)\in\tinterval{0,1}^2:\
      p\in\roitv{\sprice{i}^{\star},\sprice{i}^{\star}+\needlewidth}
      \text{ or }
      q\in\loitv{\bprice{i}^{\star}-\needlewidth,\bprice{i}^{\star}}},
    \qquad
    \badset_i\deq\tinterval{0,1}^2\setminus\goodset_i.
\]

\begin{alphalemma}
\label{lem:alpha-lower-gain}
\begin{flushleft}
For every $i\in\set{\chenfamilysize,\ldots,2\chenfamilysize-1}$,
$(p,q)\in\tinterval{0,1}^2$, and round $\ti$,
\[
    \expect[(\sval{\ti},\bval{\ti})\sim\mathcal D^i]{\GFT_{\ti}(p,q)}
    \le
    \begin{cases}
        \tfrac{d^2}4, & (p,q)\in\badset_i,\\[1mm]
        \tfrac{d^2}4(3-2d-4i\needlewidth), & (p,q)\in\goodset_i,
    \end{cases}
\]
and
\[
    \expect[(\sval{\ti},\bval{\ti})\sim\widetilde{\mathcal D}^i]{\GFT_{\ti}(p,q)}
    \le
    \begin{cases}
        \tfrac{\zeta d^2}4, & (p,q)\in\badset_i,\\[1mm]
        \tfrac{\zeta d^2}4(2-2d-4i\needlewidth), & (p,q)\in\goodset_i.
    \end{cases}
\]
\end{flushleft}
\end{alphalemma}

\begin{proof}
Under $\mathcal D^i$, a traded earning point contributes at most
$\tfrac{d^2}4$, the central point $\tfrac{d^2}4$, and a point of mass
$\tfrac14$ at most $-\tfrac{d^2}4$.
\begin{itemize}[leftmargin=*]
\item \emph{For $(p,q)\in\goodset_i$.}  Only the two earning points and the
  central point contribute positively, so no action earns more than their
  total contribution
  \[
      \tfrac{d^2}2(1-\sprice{i}^{\star})+\tfrac{d^2}2\bprice{i}^{\star}
      +\tfrac d4\cdot d
      =\tfrac{d^2}4(3-2d-4i\needlewidth).
  \]
\item \emph{For $(p,q)\in\badset_i$.}  If $(p,q)$ does not trade the
  central point, it trades at most one earning point.  Otherwise
  $p\ge\tfrac12$ and $q\le\tfrac12+d$.  If $(p,q)$ then trades
  $(\sprice{i}^{\star},1)$, it has $p\ge\sprice{i}^{\star}+\needlewidth$ as
  $(p,q)\notin\goodset_i$, so it also trades
  $(\sprice{i}^{\star}+\needlewidth,\tfrac12+d)$; symmetrically, if it trades
  $(0,\bprice{i}^{\star})$, it also trades
  $(\tfrac12,\bprice{i}^{\star}-\needlewidth)$.  Each of these pairs
  contributes nonpositively.  Either way, $(p,q)$ earns at most
  $\tfrac{d^2}4$.
\end{itemize}
The bounds under $\widetilde{\mathcal D}^i$ follow in the same way.
\end{proof}

Let
\[
    \textstyle T_{\goodset_i}\deq\sum_{\ti=1}^{\tiH}\indicator{(\sprice{\ti},\bprice{\ti})\in\goodset_i}
\]
be the number of rounds in which an algorithm posts a good action.

\begin{alphalemma}
\label{lem:alpha-lower-learning}
\begin{flushleft}
Every algorithm satisfies $\expect{T_{\goodset_i}}<\tfrac23$, even with
two-bit feedback and without the budget constraint.
\end{flushleft}
\end{alphalemma}

\begin{proof}
Fix the algorithm's internal randomization.  The action of round $\ti$ is
then determined by
the two-bit feedback of the previous rounds, so it takes at most
$4^{\ti-1}$ values.  The intervals
$\roitv{\sprice{i}^{\star},\sprice{i}^{\star}+\needlewidth}$ are pairwise
disjoint, and so are
$\loitv{\bprice{i}^{\star}-\needlewidth,\bprice{i}^{\star}}$, so each of
these values lies in $\goodset_i$ for at most two indices $i$.  Since
$i$ is uniform and independent of this randomization,
\begin{align*}
\expect{T_{\goodset_i}}
&\textstyle=\sum_{\ti=1}^{\tiH}\prob{(\sprice{\ti},\bprice{\ti})\in\goodset_i}
\\
&\textstyle\le\sum_{\ti=1}^{\tiH}\tfrac{2\cdot4^{\ti-1}}{\chenfamilysize}
\\
\mr{$\chenfamilysize=4^{\tiH}$}
&\textstyle<\tfrac23.
\end{align*}
Averaging over the internal randomization proves the claim.
\end{proof}

Since $\tiH\distributionbenchmark$ is the gains-from-trade of the best
feasible distribution, we bound it from below by those of a feasible
distribution on the hidden pair and the earning pairs
$(\sprice{i}^{\star},1)$ and $(0,\bprice{i}^{\star})$.  For each realized
valuation sequence, let
\[
    \comparatordist_i(a)\deq
    \begin{cases}
        w_i, & a=(\sprice{i}^{\star},\bprice{i}^{\star}),\\[1mm]
        \tfrac{1-w_i}2, & a\in\set{(\sprice{i}^{\star},1),(0,\bprice{i}^{\star})}.
    \end{cases}
\]
With
\[
    \textstyle X\deq\sum_{\ti=1}^{\tiH}(1+\intrate)^{\tiH-\ti}
      \bigl(\Profit_{\ti}(\sprice{i}^{\star},1)+\Profit_{\ti}(0,\bprice{i}^{\star})\bigr),
    \qquad
    Y\deq X-2\sum_{\ti=1}^{\tiH}(1+\intrate)^{\tiH-\ti}
      \Profit_{\ti}(\sprice{i}^{\star},\bprice{i}^{\star}),
\]
the expected account of $\comparatordist_i$ equals $\tfrac12(X-w_iY)$.
Since the earning pairs only earn and the hidden pair only subsidizes,
$0\le X\le Y$, and we set $w_i\deq\tfrac XY$ if $Y>0$ and $w_i\deq0$
otherwise.  Then $w_i\in\tinterval{0,1}$ and the account is zero, so
$\comparatordist_i$ is feasible.

\begin{alphalemma}
\label{lem:alpha-lower-benchmark}
\begin{flushleft}
For the distributions $\comparatordist_i$ constructed above,
\[
    \textstyle\liminf_{\tiH\to\infty}\tfrac1{\tiH}
    \expect{\sum_{\ti=1}^{\tiH}\expect[a\sim\comparatordist_i]{\GFT_{\ti}(a)}}
    \ge\tfrac{d^3(1+\zeta)}4\bigl(\critfactor{\rateexponent}-14d\bigr).
\]
\end{flushleft}
\end{alphalemma}

\begin{proof}
Condition on $i$; all estimates below are uniform in $i$.  By the
definition of $\comparatordist_i$,
\begin{align*}
    &\textstyle\sum_{\ti=1}^{\tiH}\expect[a\sim\comparatordist_i]{\GFT_{\ti}(a)}
    \\*
    &\textstyle\qquad=w_i
      \sum_{\ti=1}^{\tiH}\GFT_{\ti}(\sprice{i}^{\star},\bprice{i}^{\star})
    +\tfrac{1-w_i}2\sum_{\ti=1}^{\tiH}\bigl(\GFT_{\ti}(\sprice{i}^{\star},1)+\GFT_{\ti}(0,\bprice{i}^{\star})\bigr).
\end{align*}
The per-round means are
\begin{gather*}
    \expect{\GFT_{\ti}(\sprice{i}^{\star},\bprice{i}^{\star})}
    =\begin{cases}
        \tfrac{\zeta d^2}2(1-d-2i\needlewidth), & \ti\le\floor{d\tiH},\\[1mm]
        \tfrac{d^2}2(1-d-2i\needlewidth)+\tfrac{d^2}4, & \ti>\tiH-\floor{d\tiH},\\[1mm]
        0, & \text{otherwise},
    \end{cases}
    \\
    \expect{\GFT_{\ti}(\sprice{i}^{\star},1)+\GFT_{\ti}(0,\bprice{i}^{\star})}
    =\begin{cases}
        \tfrac{\zeta d^2}2(1-d-2i\needlewidth), & \ti\le\floor{d\tiH},\\[1mm]
        \tfrac{d^2}2(1-d-2i\needlewidth), & \ti>\tiH-\floor{d\tiH},\\[1mm]
        0, & \text{otherwise},
    \end{cases}
    \\
    -\expect{\Profit_{\ti}(\sprice{i}^{\star},\bprice{i}^{\star})}
    =\begin{cases}
        (d+2i\needlewidth)\zeta d^2, & \ti\le\floor{d\tiH},\\[1mm]
        (d+2i\needlewidth)\bigl(d^2+\tfrac d4\bigr), & \ti>\tiH-\floor{d\tiH},\\[1mm]
        0, & \text{otherwise}.
    \end{cases}
\end{gather*}
Let $x_{\tiH}\deq(1+\intrate)^{\tiH-\floor{d\tiH}}$.  We now show that, for
every $\rateexponent\ge0$,
\[
    \textstyle\expect{w_i}\ge(1-5d)\frac{\zeta x_{\tiH}+1}{\zeta x_{\tiH}+2}-\smallO{}{1}.
\]
\begin{itemize}[leftmargin=*]
\item \emph{For $\rateexponent\ge1$.}  Let
  $\bar w_i\deq\tfrac{\expect{X}}{\expect{Y}}$.  As
  $(1+\intrate)^{\tiH-\ti}\in\tinterval{1,\ee}$, $X$ and $Y$ are sums of
  $\tiH$ independent terms in $\tinterval{0,4\ee}$, and
  $\expect{Y}\ge\expect{X}\ge\tfrac{d^2}2(1-d-2i\needlewidth)\floor{d\tiH}$.
  Moreover, the discount weight of round $\ti\le\floor{d\tiH}$ is
  $x_{\tiH}$ times that of round $\ti+\tiH-\floor{d\tiH}$.  Hence
  \begin{align*}
      \expect{w_i}
      &\ge\bar w_i-\expect{|w_i-\bar w_i|}
      \\
      &\textstyle\stepmark{a}{\ge}\bar w_i
        -\frac{\expect{|X-\expect{X}|}+\expect{|Y-\expect{Y}|}}{\expect{Y}}
      \\
      &\textstyle\stepmark{b}{\ge}\frac{(1-d-2i\needlewidth)(\zeta x_{\tiH}+1)}
            {(1+3d+6i\needlewidth)(\zeta x_{\tiH}+1)+1+2i\needlewidth/d}
        -\frac{4\ee\sqrt\tiH}{\frac{d^2}2(1-d-2i\needlewidth)\floor{d\tiH}}
      \\
      &\textstyle\stepmark{c}{\ge}(1-5d)\frac{\zeta x_{\tiH}+1}{\zeta x_{\tiH}+2}
        -\bO{\tiH^{-1/2}},
  \end{align*}
  where (a) uses $0\le X\le Y$, (b) uses the per-round means and
  $\mathrm{Var}(X),\mathrm{Var}(Y)\le4\ee^2\tiH$, and (c) uses
  $i\needlewidth<2d^2$ and $d\le\tfrac1{50}$.
\item \emph{For $0\le\rateexponent<1$.}  The hidden pair trades exactly at
  the earning points and the central point, each time with profit
  $\bprice{i}^{\star}-\sprice{i}^{\star}=-(d+2i\needlewidth)$, while
  $(\sprice{i}^{\star},1)$ and $(0,\bprice{i}^{\star})$ trade only at
  their own earning points, with profits $1-\sprice{i}^{\star}$ and
  $\bprice{i}^{\star}$.  The central point occurs only in the last
  $\floor{d\tiH}$ rounds.  Hence
  \begin{align*}
      Y-X
      &\textstyle=-2\sum_{\ti=1}^{\tiH}(1+\intrate)^{\tiH-\ti}
        \Profit_{\ti}(\sprice{i}^{\star},\bprice{i}^{\star})
      \\*
      &\textstyle=2(d+2i\needlewidth)\sum_{\ti=1}^{\tiH}(1+\intrate)^{\tiH-\ti}
        \Bigl(\frac{\Profit_{\ti}(\sprice{i}^{\star},1)}{1-\sprice{i}^{\star}}
        +\frac{\Profit_{\ti}(0,\bprice{i}^{\star})}{\bprice{i}^{\star}}
        +\indicator{(\sval{\ti},\bval{\ti})=(\tfrac12,\tfrac12+d)}\Bigr)
      \\*
      &\textstyle\stepmark{a}{\le}2(d+2i\needlewidth)\sum_{\ti=1}^{\tiH}(1+\intrate)^{\tiH-\ti}
        \Bigl(\frac{\Profit_{\ti}(\sprice{i}^{\star},1)+\Profit_{\ti}(0,\bprice{i}^{\star})}{1-\sprice{i}^{\star}}
        +\indicator{(\sval{\ti},\bval{\ti})=(\tfrac12,\tfrac12+d)}\Bigr)
      \\*
      &\textstyle=2(d+2i\needlewidth)\Bigl(\frac X{1-\sprice{i}^{\star}}
        +\sum_{\ti=1}^{\tiH}(1+\intrate)^{\tiH-\ti}
        \indicator{(\sval{\ti},\bval{\ti})=(\tfrac12,\tfrac12+d)}\Bigr)
      \\*
      &\textstyle\le2(d+2i\needlewidth)
        \Bigl(\frac X{1-\sprice{i}^{\star}}+\floor{d\tiH}(1+\intrate)^{\floor{d\tiH}}\Bigr),
  \end{align*}
  where (a) uses $\bprice{i}^{\star}\ge1-\sprice{i}^{\star}$.
  If an earning point occurs in the first $\floor{d\tiH}$ rounds, which has
  probability at least $1-\ee^{-\zeta d^2\floor{d\tiH}}$, then
  $X\ge(1-\sprice{i}^{\star})x_{\tiH}$, and
  \begin{align*}
      w_i=\Bigl(1+\tfrac{Y-X}X\Bigr)^{-1}
      &\textstyle\ge\Bigl[1+\frac{2(d+2i\needlewidth)}{1-\sprice{i}^{\star}}
         \bigl(1+\floor{d\tiH}(1+\intrate)^{-(\tiH-2\floor{d\tiH})}\bigr)\Bigr]^{-1}
      \\
      &\textstyle\stepmark{a}{\ge}\Bigl[1+\frac{2(d+2i\needlewidth)}{1-\sprice{i}^{\star}}
         \bigl(1+d\tiH\,\ee^{-(1-2d)\tiH^{1-\rateexponent}/2}\bigr)\Bigr]^{-1}
      \\
      &\textstyle\stepmark{b}{\ge}\frac1{1+5d}-\smallO{}{1}
      \ge1-5d-\smallO{}{1},
  \end{align*}
  where (a) uses
  $\log(1+\tiH^{-\rateexponent})\ge\tfrac12\tiH^{-\rateexponent}$, and (b)
  uses $\tfrac{2(d+2i\needlewidth)}{1-\sprice{i}^{\star}}\le5d$.
  Hence $\expect{w_i}\ge1-5d-\smallO{}{1}$, and the bound follows.
\end{itemize}
Now let
\[
    \textstyle f(x)\deq2+\frac{\zeta(x-1)}{(\zeta x+2)(\zeta+1)}.
\]
For every $\rateexponent\ge0$,
\begin{align*}
    &\textstyle\tfrac1{\tiH}\expect{\sum_{\ti=1}^{\tiH}\expect[a\sim\comparatordist_i]{\GFT_{\ti}(a)}}
    \\*
    &\textstyle\qquad\stepmark{a}{\ge}\tfrac{\expect{w_i}}{\tiH}\,
      \expect{\sum_{\ti=1}^{\tiH}\GFT_{\ti}(\sprice{i}^{\star},\bprice{i}^{\star})}
      +\tfrac{1-\expect{w_i}}{2\tiH}\,
      \expect{\sum_{\ti=1}^{\tiH}\bigl(\GFT_{\ti}(\sprice{i}^{\star},1)+\GFT_{\ti}(0,\bprice{i}^{\star})\bigr)}
      -\bO{\tiH^{-1/2}}
    \\*
    &\textstyle\qquad\stepmark{b}{\ge}\tfrac{d^3(1+\zeta)}4
      \bigl((1+\expect{w_i})(1-d-2i\needlewidth)+\tfrac{\expect{w_i}}{1+\zeta}\bigr)
      -\bO{\tiH^{-1/2}}
    \\*
    &\textstyle\qquad\stepmark{c}{\ge}\tfrac{d^3(1+\zeta)}4
      \bigl(f(x_{\tiH})-13d\bigr)-\smallO{}{1},
\end{align*}
where (a) uses $w_i\in\tinterval{0,1}$ and that both sums are sums of
$\tiH$ independent terms in $\tinterval{0,1}$, (b) uses the per-round
means, and (c) uses the bound on $\expect{w_i}$ above,
$f(x_{\tiH})=1+\tfrac{\zeta+2}{\zeta+1}\cdot\tfrac{\zeta x_{\tiH}+1}{\zeta x_{\tiH}+2}$,
and $i\needlewidth<2d^2\le\tfrac d{25}$.
Hence, to prove the lemma, it suffices to show that
\[
    \textstyle\lim_{\tiH\to\infty}f(x_{\tiH})\ge\critfactor{\rateexponent}-d.
\]
\begin{itemize}[leftmargin=*]
\item \emph{For $\rateexponent>1$.}  Here $x_{\tiH}\to1$, and $f(1)=2$.
\item \emph{For $\rateexponent=1$.}  Here $x_{\tiH}\to\ee^{1-d}$ and
  $\zeta=\sqrt{\tfrac2\ee}$.  Since
  $\tfrac{\dd}{\dd x}f(\ee^x)
  =\tfrac{\zeta\ee^x(\zeta+2)}{(\zeta\ee^x+2)^2(\zeta+1)}\le\tfrac14$, we
  have
  \[
      \textstyle f(\ee^{1-d})
      \ge f(\ee)-\tfrac d4
      =2+\frac{\ee-1}{(\sqrt\ee+\sqrt2)^2}-\tfrac d4
      =\criticalfactor-\tfrac d4.
  \]
\item \emph{For $0\le\rateexponent<1$.}  Here $x_{\tiH}\to\infty$, and
  with $\zeta=d$, $\lim_{x\to\infty}f(x)=2+\tfrac1{1+d}\ge3-d$.
\end{itemize}
Averaging over $i$ proves the claim.
\end{proof}

\begin{proof}[Proof of the lower bounds in \Cref{thm:alpha-main}]
Fix an algorithm.  Given $i$, the valuations of each round are independent of
the history and of the action in that round, so bounding each round
conditionally on $i$ and then averaging over $i$ gives
\begin{align*}
    \textstyle\expect{\sum_{\ti=1}^{\tiH}\GFT_{\ti}(\sprice{\ti},\bprice{\ti})}
    &\textstyle=\expect{\sum_{\ti=1}^{\tiH}\GFT_{\ti}(\sprice{\ti},\bprice{\ti})
      \indicator{(\sprice{\ti},\bprice{\ti})\in\badset_i}}
    \\*
    &\textstyle\quad+\expect{\sum_{\ti=1}^{\tiH}\GFT_{\ti}(\sprice{\ti},\bprice{\ti})
      \indicator{(\sprice{\ti},\bprice{\ti})\in\goodset_i}}
    \\
    &\textstyle\le\tfrac{(1+\zeta)d^2}4\floor{d\tiH}+\tfrac{3d^2}4\,\expect{T_{\goodset_i}}
    \tag*{\Cref{lem:alpha-lower-gain}}
    \\
    &\textstyle<\tfrac{(1+\zeta)d^2}4\floor{d\tiH}+\tfrac{d^2}2.
    \tag*{\Cref{lem:alpha-lower-learning}}
\end{align*}
Since $\comparatordist_i$ is feasible, \Cref{lem:alpha-lower-benchmark}
and the choice of $d$ give
\begin{align*}
    \textstyle\liminf_{\tiH\to\infty}\tfrac1{\tiH}
    \expect{\tiH\distributionbenchmark
       -\regretfactor\sum_{\ti=1}^{\tiH}\GFT_{\ti}(\sprice{\ti},\bprice{\ti})}
    &\textstyle\ge\tfrac{d^3(1+\zeta)}4
       \bigl(\critfactor{\rateexponent}-14d-\regretfactor\bigr)
    \\*
    &\textstyle=\tfrac{43d^3(1+\zeta)}{200}
       \bigl(\critfactor{\rateexponent}-\regretfactor\bigr)>0.
\end{align*}
Hence, for each algorithm, some valuation sequence in the support of the prior
yields this bound.
\end{proof}

\newpage
\printbibliography

@article{cesabianchi2021extended,
  title={Bilateral Trade: A Regret Minimization Perspective},
  author={Cesa-Bianchi, Nicol\`o and Cesari, Tommaso and Colomboni, Roberto and Fusco, Federico and Leonardi, Stefano},
  journal={Mathematics of Operations Research},
  volume={49},
  number={1},
  pages={171--203},
  year={2024},
  eprint={2109.12974},
  archiveprefix={arXiv}
}

@inproceedings{cesabianchi2023colt,
  title={Repeated Bilateral Trade Against a Smoothed Adversary},
  author={Cesa-Bianchi, Nicol\`o and Cesari, Tommaso and Colomboni, Roberto and Fusco, Federico and Leonardi, Stefano},
  booktitle={Proceedings of the 36th Conference on Learning Theory},
  series={Proceedings of Machine Learning Research},
  volume={195},
  pages={1095--1130},
  year={2023},
  url={https://proceedings.mlr.press/v195/cesa-bianchi23a.html}
}

@inproceedings{azar2022,
  title={An {$\alpha$}-Regret Analysis of Adversarial Bilateral Trade},
  author={Azar, Yossi and Fiat, Amos and Fusco, Federico},
  booktitle={Advances in Neural Information Processing Systems},
  volume={35},
  pages={1685--1697},
  year={2022}
}

@inproceedings{bernasconi2024,
  title={No-Regret Learning in Bilateral Trade via Global Budget Balance},
  author={Bernasconi, Martino and Castiglioni, Matteo and Celli, Andrea and Fusco, Federico},
  booktitle={Proceedings of the 56th Annual ACM Symposium on Theory of Computing},
  pages={247--258},
  year={2024},
  doi={10.1145/3618260.3649653}
}

@inproceedings{bacchiocchi2025,
  title={Online Bilateral Trade With Minimal Feedback: Don't Waste Seller's Time},
  author={Bacchiocchi, Francesco and Castiglioni, Matteo and Colomboni, Roberto and Marchesi, Alberto},
  booktitle={Advances in Neural Information Processing Systems},
  volume={38},
  year={2025},
  url={https://openreview.net/forum?id=tH8Q2TRSZf}
}

@misc{lunghi2025many,
  title={Online Two-Sided Markets: Many Buyers Enhance Learning},
  author={Lunghi, Anna and Castiglioni, Matteo and Marchesi, Alberto},
  year={2025},
  eprint={2503.01529},
  archiveprefix={arXiv}
}

@inproceedings{chen2025,
  title={Tight Regret Bounds for Fixed-Price Bilateral Trade},
  author={Chen, Houshuang and Jin, Yaonan and Lu, Pinyan and Zhang, Chihao},
  booktitle={International Colloquium on Automata, Languages, and Programming},
  series={LIPIcs},
  volume={374},
  pages={57},
  year={2026},
  eprint={2504.04349},
  archiveprefix={arXiv}
}

@misc{jin2026,
  title={Tight Regret Bounds for Bilateral Trade under Semi Feedback},
  author={Jin, Yaonan},
  year={2026},
  eprint={2601.16412},
  archiveprefix={arXiv}
}

@inproceedings{lunghi2026benchmark,
  title={A Stronger Benchmark for Online Bilateral Trade: From Fixed Prices to Distributions},
  author={Lunghi, Anna and Piccinato, Mattia and Castiglioni, Matteo and Marchesi, Alberto},
  booktitle={International Conference on Machine Learning},
  year={2026},
  note={Accepted},
  url={https://openreview.net/forum?id=M52jcbntdB}
}

@misc{lunghi2026perturbed,
  title={Regret Minimization in Bilateral Trade With Perturbed Markets},
  author={Lunghi, Anna and Castiglioni, Matteo and Marchesi, Alberto},
  year={2026},
  eprint={2605.10475},
  archiveprefix={arXiv}
}

@misc{lunghi2025,
  author={Lunghi, Anna and Castiglioni, Matteo and Marchesi, Alberto},
  title={Better Regret Rates in Bilateral Trade via Sublinear Budget Violation},
  year={2025},
  eprint={2507.11419},
  archiveprefix={arXiv},
  primaryclass={cs.GT},
  note={arXiv:2507.11419v1}
}

@inproceedings{deng2022approximately,
  title={Approximately Efficient Bilateral Trade},
  author={Deng, Yuan and Mao, Jieming and Sivan, Balasubramanian and Wang, Kangning},
  booktitle={Proceedings of the 54th Annual ACM SIGACT Symposium on Theory of Computing},
  pages={718--721},
  year={2022},
  doi={10.1145/3519935.3520054}
}

@book{lattimore2020bandit,
  title={Bandit Algorithms},
  author={Lattimore, Tor and Szepesv\'ari, Csaba},
  publisher={Cambridge University Press},
  year={2020}
}

@inproceedings{babaioff2024,
  author    = {Babaioff, Moshe and Frey, Amitai and Nisan, Noam},
  title     = {Learning to Maximize Gains From Trade in Small Markets},
  booktitle = {Proceedings of the 25th ACM Conference on Economics and Computation},
  pages     = {195},
  publisher = {ACM},
  year      = {2024},
  doi       = {10.1145/3670865.3673463},
  url       = {https://doi.org/10.1145/3670865.3673463}
}

@inproceedings{bachoc2024fair,
  author    = {Bachoc, Fran{\c{c}}ois and Cesa{-}Bianchi, Nicol{\`{o}} and
               Cesari, Tommaso and Colomboni, Roberto},
  title     = {Fair Online Bilateral Trade},
  booktitle = {Advances in Neural Information Processing Systems 37},
  pages     = {37241--37263},
  year      = {2024},
  url       = {https://proceedings.neurips.cc/paper_files/paper/2024/file/419ce7cb77aa90868c16882ce4257a69-Paper-Conference.pdf}
}

@inproceedings{bachoc2025parametric,
  author    = {Bachoc, Fran{\c{c}}ois and Cesari, Tommaso and Colomboni, Roberto},
  title     = {A Parametric Contextual Online Learning Theory of Brokerage},
  booktitle = {Proceedings of the 42nd International Conference on Machine Learning},
  series    = {Proceedings of Machine Learning Research},
  volume    = {267},
  pages     = {2293--2309},
  publisher = {PMLR},
  year      = {2025},
  url       = {https://proceedings.mlr.press/v267/bachoc25a.html}
}

@inproceedings{bachoc2025tight,
  author    = {Bachoc, Fran{\c{c}}ois and Cesari, Tommaso and Colomboni, Roberto},
  title     = {A Tight Regret Analysis of Non-Parametric Repeated Contextual Brokerage},
  booktitle = {Proceedings of the 28th International Conference on Artificial
               Intelligence and Statistics},
  series    = {Proceedings of Machine Learning Research},
  volume    = {258},
  pages     = {2836--2844},
  publisher = {PMLR},
  year      = {2025},
  url       = {https://proceedings.mlr.press/v258/bachoc25a.html}
}

@inproceedings{blumrosen2016,
  author    = {Blumrosen, Liad and Mizrahi, Yehonatan},
  title     = {Approximating Gains-from-Trade in Bilateral Trading},
  booktitle = {Web and Internet Economics -- 12th International Conference, {WINE} 2016},
  pages     = {400--413},
  publisher = {Springer},
  year      = {2016},
  doi       = {10.1007/978-3-662-54110-4_28},
  url       = {https://doi.org/10.1007/978-3-662-54110-4_28}
}

@inproceedings{bolic2024,
  author    = {Boli{\'c}, Natasa and Cesari, Tommaso and Colomboni, Roberto},
  title     = {An Online Learning Theory of Brokerage},
  booktitle = {Proceedings of the 23rd International Conference on Autonomous Agents
               and Multiagent Systems},
  pages     = {216--224},
  year      = {2024}
}

@inproceedings{brustle2017,
  author    = {Brustle, Johannes and Cai, Yang and Wu, Fa and Zhao, Mingfei},
  title     = {Approximating Gains from Trade in Two-sided Markets via Simple Mechanisms},
  booktitle = {Proceedings of the 2017 ACM Conference on Economics and Computation},
  pages     = {589--590},
  publisher = {ACM},
  year      = {2017},
  doi       = {10.1145/3033274.3085148},
  url       = {https://doi.org/10.1145/3033274.3085148}
}

@article{chatterjee1983,
  title={Bargaining under Incomplete Information},
  author={Chatterjee, Kalyan and Samuelson, William},
  journal={Operations Research},
  volume={31},
  number={5},
  pages={835--851},
  year={1983},
  doi={10.1287/opre.31.5.835}
}

@inproceedings{colinibaldeschi2017,
  author    = {Colini{-}Baldeschi, Riccardo and Goldberg, Paul W. and
               de Keijzer, Bart and Leonardi, Stefano and Turchetta, Stefano},
  title     = {Fixed Price Approximability of the Optimal Gain from Trade},
  booktitle = {Web and Internet Economics -- 13th International Conference, {WINE} 2017},
  pages     = {146--160},
  publisher = {Springer},
  year      = {2017},
  doi       = {10.1007/978-3-319-71924-5_11},
  url       = {https://doi.org/10.1007/978-3-319-71924-5_11}
}

@inproceedings{digregorio2025,
  author    = {Di Gregorio, Simone and D{\"u}tting, Paul and
               Fusco, Federico and Schwiegelshohn, Chris},
  title     = {Nearly Tight Regret Bounds for Profit Maximization in Bilateral Trade},
  booktitle = {2025 IEEE 66th Annual Symposium on Foundations of Computer Science},
  pages     = {1570--1594},
  publisher = {IEEE},
  year      = {2025},
  doi       = {10.1109/FOCS63196.2025.00083},
  url       = {https://doi.org/10.1109/FOCS63196.2025.00083}
}

@inproceedings{dutting2021,
  author    = {D{\"u}tting, Paul and Fusco, Federico and Lazos, Philip and
               Leonardi, Stefano and Reiffenh{\"a}user, Rebecca},
  title     = {Efficient Two-Sided Markets with Limited Information},
  booktitle = {Proceedings of the 53rd Annual ACM SIGACT Symposium on Theory of Computing},
  pages     = {1452--1465},
  publisher = {ACM},
  year      = {2021},
  doi       = {10.1145/3406325.3451076},
  url       = {https://doi.org/10.1145/3406325.3451076}
}

@article{feng2026,
  author        = {Feng, Yiding and Ma, Mengfan and Peng, Bo and Wan, Zongqi},
  title         = {Searching for Optimal Prices in Two-Sided Markets},
  journal       = {arXiv preprint arXiv:2602.11691},
  year          = {2026},
  eprint        = {2602.11691},
  archivePrefix = {arXiv},
  primaryClass  = {cs.GT},
  url           = {https://arxiv.org/abs/2602.11691}
}

@inproceedings{gaucher2025,
  author    = {Gaucher, Solenne and Bernasconi, Martino and Castiglioni, Matteo and
               Celli, Andrea and Perchet, Vianney},
  title     = {Feature-Based Online Bilateral Trade},
  booktitle = {The Thirteenth International Conference on Learning Representations},
  year      = {2025},
  url       = {https://openreview.net/forum?id=xnF2U0ro7b}
}

@article{hagerty1987,
  title={Robust Trading Mechanisms},
  author={Hagerty, Kathleen M. and Rogerson, William P.},
  journal={Journal of Economic Theory},
  volume={42},
  number={1},
  pages={94--107},
  year={1987},
  doi={10.1016/0022-0531(87)90104-9}
}

@inproceedings{kang2022,
  author    = {Kang, Zi Yang and Pernice, Francisco and Vondr{\'a}k, Jan},
  title     = {Fixed-Price Approximations in Bilateral Trade},
  booktitle = {Proceedings of the 2022 Annual ACM-SIAM Symposium on Discrete Algorithms},
  pages     = {2964--2985},
  publisher = {Society for Industrial and Applied Mathematics},
  year      = {2022},
  url       = {https://arxiv.org/abs/2107.14327}
}

@article{myerson1983,
  title={Efficient Mechanisms for Bilateral Trading},
  author={Myerson, Roger B. and Satterthwaite, Mark A.},
  journal={Journal of Economic Theory},
  volume={29},
  number={2},
  pages={265--281},
  year={1983},
  doi={10.1016/0022-0531(83)90048-0}
}

@misc{nobel2026,
  title={The {Prize in Economic Sciences} 2026},
  author={{Royal Swedish Academy of Sciences}},
  howpublished={Press release},
  month=oct,
  year={2026},
  url={https://www.kva.se/app/uploads/2026/10/prize-economic-sciences-2026-press-release_nbi239fjub.pdf}
}

@misc{airbnb2023form10k,
  title={Annual Report on {Form 10-K} for the Fiscal Year Ended December 31, 2023},
  author={{Airbnb, Inc.}},
  howpublished={Filed with the U.S. Securities and Exchange Commission},
  month=feb,
  year={2024},
  note={Filed February 16, 2024},
  url={https://www.sec.gov/Archives/edgar/data/1559720/000155972024000006/abnb-20231231.htm}
}

@misc{shopify2025form10k,
  author       = {{Shopify Inc.}},
  title        = {Annual Report on Form 10-K for the Fiscal Year Ended December 31, 2025},
  year         = {2026},
  month        = feb,
  howpublished = {U.S. Securities and Exchange Commission},
  url          = {https://www.sec.gov/Archives/edgar/data/1594805/000159480526000007/shop-20251231.htm}
}

@book{woodford2003interest,
  title={Interest and Prices: Foundations of a Theory of Monetary Policy},
  author={Woodford, Michael},
  publisher={Princeton University Press},
  year={2003},
  isbn={9780691010496},
  url={https://www.jstor.org/stable/j.ctv30pnvmf}
}

@article{blum2006market,
  author  = {Blum, Avrim and Sandholm, Tuomas and Zinkevich, Martin},
  title   = {Online Algorithms for Market Clearing},
  journal = {Journal of the ACM},
  volume  = {53},
  number  = {5},
  pages   = {845--879},
  year    = {2006},
  doi     = {10.1145/1183907.1183913},
  url     = {https://doi.org/10.1145/1183907.1183913}
}

@article{rubinstein1982,
  author  = {Rubinstein, Ariel},
  title   = {Perfect Equilibrium in a Bargaining Model},
  journal = {Econometrica},
  volume  = {50},
  number  = {1},
  pages   = {97--109},
  year    = {1982},
  doi     = {10.2307/1912531}
}

@article{athey2007efficiency,
  author  = {Athey, Susan and Miller, David A.},
  title   = {Efficiency in Repeated Trade with Hidden Valuations},
  journal = {Theoretical Economics},
  volume  = {2},
  number  = {3},
  pages   = {299--354},
  year    = {2007}
}

@article{athey2013efficient,
  author  = {Athey, Susan and Segal, Ilya},
  title   = {An Efficient Dynamic Mechanism},
  journal = {Econometrica},
  volume  = {81},
  number  = {6},
  pages   = {2463--2485},
  year    = {2013},
  doi     = {10.3982/ECTA6995}
}

@article{skrzypacz2015mechanisms,
  author  = {Skrzypacz, Andrzej and Toikka, Juuso},
  title   = {Mechanisms for Repeated Trade},
  journal = {American Economic Journal: Microeconomics},
  volume  = {7},
  number  = {4},
  pages   = {252--293},
  year    = {2015},
  doi     = {10.1257/mic.20140173}
}

@article{gittins1979bandit,
  author  = {Gittins, John C.},
  title   = {Bandit Processes and Dynamic Allocation Indices},
  journal = {Journal of the Royal Statistical Society: Series B (Methodological)},
  volume  = {41},
  number  = {2},
  pages   = {148--177},
  year    = {1979}
}

@inproceedings{amin2013learning,
  author    = {Amin, Kareem and Rostamizadeh, Afshin and Syed, Umar},
  title     = {Learning Prices for Repeated Auctions with Strategic Buyers},
  booktitle = {Advances in Neural Information Processing Systems},
  volume    = {26},
  pages     = {1169--1177},
  year      = {2013}
}

@book{puterman1994markov,
  author    = {Puterman, Martin L.},
  title     = {Markov Decision Processes: Discrete Stochastic Dynamic Programming},
  publisher = {John Wiley \& Sons},
  year      = {1994}
}

@book{bellman1957dynamic,
  author    = {Bellman, Richard},
  title     = {Dynamic Programming},
  publisher = {Princeton University Press},
  year      = {1957}
}

@book{howard1960dynamic,
  author    = {Howard, Ronald A.},
  title     = {Dynamic Programming and Markov Processes},
  publisher = {MIT Press},
  year      = {1960}
}

@article{blackwell1965discounted,
  author  = {Blackwell, David},
  title   = {Discounted Dynamic Programming},
  journal = {The Annals of Mathematical Statistics},
  volume  = {36},
  number  = {1},
  pages   = {226--235},
  year    = {1965},
  doi     = {10.1214/aoms/1177700285}
}

@article{whittle1980multi,
  author  = {Whittle, Peter},
  title   = {Multi-Armed Bandits and the Gittins Index},
  journal = {Journal of the Royal Statistical Society: Series B (Methodological)},
  volume  = {42},
  number  = {2},
  pages   = {143--149},
  year    = {1980}
}

@book{sutton2018reinforcement,
  author    = {Sutton, Richard S. and Barto, Andrew G.},
  title     = {Reinforcement Learning: An Introduction},
  edition   = {2},
  publisher = {MIT Press},
  year      = {2018}
}

@article{azar2013minimax,
  author  = {Azar, Mohammad Gheshlaghi and Munos, R{\'e}mi and Kappen, Hilbert J.},
  title   = {Minimax {PAC} Bounds on the Sample Complexity of Reinforcement Learning with a Generative Model},
  journal = {Machine Learning},
  volume  = {91},
  number  = {3},
  pages   = {325--349},
  year    = {2013},
  doi     = {10.1007/s10994-013-5368-1}
}

\newpage
\appendix
\crefalias{section}{appendix}
\crefalias{subsection}{appendix}
\section{Proofs and Algorithms for \texorpdfstring{Section~\ref*{sec:corr-upper}}{Section 4}}
\label{app:corr-upper}

\corrub*

\begin{proof}
Write $I=\tinterval{p,q}$.  Since $p\le q$, the event
$\set{s\le p,\ q\le b}$ is the intersection of $\set{s\le p\le b}$ and
$\set{s\le q\le b}$, so inclusion--exclusion gives
\begin{equation}
\label{eq:corr-ub-integrand}
\textstyle\GFT^{\scoremark}(I)
=\expect{(b-s)\indicator{\set{s\le p\le b}\cup\set{s\le q\le b}}
+(q-p)\indicator{p<s\le q,\ p\le b<q}}.
\end{equation}
Moreover,
\[
\textstyle\set{s\le q,\ p\le b}
=\bigl(\set{s\le p\le b}\cup\set{s\le q\le b}\bigr)
\sqcup\set{p<s\le q,\ p\le b<q},
\]
where $b-s\ge0$ on the union and $\card{b-s}\le q-p$ on the rectangle.

\medskip\noindent\textbf{(i) Upper bound.}
\par\nopagebreak\noindent
For $u,v\in I$,
\begin{align*}
\textstyle\expect{\GFT_{\ti}(u,v)}
&\textstyle\stepmark{a}{\le}\expect{(b-s)\indicator{\set{s\le p\le b}\cup\set{s\le q\le b}}
+(q-p)\indicator{p<s\le q,\ p\le b<q}}
\\*
&\textstyle=\GFT^{\scoremark}(I),
\tag*{\cref{eq:corr-ub-integrand}}
\end{align*}
where (a) holds pointwise by the decomposition, since
$\set{s\le u,\ v\le b}\subseteq\set{s\le q,\ p\le b}$.

\medskip\noindent\textbf{(ii) Monotonicity.}
\par\nopagebreak\noindent
For $\tinterval{u,v}\subseteq I$,
\begin{align*}
&\textstyle\GFT^{\scoremark}(\tinterval{u,v})
\\*
&\textstyle\qquad=\expect{(b-s)\indicator{\set{s\le u\le b}\cup\set{s\le v\le b}}
+(v-u)\indicator{u<s\le v,\ u\le b<v}}
\tag*{\cref{eq:corr-ub-integrand}}
\\*
&\textstyle\qquad\stepmark{b}{\le}\expect{(b-s)\indicator{\set{s\le p\le b}\cup\set{s\le q\le b}}
+(q-p)\indicator{p<s\le q,\ p\le b<q}}
\\*
&\textstyle\qquad=\GFT^{\scoremark}(I),
\tag*{\cref{eq:corr-ub-integrand}}
\end{align*}
where (b) holds pointwise by the decomposition, since $v-u\le q-p$,
$\set{s\le u\le b}\cup\set{s\le v\le b}\subseteq\set{s\le q,\ p\le b}$ and
$\set{u<s\le v,\ u\le b<v}\subseteq\set{p<s\le q,\ p\le b<q}$.

\medskip\noindent\textbf{(iii) Tightness.}
\par\nopagebreak\noindent
Since $\GFT_{\ti}(q,p)=(b-s)\indicator{s\le q,\ p\le b}$, the decomposition
and \cref{eq:corr-ub-integrand} give
\[
\textstyle\GFT^{\scoremark}(I)-\expect{\GFT_{\ti}(q,p)}
=\expect{(q-p-b+s)\indicator{p<s\le q,\ p\le b<q}}.
\]
Since $0\le q-p-b+s\le2(q-p)$ on this rectangle, this proves
\Cref{prop:corr-ub-cover}.
\end{proof}

\begin{algorithm}[tp]
\caption{$\fbest$, $\safepost$, $\gftest$ and $\gftplusest$}
\label{alg:corr-subroutines}
\begin{algorithmic}[1]
\linespread{0.9}\selectfont
    \Function{FB-Est}{$N$}
        \For{$\ti=1,\ldots,\tiH$}
            \State draw $\auxprice\sim\Unif\tinterval{0,1}$ and post $(\auxprice,\auxprice)$
            \If{$\sum_{j=1}^{\ti}\tradeindicator{j}=N$}
                \State \Return $\frac{N}{\ti}$
            \EndIf
        \EndFor
        \State \Return $0$
    \EndFunction
    \Statex {\color{gray}\rule[0.5ex]{\linewidth}{0.4pt}}
    \Function{Safe-Post}{$(p,q),\widehat\firstbest$}
        \While{$(1+\intrate)\account{\ti}<p-q$}
            \State draw $\auxprice\sim\Unif[0,1-\frac13\widehat\firstbest]$ and post
                $(\auxprice,\auxprice+\frac13\widehat\firstbest)$ \Comment{$\Tsave$}
                \label{algline:safepost-save}
        \EndWhile
        \State post $(p,q)$
        \State \Return $\tradeindicator{\ti}$
    \EndFunction
    \Statex {\color{gray}\rule[0.5ex]{\linewidth}{0.4pt}}
    \Function{GFT-Est}{$(p,q),N$}
        \State $\widehat{\GFT}(p,q)\gets0$
        \For{$i=1,\ldots,N$}
            \State draw $\auxprice\sim\Unif\tinterval{0,1}$
            \If{$\auxprice<p$}
                \State post $(\auxprice,q)$
            \ElsIf{$\auxprice\le q$}
                \State post $(p,q)$
            \Else
                \State post $(p,\auxprice)$
            \EndIf
            \State $\widehat{\GFT}(p,q)\mathrel{+}=\frac1N\tradeindicator{\ti}$
        \EndFor
        \State \Return $\widehat{\GFT}(p,q)$
    \EndFunction
    \Statex {\color{gray}\rule[0.5ex]{\linewidth}{0.4pt}}
    \Function{GFT$^{\scoremark}$-Est}{$[p,q],\widehat\firstbest,k$}
        \ForAll{$(u,v)\in\set{p,q}^2$}
            \State $\hprob{s\le u,\ b\ge v}\gets\frac1k\sum_{i=1}^k\safepost((u,v),\widehat\firstbest)$
        \EndFor
        \State \Return $\gftest((p,p),k)+\gftest((q,q),k)-\gftest((p,q),k)$
        \Statex \hspace{\algorithmicindent}\hspace{\algorithmicindent}\hspace{\algorithmicindent}$
            +(q-p)\,\hprob{p<s\le q,\ p\le b<q}$
    \EndFunction
\end{algorithmic}
\end{algorithm}

\begin{algorithm}[tp]
\caption{\textsc{Pair-Intervals}}
\label{alg:corr-pair-intervals}
\begin{algorithmic}[1]
\linespread{0.9}\selectfont
    \Function{Pair-Intervals}{$\mathcal S,\widehat\firstbest,k$}
        \State Sort $\mathcal S$ by left endpoint: $I_1,\ldots,I_n$.
        \State $L\gets\ceil{\frac{16}{\intrate\tiH}}+2$
        \State $\mathcal P\gets\set{(I_{(i-1)L+1},I_{iL}):1\le i\le \frac{n}{L},\
            \max I_{iL}-\min I_{(i-1)L+1}\le\widehat\firstbest}$
            \label{algline:corr-blocks}
        \ForAll{$(\tinterval{p_1,q_1},\tinterval{p_2,q_2})\in\mathcal P$}
            \State $\widehat{\GFT}(q_1,p_2)\gets\gftest((q_1,p_2),k)$
            \ForAll{$(u,v)\in\set{q_1,q_2}\times\set{p_1,p_2}$}
                \State $\hprob{s\le u,\ b\ge v}\gets\frac1k\sum_{i=1}^k\safepost((u,v),\widehat\firstbest)$
            \EndFor
            \State $\widehat{\GFT}(q_i,p_i)\gets\widehat{\GFT}{}^{\scoremark}(\tinterval{p_i,q_i})$
                for $i=1,2$
            \State $\widehat{\GFT}(q_2,p_1)\gets\widehat{\GFT}{}^{\scoremark}(\tinterval{p_1,q_1})
                +\widehat{\GFT}{}^{\scoremark}(\tinterval{p_2,q_2})-\widehat{\GFT}(q_1,p_2)$
            \Statex \hspace{\algorithmicindent}\hspace{\algorithmicindent}\hspace{\algorithmicindent}$
                -(q_2-p_1)\bigl(\hprob{q_1<s\le q_2,\ p_1\le b<p_2}+4\accscale\bigr)$
        \EndFor
        \State $\mathcal Q_E\gets\set{(q_1,p_2):(\tinterval{p_1,q_1},\tinterval{p_2,q_2})\in\mathcal P}$
        \State $\mathcal Q_S\gets\set{(q_1,p_1),(q_2,p_2),(q_2,p_1):(\tinterval{p_1,q_1},\tinterval{p_2,q_2})\in\mathcal P}$
        \State $\widehat{\Profit}(p,q)\gets(q-p)\,\hprob{s\le p,\ b\ge q}$
            for all $(p,q)\in\mathcal Q_E\cup\mathcal Q_S$
        \State \Return $(\mathcal Q_E,\mathcal Q_S,\widehat{\GFT}(\cdot),\widehat{\Profit}(\cdot))$
    \EndFunction
\end{algorithmic}
\end{algorithm}

\section{Proofs for \texorpdfstring{Section~\ref*{sec:lower}}{Section 5}}
\label{app:lower}

\subsection{Proofs for \texorpdfstring{Section~\ref*{sec:adv-lower}}{Section 5.1}}
\label{app:adv-lower}

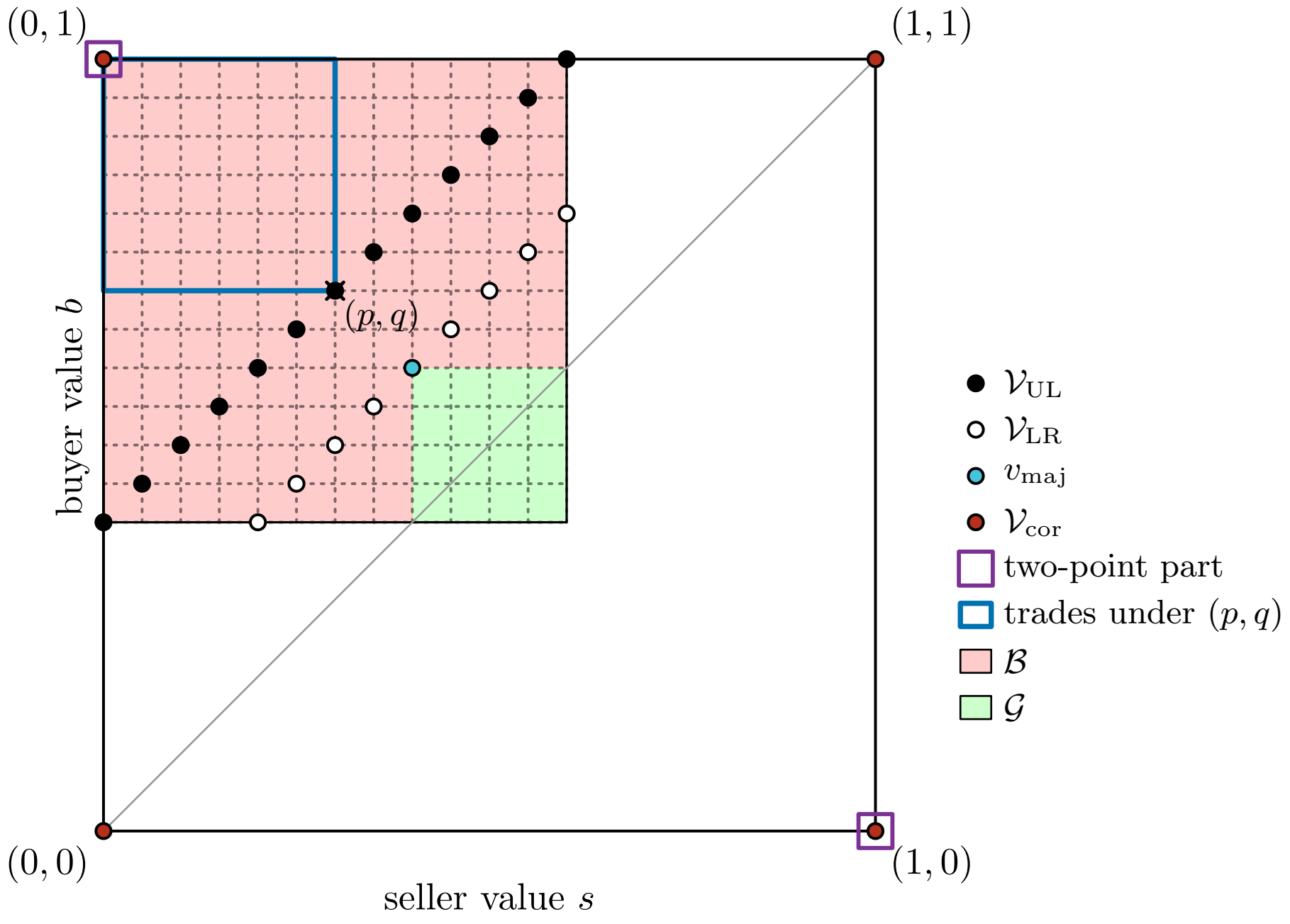
\begin{figure}[!t]
\centering
\begin{tikzpicture}[cor picture,font=\small]
\pgfmathtruncatemacro{\K}{4}
\pgfmathtruncatemacro{\Umax}{3*\K}
\pgfmathtruncatemacro{\Lmax}{2*\K}
\begin{scope}[x=7.2cm,y=7.2cm]
    \path[cor bad region] (0,.4) rectangle (.4,1);
    \path[cor bad region] (.4,.6) rectangle (.6,1);
    \path[cor good region] (.4,.4) rectangle (.6,.6);
    \draw[cor grid,step=.05] (0,.4) grid (.6,1);
    \draw[corblue,line width=1.4pt] (0,.7) rectangle (.3,1);
    \draw[cor axis,line width=.6pt] (0,.4) rectangle (.6,1);
    \draw[cor diagonal,line width=.5pt,draw=black!40] (0,0) -- (1,1);
    \draw[cor axis] (0,0) rectangle (1,1);

    \foreach \j in {0,...,\Umax} {
        \node[cor upper large] at ({\j/(5*\K)},{.4+\j/(5*\K)}) {};
    }
    \foreach \j in {0,...,\Lmax} {
        \ifnum\j=\K\relax
        \else
            \node[cor lower large] at ({.2+\j/(5*\K)},{.4+\j/(5*\K)}) {};
        \fi
    }
    \node[cor majority large] at (.4,.6) {};
    \foreach \c in {(0,0),(1,1),(0,1),(1,0)} {
        \node[cor corner large] at \c {};
    }
    \foreach \c in {(0,1),(1,0)} {
        \node[rectangle,draw=corviolet,line width=1.1pt,minimum size=9pt,inner sep=0] at \c {};
    }
    \draw[black,line width=.9pt] (.3-.011,.7-.011) -- (.3+.011,.7+.011)
                                 (.3-.011,.7+.011) -- (.3+.011,.7-.011);
    \node[cor legend,anchor=north west,inner sep=2pt] at (.3,.7) {$(p,q)$};

    \node[anchor=north east] at (0,0) {$(0,0)$};
    \node[anchor=south east] at (0,1) {$(0,1)$};
    \node[anchor=north west] at (1,0) {$(1,0)$};
    \node[anchor=south west] at (1,1) {$(1,1)$};
    \node[below=10pt] at (.5,0) {seller value $s$};
    \node[above=10pt,rotate=90] at (0,.5) {buyer value $b$};
\end{scope}

\begin{scope}[x=7.2cm,y=7.2cm]
    \foreach \sty/\lab [count=\i] in {
        cor upper large/\uppersupport,
        cor lower large/\lowersupport,
        cor majority large/\majorityatom,
        cor corner large/\cornersupport}{
        \node[\sty] at (1.13,{.64-.06*\i}) {};
        \node[cor legend,anchor=west] at (1.16,{.64-.06*\i}) {$\lab$};
    }
    \node[rectangle,draw=corviolet,line width=1.1pt,minimum size=9pt,inner sep=0] at (1.13,.34) {};
    \node[cor legend,anchor=west] at (1.16,.34) {two-point part};
    \draw[corblue,line width=1.4pt] (1.11,.265) rectangle (1.15,.295);
    \node[cor legend,anchor=west] at (1.16,.28) {trades under $(p,q)$};
    \path[cor bad region] (1.11,.205) rectangle (1.15,.235);
    \draw[black,line width=.5pt] (1.11,.205) rectangle (1.15,.235);
    \node[cor legend,anchor=west] at (1.16,.22) {$\badset$};
    \path[cor good region] (1.11,.145) rectangle (1.15,.175);
    \draw[black,line width=.5pt] (1.11,.145) rectangle (1.15,.175);
    \node[cor legend,anchor=west] at (1.16,.16) {$\goodset$};
\end{scope}
\end{tikzpicture}
\caption{The valuations of round $\ti$ in possibility $k$.}
\label{fig:adv-hard-instance}
\end{figure}

We modify the construction of \citet[Theorem~3]{chen2025}.  We refer the reader
to their Section~5 for its details, whose notation we follow, and describe
only the changes.  Their parameters are taken with the horizon $\tiH$ replaced
by $\widetilde T\deq\sum_{\ti}(1+\intrate)^{-\ti}$, that is,
$\chenfamilysize\deq\ceil{\widetilde T^{1/4}}$ and
$\chenperturb\deq\frac{0.1}{5\chenfamilysize+2}$.  Let $\refprob$ be the trade
probability of the action $(\frac12,\frac12)$, which is the same under every
$\mathcal D^k$.  In possibility $k$, the valuations are drawn independently
across rounds, in round $\ti$ from
\[
    \textstyle\mathcal D^k_{\ti}
    \deq(1+\intrate)^{-\ti}\mathcal D^k
    +\bigl(1-(1+\intrate)^{-\ti}\bigr)
    \bigl(\refprob\Dirac_{(0,1)}+(1-\refprob)\Dirac_{(1,0)}\bigr),
\]
as illustrated in \Cref{fig:adv-hard-instance} for $\chenfamilysize=4$.  As
in \Cref{sec:corr-lower}, the valuations of round $\ti$ are independent of
the history and of $(\sprice{\ti},\bprice{\ti})$.

Every $(p,q)\in\actiongrid$ has $p\le\frac35$ and $q\ge\frac25$, so under the
two-point part it trades exactly at $(0,1)$, whose gains-from-trade are $1$,
as does every fixed price; the blue rectangle of \Cref{fig:adv-hard-instance}
shows this for one action.  The two-point part therefore adds the same trade
probability and the same expected gains-from-trade to all of them: for every
$(p,q)\in\actiongrid\cup\set{(z,z):z\in\tinterval{0,1}}$,
\begin{align}
\textstyle\prob[\mathcal D^k_{\ti}]{s\le p,\ q\le b}
&\textstyle=(1+\intrate)^{-\ti}\prob[\mathcal D^k]{s\le p,\ q\le b}
  +\bigl(1-(1+\intrate)^{-\ti}\bigr)\refprob,
\label{eq:adv-lb-scaling}
\\
\textstyle\expect[\mathcal D^k_{\ti}]{(b-s)\indicator{s\le p,\ q\le b}}
&\textstyle=(1+\intrate)^{-\ti}\expect[\mathcal D^k]{(b-s)\indicator{s\le p,\ q\le b}}
  +\bigl(1-(1+\intrate)^{-\ti}\bigr)\refprob.
\label{eq:adv-lb-gft}
\end{align}
Subtracting \cref{eq:adv-lb-gft} at $(p,q)$ from its maximum over fixed
prices, the one-round regret of every $(p,q)\in\actiongrid$ satisfies
\begin{align}
\textstyle\reg_{\mathcal D^k_{\ti}}(p,q)
&\textstyle=\max_z\expect[\mathcal D^k_{\ti}]{(b-s)\indicator{s\le z\le b}}
  -\expect[\mathcal D^k_{\ti}]{(b-s)\indicator{s\le p,\ q\le b}}
\notag\\*
&\textstyle=(1+\intrate)^{-\ti}\bigl(\max_z\expect[\mathcal D^k]{(b-s)\indicator{s\le z\le b}}
  -\expect[\mathcal D^k]{(b-s)\indicator{s\le p,\ q\le b}}\bigr)
\notag\\*
&\textstyle=(1+\intrate)^{-\ti}\reg_{\mathcal D^k}(p,q).
\label{eq:adv-lb-regret}
\end{align}
By \cref{eq:adv-lb-scaling}, the budget constraint relaxes to a weighted
balance of spreads, which replaces their Lemma~13.

\begin{adversariallemma}
\label{lem:adv-weighted-price-balance}
\begin{flushleft}
Every feasible algorithm posting actions in $\actiongrid$ satisfies
$\mathbb E^k\bigl[\sum_{\ti}(1+\intrate)^{-\ti}(\bprice{\ti}-\sprice{\ti})\bigr]\ge0$
in every possibility $k$.
\end{flushleft}
\end{adversariallemma}

\begin{proof}
Since
$\account{\tiH+1}=\sum_{\ti}(1+\intrate)^{\tiH-\ti}(\bprice{\ti}-\sprice{\ti})\tradeindicator{\ti}$,
feasibility gives
$\sum_{\ti}(1+\intrate)^{-\ti}(\bprice{\ti}-\sprice{\ti})\tradeindicator{\ti}\ge0$.
By the proof of their Lemma~13, every $(p,q)\in\actiongrid$ satisfies
$(q-p)\bigl(\prob[\mathcal D^k]{s\le p,\ q\le b}-\refprob\bigr)\le0$.
Hence
\begin{align*}
0
&\textstyle\le\mathbb E^k\bigl[\sum_{\ti}(1+\intrate)^{-\ti}
  (\bprice{\ti}-\sprice{\ti})\tradeindicator{\ti}\bigr]
\\*
&\textstyle=\mathbb E^k\bigl[\sum_{\ti}(1+\intrate)^{-\ti}
  (\bprice{\ti}-\sprice{\ti})
  \prob[\mathcal D^k_{\ti}]{s\le \sprice{\ti},\ \bprice{\ti}\le b}\bigr]
\\*
&\textstyle=\mathbb E^k\bigl[\sum_{\ti}(1+\intrate)^{-\ti}
  (\bprice{\ti}-\sprice{\ti})\bigl(\refprob+(1+\intrate)^{-\ti}
  \bigl(\prob[\mathcal D^k]{s\le \sprice{\ti},\ \bprice{\ti}\le b}-\refprob\bigr)\bigr)\bigr]
\tag*{\cref{eq:adv-lb-scaling}}
\\*
&\textstyle\le\refprob\,\mathbb E^k\bigl[\sum_{\ti}(1+\intrate)^{-\ti}
  (\bprice{\ti}-\sprice{\ti})\bigr],
\end{align*}
and $\refprob>0$.
\end{proof}

For $S\subseteq\actiongrid$, let
$\widetilde T_S\deq\sum_{\ti}(1+\intrate)^{-\ti}\indicator{(\sprice{\ti},\bprice{\ti})\in S}$
be their count $T_S$ with round $\ti$ weighted by $(1+\intrate)^{-\ti}$, so
that $\widetilde T_{\actiongrid}=\widetilde T$.  Their Lemma~14 carries over to
these counts: for every algorithm posting actions in $\actiongrid$ and every
$k\in[\chenfamilysize]$,
\begin{equation}
\label{eq:adv-lb-change-of-measure}
    \textstyle\mathbb E^k\bigl[\widetilde T_{\goodrow{k}}\bigr]
    -\mathbb E^0\bigl[\widetilde T_{\goodrow{k}}\bigr]
    \le4\chenperturb\widetilde T\sqrt{\mathbb E^0\bigl[\widetilde T_{\infoset{k}}\bigr]},
\end{equation}
since the feedback law under $\mathcal D^k_{\ti}$ mixes that under
$\mathcal D^k$ with one that does not depend on $k$, so by the joint convexity
of relative entropy, the per-round divergences of their Lemma~16 acquire at
most the factor $(1+\intrate)^{-\ti}$.

\begin{proof}[Proof of \Cref{thm:adv-lower}]
The bound $\bomega{\tiH^{5/7}}$ follows from
\citet[Theorem~5.5]{bernasconi2024}: none of their Lemmas~5.6--5.12 uses the
budget constraint, and on their support the benchmark of their Lemma~5.6 is
attained by a fixed price.  It remains to show
$\bomega{\tiH^{3\min\set{\rateexponent,1}/4}}$ for $\rateexponent>\frac{20}{21}$,
where it is the larger of the two bounds; the two then give the three cases.

By their Lemma~12, which still preserves feasibility since
$\account{\tiH+1}=\sum_{\ti}(1+\intrate)^{\tiH-\ti}\Profit_{\ti}$ is
increasing in every $\Profit_{\ti}$, we may consider feasible algorithms that
post actions in $\actiongrid$.  By \cref{eq:adv-lb-gft}, a single fixed
price maximizes $\expect[\mathcal D^k_{\ti}]{(b-s)\indicator{s\le z\le b}}$ in
every round $\ti$, so the expected regret in possibility $k$ satisfies
\begin{align*}
\textstyle\mathbb E^k\Bigl[\tiH\cdot\fixedpricebenchmark
  -\sum_{\ti}\GFT_{\ti}(\sprice{\ti},\bprice{\ti})\Bigr]
&\textstyle\ge\max_z\sum_{\ti}\expect[\mathcal D^k_{\ti}]{(b-s)\indicator{s\le z\le b}}
  -\mathbb E^k\Bigl[\sum_{\ti}\GFT_{\ti}(\sprice{\ti},\bprice{\ti})\Bigr]
\\*
&\textstyle=\mathbb E^k\Bigl[\sum_{\ti}
  \reg_{\mathcal D^k_{\ti}}(\sprice{\ti},\bprice{\ti})\Bigr]
\\*
&\textstyle=\mathbb E^k\Bigl[\sum_{\ti}(1+\intrate)^{-\ti}
  \reg_{\mathcal D^k}(\sprice{\ti},\bprice{\ti})\Bigr],
\tag*{\cref{eq:adv-lb-regret}}
\end{align*}
which we denote by $\reg_{\mathcal D^k}$.  In possibility $0$, since
$\actiongrid=\goodset\cup\badset$,
\begin{align*}
\reg_{\mathcal D^0}
&\textstyle=\mathbb E^0\Bigl[\sum_{\ti}(1+\intrate)^{-\ti}
  \reg_{\mathcal D^0}(\sprice{\ti},\bprice{\ti})
  \bigl(\indicator{(\sprice{\ti},\bprice{\ti})\in\badset}
  +\indicator{(\sprice{\ti},\bprice{\ti})\in\goodset}\bigr)\Bigr]
\\*
&\textstyle\stepmark{a}{\ge}0.1\,\mathbb E^0\bigl[\widetilde T_{\badset}\bigr]
  +3\chenperturb\chenfamilysize\,\mathbb E^0\Bigl[\sum_{\ti}(1+\intrate)^{-\ti}
  (\bprice{\ti}-\sprice{\ti})\indicator{(\sprice{\ti},\bprice{\ti})\in\goodset}\Bigr]
\\*
&\textstyle\ge0.1\,\mathbb E^0\bigl[\widetilde T_{\badset}\bigr]
  -3\chenperturb\chenfamilysize\,\mathbb E^0\Bigl[\sum_{\ti}(1+\intrate)^{-\ti}
  (\bprice{\ti}-\sprice{\ti})\indicator{(\sprice{\ti},\bprice{\ti})\in\badset}\Bigr]
\tag*{\Cref{lem:adv-weighted-price-balance}}
\\*
&\textstyle\stepmark{b}{\ge}0.04\,\mathbb E^0\bigl[\widetilde T_{\badset}\bigr],
\end{align*}
where (a) uses their Proposition~3, and (b) uses $0<q-p\le1$ on $\badset$ and
$\chenperturb\chenfamilysize=\frac{0.1\chenfamilysize}{5\chenfamilysize+2}\le0.02$.
In possibility $k\in[\chenfamilysize]$, the same steps apply, and their
Proposition~3 adds $0.2\chenperturb$ in every round with an action in
$\goodset\setminus\goodrow{k}$.  Hence
\begin{align*}
\reg_{\mathcal D^k}
&\textstyle\ge0.04\,\mathbb E^k\bigl[\widetilde T_{\badset}\bigr]
  +0.2\chenperturb\,\mathbb E^k\bigl[\widetilde T_{\goodset\setminus\goodrow{k}}\bigr]
\\*
\mr{$\widetilde T_{\goodset\setminus\goodrow{k}}=\widetilde T-\widetilde T_{\badset}-\widetilde T_{\goodrow{k}}$}
&\textstyle=0.04\,\mathbb E^k\bigl[\widetilde T_{\badset}\bigr]
  +0.2\chenperturb\bigl(\widetilde T-\mathbb E^k\bigl[\widetilde T_{\badset}\bigr]
  -\mathbb E^k\bigl[\widetilde T_{\goodrow{k}}\bigr]\bigr)
\\*
\mr{$0.04\ge0.2\chenperturb$}
&\textstyle\ge0.2\chenperturb\bigl(\widetilde T-\mathbb E^k\bigl[\widetilde T_{\goodrow{k}}\bigr]\bigr)
\\*
&\textstyle\ge0.2\chenperturb\Bigl(\widetilde T-\mathbb E^0\bigl[\widetilde T_{\goodrow{k}}\bigr]
  -4\chenperturb\widetilde T\sqrt{\mathbb E^0\bigl[\widetilde T_{\infoset{k}}\bigr]}\Bigr).
\tag*{\cref{eq:adv-lb-change-of-measure}}
\end{align*}
The rows $\goodrow{k}$ are disjoint, and every action lies in at most two of
the sets $\infoset{k}$, so pathwise
$\sum_k\widetilde T_{\goodrow{k}}\le\widetilde T$ and
$\sum_k\widetilde T_{\infoset{k}}\le2\widetilde T_{\badset}$, as in
\cref{eq:corr-occupation}.  Averaging over $k\in[\chenfamilysize]$ therefore
gives
\begin{align*}
\textstyle\frac1{\chenfamilysize}\sum_k\reg_{\mathcal D^k}
&\textstyle\ge0.2\chenperturb\Bigl(\widetilde T
  -\frac1{\chenfamilysize}\sum_k\mathbb E^0\bigl[\widetilde T_{\goodrow{k}}\bigr]
  -\frac{4\chenperturb\widetilde T}{\chenfamilysize}\sum_k
  \sqrt{\mathbb E^0\bigl[\widetilde T_{\infoset{k}}\bigr]}\Bigr)
\\*
\mr{$\sum_k\widetilde T_{\goodrow{k}}\le\widetilde T$}
&\textstyle\ge0.2\chenperturb\Bigl(\widetilde T-\frac{\widetilde T}{\chenfamilysize}
  -\frac{4\chenperturb\widetilde T}{\chenfamilysize}\sum_k
  \sqrt{\mathbb E^0\bigl[\widetilde T_{\infoset{k}}\bigr]}\Bigr)
\\*
\mr{Cauchy--Schwarz inequality}
&\textstyle\ge0.2\chenperturb\Bigl(\widetilde T-\frac{\widetilde T}{\chenfamilysize}
  -4\chenperturb\widetilde T\sqrt{\tfrac1{\chenfamilysize}\sum_k
  \mathbb E^0\bigl[\widetilde T_{\infoset{k}}\bigr]}\Bigr)
\\*
\mr{$\sum_k\widetilde T_{\infoset{k}}\le2\widetilde T_{\badset}$}
&\textstyle\ge0.2\chenperturb\widetilde T\Bigl(1-\frac1{\chenfamilysize}
  -4\chenperturb\sqrt{\tfrac2{\chenfamilysize}\mathbb E^0\bigl[\widetilde T_{\badset}\bigr]}\Bigr).
\end{align*}
Combining the two bounds with $\chenfamilysize=\ceil{\widetilde T^{1/4}}$ and
$\chenperturb=\frac{0.1}{5\chenfamilysize+2}$, it is easy to verify through
elementary algebra that
\[
    \textstyle\max\set{\reg_{\mathcal D^0},\
    \frac1{\chenfamilysize}\sum_k\reg_{\mathcal D^k}}
    \ge\max\set{0.04\,\mathbb E^0\bigl[\widetilde T_{\badset}\bigr],\
    0.2\chenperturb\widetilde T\Bigl(1-\frac1{\chenfamilysize}
    -4\chenperturb\sqrt{\tfrac2{\chenfamilysize}\mathbb E^0\bigl[\widetilde T_{\badset}\bigr]}\Bigr)}
    =\bomega{\widetilde T^{3/4}}.
\]
Since $\widetilde T=\btheta{\tiH^{\min\set{\rateexponent,1}}}$, this is
$\bomega{\tiH^{3\min\set{\rateexponent,1}/4}}$ for
$\rateexponent>\frac{20}{21}$, which proves \Cref{thm:adv-lower}.
\end{proof}

\subsection{Proofs for \texorpdfstring{Section~\ref*{sec:corr-lower}}{Section 5.2}}
\label{app:corr-iid}

\corrprojection*

\begin{proof}
We transform $\learner$ into $\learner'$ explicitly.  In every round,
$\learner'$ takes the action $(p,q)$ chosen by $\learner$, posts $(p',q')$,
which lies in $\actiongrid$ by construction, and feeds $\learner$ the two-bit
feedback
\[
    \bigl(\max\set{\indicator{s\le p'},\,\indicator{p=1}},\
    \max\set{\indicator{q'\le b},\,\indicator{q=0}}\bigr).
\]

Apart from the corners, every support point lies on the multiples of
$\frac1{5\chenfamilysize}$ in $\tinterval{0,\frac35}\times\tinterval{\frac25,1}$,
so on the same realization $(s,b)$, the actions $(p,q)$ and $(p',q')$ induce
the same trade outcome except at the corners $(0,0)$, $(1,0)$ and $(1,1)$.
At each of these three points $s\ge b$, and a trade under $(p,q)$ requires
$p=1$ or $q=0$, so both its gains-from-trade $b-s$ and its profit $q-p$ are
nonpositive.  At every other point, a trade has the same gains-from-trade
under both actions and a profit $q'-p'\ge q-p$, since $p'\le p$ and
$q'\ge q$.

Moreover, the feedback above identically simulates the two-bit feedback
$\indicator{s\le p}$ and $\indicator{q\le b}$ of $\learner$ on the same
realization.  A coupling argument then gives, in every round,
\[
    \GFT_{\ti}(p',q')\ge\GFT_{\ti}(p,q),
    \qquad
    \Profit_{\ti}(p',q')\ge\Profit_{\ti}(p,q).
\]
Since $\account{\tiH+1}=\sum_{\ti}(1+\intrate)^{\tiH-\ti}\Profit_{\ti}$ is
increasing in every $\Profit_{\ti}$, $\learner'$ is feasible whenever
$\learner$ is, and since both face the same benchmark
$\fixedpricebenchmark$, its regret is no larger.
\end{proof}

\corrgaps*

\begin{proof}
\leavevmode
\begin{itemize}[leftmargin=*]
\item \emph{For $(p,q)\in\goodset$.}  Write
  $(p,q)=\bigl(\frac25+\frac{i}{5\chenfamilysize},
  \frac25+\frac{j}{5\chenfamilysize}\bigr)$ with $0\le i,j\le\chenfamilysize$,
  so that $p-q=\frac{i-j}{5\chenfamilysize}$.  The action trades
  $\uppersupport[r]$ for $j\le r\le2\chenfamilysize+i$, $\lowersupport[r]$
  for $j\le r\le\chenfamilysize+i$, which include
  $\majorityatom=\lowersupport[\chenfamilysize]$, the corner $(0,1)$, and
  $\revsupport[r]$ for $j\le r\le i-\revspanindex$, whose surpluses are
  $\frac25$, $\frac15$, $1$ and $-\frac{\revspanindex}{5\chenfamilysize}$.
  Hence, substituting $\mathcal D^0(\majorityatom)$,
  \begin{align*}
  \textstyle\expect[\mathcal D^0]{\GFT_{\ti}(p,q)}
  &\textstyle=\frac25\chenperturb(2\chenfamilysize+i-j+1)
    +\frac15\chenperturb(\chenfamilysize+i-j)
    +\frac15\mathcal D^0(\majorityatom)
  \\*
  &\textstyle\quad+\frac1{10}
    -\frac{\revspanindex}{5\chenfamilysize}\cdot\frac{\revmass}{\revspanindex}
    \plus{i-j-\revspanindex+1}
  \\*
  &\textstyle=\frac{11}{50}+\frac{\chenperturb}5
    -\frac{\chenfamilysize-\revspanindex+1}{5\revspanindex}\revmass
    +3\chenperturb\chenfamilysize(p-q)
    -\revmass\plus{p-q-\frac{\revspanindex-1}{5\chenfamilysize}}.
  \end{align*}
  In possibility $k\ge1$, $\mathcal D^k$ moves mass $\chenperturb$ from
  $\uppersupport[k-1]$ and $\lowersupport[k]$ to $\uppersupport[k]$ and
  $\lowersupport[k-1]$, which changes these gains-from-trade by
  \[
      \textstyle\chenperturb\bigl(\frac25\indicator{j\le k}-\frac25\indicator{j\le k-1}
      +\frac15\indicator{j\le k-1}-\frac15\indicator{j\le k}\bigr)
      =\frac{\chenperturb}5\indicator{j=k}.
  \]
  Hence, in every possibility $k$,
  \begin{align}
  \textstyle\expect[\mathcal D^k]{\GFT_{\ti}(p,q)}
  &\textstyle=\frac{11}{50}+\frac{\chenperturb}5
    -\frac{\chenfamilysize-\revspanindex+1}{5\revspanindex}\revmass
    +3\chenperturb\chenfamilysize(p-q)
  \notag\\*
  &\textstyle\quad-\revmass\plus{p-q-\frac{\revspanindex-1}{5\chenfamilysize}}
    +\frac{\chenperturb}5\indicator{k\ge1,\ j=k}.
  \label{eq:corr-gaps-good}
  \end{align}
\item \emph{For $(p,q)\in\badset$.}  The action, like a fixed price outside
  $\tinterval{\frac25,\frac35}$, misses $\majorityatom=(\frac25,\frac35)$, so
  it trades at most the corner $(0,1)$, the two chains and points with
  $b\le s$.  Hence, in every possibility $k$,
  \begin{align}
  \textstyle\expect[\mathcal D^k]{\GFT_{\ti}(p,q)}
  &\textstyle\le\frac1{10}+\frac25(3\chenfamilysize+1)\chenperturb
    +\frac15(2\chenfamilysize+1)\chenperturb
  \notag\\*
  \mathllap{\mr{$\chenperturb\chenfamilysize\le\frac1{100}$, $\chenperturb\le\frac1{800}$}}
  &\textstyle\le\frac1{10}+\frac85\cdot\frac1{100}+\frac35\cdot\frac1{800}
  <\frac3{25}.
  \label{eq:corr-gaps-bad}
  \end{align}
\end{itemize}
Substituting $p=q$ into \cref{eq:corr-gaps-good} gives
\begin{equation}
\label{eq:corr-gaps-best}
    \textstyle\max_z\expect[\mathcal D^k]{\GFT_{\ti}(z,z)}
    =\frac{11}{50}+\frac{\chenperturb}5
    -\frac{\chenfamilysize-\revspanindex+1}{5\revspanindex}\revmass
    +\frac{\chenperturb}5\indicator{k\ge1},
\end{equation}
since a fixed price between two grid points trades the same points as a good
action with $q-p=\frac1{5\chenfamilysize}$, which earns less, and one outside
$\tinterval{\frac25,\frac35}$ earns less than $\frac3{25}$ by
\cref{eq:corr-gaps-bad}, below the right side.
Subtracting \cref{eq:corr-gaps-good} from \cref{eq:corr-gaps-best} gives, for
$(p,q)\in\goodset$,
\begin{align*}
\textstyle\reg_{\mathcal D^k}(p,q)
&\textstyle=3\chenperturb\chenfamilysize(q-p)
  +\revmass\plus{p-q-\frac{\revspanindex-1}{5\chenfamilysize}}
  +\frac{\chenperturb}5\bigl(\indicator{k\ge1}-\indicator{k\ge1,\ j=k}\bigr)
\\*
&\textstyle=3\chenperturb\chenfamilysize(q-p)
  +\revmass\plus{p-q-\frac{\revspanindex-1}{5\chenfamilysize}}
  +\frac{\chenperturb}5\indicator{k\ge1,\ (p,q)\notin\goodrow{k}},
\end{align*}
and for $(p,q)\in\badset$,
\begin{align*}
\textstyle\reg_{\mathcal D^k}(p,q)
&\textstyle\ge\frac{11}{50}+\frac{\chenperturb}5
  -\frac{\chenfamilysize-\revspanindex+1}{5\revspanindex}\revmass-\frac3{25}
\tag*{\cref{eq:corr-gaps-bad,eq:corr-gaps-best}}
\\*
\mathllap{\mr{$\mathcal D^0(\revsupport)\le\frac1{10}$}}
&\textstyle\ge\frac{11}{50}-\frac1{50}-\frac3{25}=\frac2{25}.
\end{align*}
\end{proof}

\section{Algorithms for \texorpdfstring{Section~\ref*{sec:alpha}}{Section 6}}
\label{app:alpha}

\begin{algorithm}[H]
\caption{$\gftest$}
\label{alg:alpha-gft-est}
\begin{algorithmic}[1]
\linespread{0.9}\selectfont
    \Function{GFT-Est}{$(p,q)$}
        \State draw $D\sim\Unif\set{0,1,2}$
        \If{$D=0$}
            \State draw $\auxprice\sim\Unif\tinterval{0,p}$ and post $(\auxprice,q)$
            \State $\widehat{\GFT}(p,q)\gets3p\,\tradeindicator{\ti}$
        \ElsIf{$D=1$}
            \State draw $\auxprice\sim\Unif\tinterval{q,1}$ and post $(p,\auxprice)$
            \State $\widehat{\GFT}(p,q)\gets3(1-q)\tradeindicator{\ti}$
        \Else
            \State post $(p,q)$
            \State $\widehat{\GFT}(p,q)\gets3(q-p)\tradeindicator{\ti}$
        \EndIf
        \State \Return $\widehat{\GFT}(p,q)$
    \EndFunction
\end{algorithmic}
\end{algorithm}

\end{document}